\documentclass[twocolumn,showpacs,superscriptaddress,prb,aps,floatfix]{revtex4-2}
\usepackage{graphicx}
\usepackage{rotating}
\usepackage{amsmath}
\usepackage[usenames,dvipsnames]{color}
\usepackage{soul}
\usepackage{float}
\usepackage{rotating}
\usepackage{booktabs}
\usepackage{multirow}
\usepackage{longtable}
\usepackage[bookmarks=false,colorlinks]{hyperref}
\hypersetup{
	linkcolor=blue, 
	citecolor=magenta, 
	filecolor=black, 
	urlcolor=black,  
}
\usepackage[T1]{fontenc}
\usepackage{bm}
\usepackage{upgreek}
\usepackage{comment}
\usepackage{amssymb}
\usepackage[normalem]{ulem} 
\makeatletter

\newcommand{\Rmnum}[1]{\expandafter\@slowromancap\romannumeral #1@}

\newcommand{\DC}[1]{\textcolor{red}{To Domenico and Cesare:}}

\hfuzz=\maxdimen
\begin{document}

\title{Stacking effects on magnetic, vibrational, and optical properties of CrSBr bilayers}

\author{Huicong Li}
\affiliation{Beijing Key Laboratory for Magneto-Photoelectrical Composite and Interface Science, The State Key Laboratory for Advanced Metals and Materials, School of Mathematics and Physics, University of Science and Technology Beijing, Beijing 100083, China}

\author{Yali Yang}
\email{ylyang@ustb.edu.cn}
\affiliation{Beijing Key Laboratory for Magneto-Photoelectrical Composite and Interface Science, The State Key Laboratory for Advanced Metals and Materials, School of Mathematics and Physics, University of Science and Technology Beijing, Beijing 100083, China}

\author{Zhonghao Xia}
\affiliation{Beijing Key Laboratory for Magneto-Photoelectrical Composite and Interface Science, The State Key Laboratory for Advanced Metals and Materials, School of Mathematics and Physics, University of Science and Technology Beijing, Beijing 100083, China}

\author{Yateng Wang}
\affiliation{Beijing Key Laboratory for Magneto-Photoelectrical Composite and Interface Science, The State Key Laboratory for Advanced Metals and Materials, School of Mathematics and Physics, University of Science and Technology Beijing, Beijing 100083, China}

\author{Jiacheng Wei}
\affiliation{Beijing Key Laboratory for Magneto-Photoelectrical Composite and Interface Science, The State Key Laboratory for Advanced Metals and Materials, School of Mathematics and Physics, University of Science and Technology Beijing, Beijing 100083, China}

\author{Jiangang He}
\email{jghe2021@ustb.edu.cn}
\affiliation{Beijing Key Laboratory for Magneto-Photoelectrical Composite and Interface Science, The State Key Laboratory for Advanced Metals and Materials, School of Mathematics and Physics, University of Science and Technology Beijing, Beijing 100083, China}

\author{Rongming Wang}
\email{rmwang@ustb.edu.cn}
\affiliation{Beijing Key Laboratory for Magneto-Photoelectrical Composite and Interface Science, The State Key Laboratory for Advanced Metals and Materials, School of Mathematics and Physics, University of Science and Technology Beijing, Beijing 100083, China}

\date{\today}
	
	
\begin{abstract}	
Layer stacking presents a promising avenue for manipulating the physical properties of two-dimensional materials. The van der Waals layered semiconductor CrSBr, which exhibits A-type antiferromagnetism and a relatively high N\'{e}el temperature, has been successfully exfoliated into atomically thin sheets. In this study, we investigate the structural, lattice dynamical, electronic, magnetic, and optical properties of four distinct stacking structures of CrSBr bilayers using first-principles calculations and Monte Carlo simulations. Our findings show that though the most energetically favorable bilayer structure retains the stacking pattern of the bulk counterpart, three other high-symmetry stacking structures can be achieved by sliding one of the layers along three distinct directions, with energy costs comparable to that observed in MoS$_2$ bilayer. All these four bilayers exhibit semiconductor behavior with A-type antiferromagnetic ordering, similar to the bulk material, and demonstrate closely aligned N\'{e}el temperatures. Moreover, these bilayers exhibit relatively low lattice thermal conductivities, pronounced anisotropy, and a strong dependence on stacking patterns. This behavior is attributed to significant phonon-phonon scattering arising from avoided crossings between acoustic and optical phonons, as well as the presence of flat optical phonon bands in the low-frequency region. While the electronic structures and optical properties of these bilayers show weak dependence on the stacking pattern for antiferromagnetic ordering, they undergo significant changes for ferromagnetic ordering, influencing the band gap, valence and conduction band splitting, and effective mass. Furthermore, we found that antiferromagnetic ordering can transition to ferromagnetic under intense visible light illumination. Thus, the integration of layer stacking and visible light illumination offers an effective means to control the heat transfer, magnetic, and optical properties of CrSBr bilayers.
\end{abstract}

\maketitle

\section{Introduction}
Two-dimensional (2D) semiconductors have garnered significant attention across various research fields due to their atomically thin structure and exceptional properties~\cite{a1,a2}, which empower promising applications such as transistors, sensors, optoelectronics, energy storage, biomedical applications, and etc. Notably, the ease of exfoliation and the absence of dangling bonds on the surface facilitate the exploration of diverse stacking patterns in few-layer van der Waals (vdWs) layered materials, thereby enabling the emergence of novel properties in known materials~\cite{a3,a4,a5,a6,PhysRevB.89.075409,a17,a28,a29,a55,doi:10.1021/acs.nanolett.8b03321,wu2019intrinsic,doi:10.1021/acs.chemrev.3c00618}. For instance, superconductivity has been observed in the twisted bilayer graphene, with a critical temperature reaching up to 1.7 K -- a phenomenon not reported in the monolayer graphene~\cite{a3}. Parallel-stacked bilayers of boron nitride (BN) exhibit ferroelectric properties, despite the fact that bulk BN does not possess ferroelectricity at all~\cite{doi:10.1126/science.abd3230}. The effective mobility of the 3R stacked WS$_2$ bilayer is 65 \% higher than that of 2H one~\cite{mobilityWS2}. The interlayer exchange interaction between antiferromagnetic (AFM) and ferromagnetic (FM) in Cr$X_3$ ($X$ = Cl, Br, and I) bilayer can be tuned by changing the interlayer stacking order~\cite{a17,doi:10.1021/acs.nanolett.8b03321,doi:10.1021/acs.nanolett.0c04794,doi:10.1021/jacs.3c10777}.

Although they are rare, 2D magnetic semiconductors represent promising candidates for quantum technology applications due to their unique combination of magnetic ordering and semiconducting properties~\cite{a7,a8,https://doi.org/10.1002/andp.201900452}. In particular, intrinsic 2D magnetic semiconductors with appropriate bandgaps and high carrier mobility hold significant potential for manipulating charge and spin carriers in electronic devices~\cite{a9}. This advancement is anticipated to enhance the operating speed, integration density, and energy efficiency of electronic devices by leveraging spin transport, thereby paving the way for the next generation of spintronic nanodevices~\cite{a10,a11}. However, the scarcity of 2D magnetic semiconductors is primarily attributed to the inherent challenge of maintaining long-range magnetic ordering as the thickness of materials is reduced~\cite{a12,a13,a14,a15}. Over past decades, there have been rapid advancements in 2D magnetic materials, although they have different stability, conductivity and magnetic transition temperature. The successful synthesis of monolayer or few-layer Cr$X_3$ ($X$ = Cl, Br, and I)\cite{a17,ghazaryan2018magnon,doi:10.1021/acs.nanolett.9b01317}, Cr$_2$Ge$_2$Te$_6$\cite{a16}, $M$PS$_3$ ($M$ = Mn, Fe, Co, and Ni)\cite{10.1038/srep20904,10.1021/acs.nanolett.6b03052}, CrSe$_2$\cite{10.1038/s41563-021-00927-2}, and CrTe$_2$\cite{WOS:000656508100006,WOS:000617499600011,WOS:000741852200021}, among others, has demonstrated the feasibility of achieving stable long-range magnetic order in 2D materials, thereby overcoming the limitations imposed by the Mermin-Wagner theorem\cite{a15}. The presence of uniaxial magnetic anisotropy introduces a spin wave excitation gap, which mitigates thermal fluctuations of the magnons, consequently facilitating the emergence of long-range magnetic ordering~\cite{a18,a19,a20}.

The vdWs layered AFM semiconductor CrSBr has recently been discovered to exhibit a relatively high N\'{e}el temperature ($T_{\rm N}$ $\sim$ 140 K) even in few-layer samples~\cite{a25,a9,doi:10.1021/acsnano.2c07316}, and a high Curie temperature ($T_{\rm C}$ $\sim$ 146 K) in the monolayer structure~\cite{a25,a30,a46,tcmono}. This characteristic has garnered significant research interest due to its magnetic field-controlled exciton effect, negative magnetoresistance, and pronounced magnetic proximity effect~\cite{a9,a23,a24,a25}. In CrSBr, the intralayer coupling is FM, while the interlayer coupling is AFM, with the easy magnetic axis oriented along the $b$ axis~\cite{a25,a26}. The air stability and magnetic semiconductor properties make CrSBr an ideal candidate for exploring and manipulating interlayer magnetic coupling~\cite{a11}. For instance, the magnetoresistance of multilayer CrSBr can be tuned by applying a magnetic field along the $a$ axis~\cite{a27}. Furthermore, CrSBr exhibits a high degree of optical anisotropy and strong coupling between its magnetic ordering and photoluminescence characteristics~\cite{a49}. These unique properties suggest that CrSBr-based heterostructures hold great promise for applications in ultra-compact spintronic devices~\cite{avsar2022highly}. However, the effects of interlayer stacking on the electronic structures, magnetic interactions, and optical properties of CrSBr bilayers have not been systematically investigated to date. Specifically, the consideration of different stacking modes in CrSBr bilayer structures is limited, and the impact of stacking on the heat transportation, magnetic and electronic properties remains unclear~\cite{a30}.

In this work, we have constructed three additional bilayer structures with distinct stacking patterns, in addition to the bilayer derived directly from the CrSBr bulk structure. These configurations are achieved by sliding one of the layers along the [100], [010], and [110] directions of the bilayer, respectively. Through first-principles calculations, we demonstrate that both the AFM ground state and the semiconducting characteristics are maintained across the various CrSBr bilayer structures. Our findings reveal that the phonon spectrum, heat transport properties, electronic structures, and both magnetic and optical characteristics exhibit clear dependence on the stacking structures. The lattice thermal conductivities and optical absorption display significant anisotropy. Furthermore, we show that under appropriate light absorption conditions, the AFM state of the CrSBr bilayer structures can be transformed into a FM state. Therefore, both stacking configurations and visible light illumination can be utilized effectively to manipulate the properties of CrSBr bilayers.

\section{Computational method}
All the DFT calculations presented in this work were conducted using the Vienna \textit{Ab initio} Simulation Package (VASP)~\cite{a31,a32,a33}. We employed the projector augmented wave (PAW) method~\cite{a34,PAW2}, the optimized B86b functional (optB86b)~\cite{a36}, and plane wave basis sets with an energy cutoff of 450 eV. To address the on-site Coulomb interaction of Cr 3$d$ electrons in CrSBr, we utilized the rotationally invariant DFT+U approach introduced by Liechtenstein \textit{et al.}~\cite{a37,a38}, with U = 2.0 eV and J = 1.0 eV. This parameterization has been previously validated to accurately reproduce both the A-type AFM state and the easy axis of magnetization~\cite{a39,a27,a30,a51}. In alignment with previous findings~\cite{a27}, our calculations indicate that a large value of U$_{\rm eff}$ (e.g., U$_{\rm eff}$ $\geqslant$ 2 eV) results in a FM ground state, as detailed in Table~\textcolor{red}{S1} of the supplementary materials, which contrasts with the experimental observation of AFM configuration. All structures were fully relaxed, ensuring that the forces on each atom were less than 0.01 eV/\AA. The total energies of the relaxed structures were calculated using a stringent convergence criterion for the energy difference between consecutive steps (10$^{-8}$ eV). During the structure optimization of the bilayer structures, the in-plane lattice parameters and atomic positions were allowed to relax freely, while the out-of-plane lattice parameter was fixed to 33 \AA, providing more than 15 \AA\ of vacuum space to mitigate spurious interactions between adjacent bilayers in the out-of-plane direction. A $\Gamma$-centered $k$-point mesh of 14 $\times$ 11 $\times$ 1 was employed to sample the Brillouin zone for the primitive cells (12 atoms). The climbing nudged elastic band (cNEB) calculations~\cite{cNEB1,cNEB2} were carried out to explore the potential transition pathways and the associated energy changes between the origin stacking structure of the bulk and the other three bilayer ones.

The second-order force constants were calculated using the finite displacement method as implemented in the Phonopy code~\cite{TOGO20151} with 6 $\times$ 5 $\times$ 1 supercell (360 atoms) and 1 $\times$ 1 $\times$ 1 $\Gamma$-centered $k$-points grid, and a 0.01 \AA\ atomic displacement. The third-order and fourth-order force constants were extracted using the compressive sensing lattice dynamics (CSLD)~\cite{PhysRevLett.113.185501,PhysRevB.100.184308,PhysRevB.100.184309}. Phonon frequency shifts at finite temperature were calculated using the Self-consistent Phonon (SCPH) theory~\cite{doi:10.1080/14786440408520575,PhysRevLett.17.753,PhysRevB.1.572}. The Peierls-Boltzmann transport equation was literately solved by sampling with a uniform 54 $\times$ 45 $\times$ 1 $q$-point mesh for $\kappa_{\rm L}$ calculations with SCPH and three and four-phonon scattering ($\kappa_{\rm 3,4ph}^{\rm SCPH}$) method~\cite{PhysRevX.10.041029}. The four-phonon scattering process is accelerated using the sampling method~\cite{guo2024sampling}. All other parameters remain consistent with the previous study~\cite{PhysRevX.10.041029}.

Frequency-dependent dielectric functions are computed using the independent-particle approximation (IPA). The optical absorption spectrum is derived from the real and imaginary components of the dielectric functions, as expressed by the formula:
$$
\begin{aligned}
	\alpha(\omega) = \omega \frac{\sqrt{2(\sqrt{(\epsilon_1(\omega))^2 + (\epsilon_2(\omega))^2} - \epsilon_2(\omega)})}{c}
\end{aligned}
$$
where $\omega$ denotes the photon frequency, $\epsilon_1(\omega)$ and $\epsilon_2(\omega)$ represent the real and imaginary parts of the dielectric function, respectively, and $c$ is the speed of light.

The photoexcited carriers under intense light illumination, with energy exceeding the bandgap of CrSBr bilayers, are simulated by promoting electrons from high-energy valence band states to low-energy conduction band states. This is achieved by modifying the occupation numbers of the valence band maximum and the conduction band minima~\cite{PhysRevLett.116.247401}. The bilayer crystal structures were fully relaxed while maintaining the occupation numbers and the fixed the out-of-plane lattice parameter.

The parallel tempering Monte Carlo (PTMC) method~\cite{MC1,MC2,PASP} was employed to simulate the temperature-dependent magnetic phase transition of CrSBr bilayer structures. A total of 288 replicas were utilized, and a 16$\times$16$\times$1 supercell was adopted for the simulations.

\begin{figure*}[htbp!]
	\centering
	\includegraphics[clip,width=0.98\linewidth]{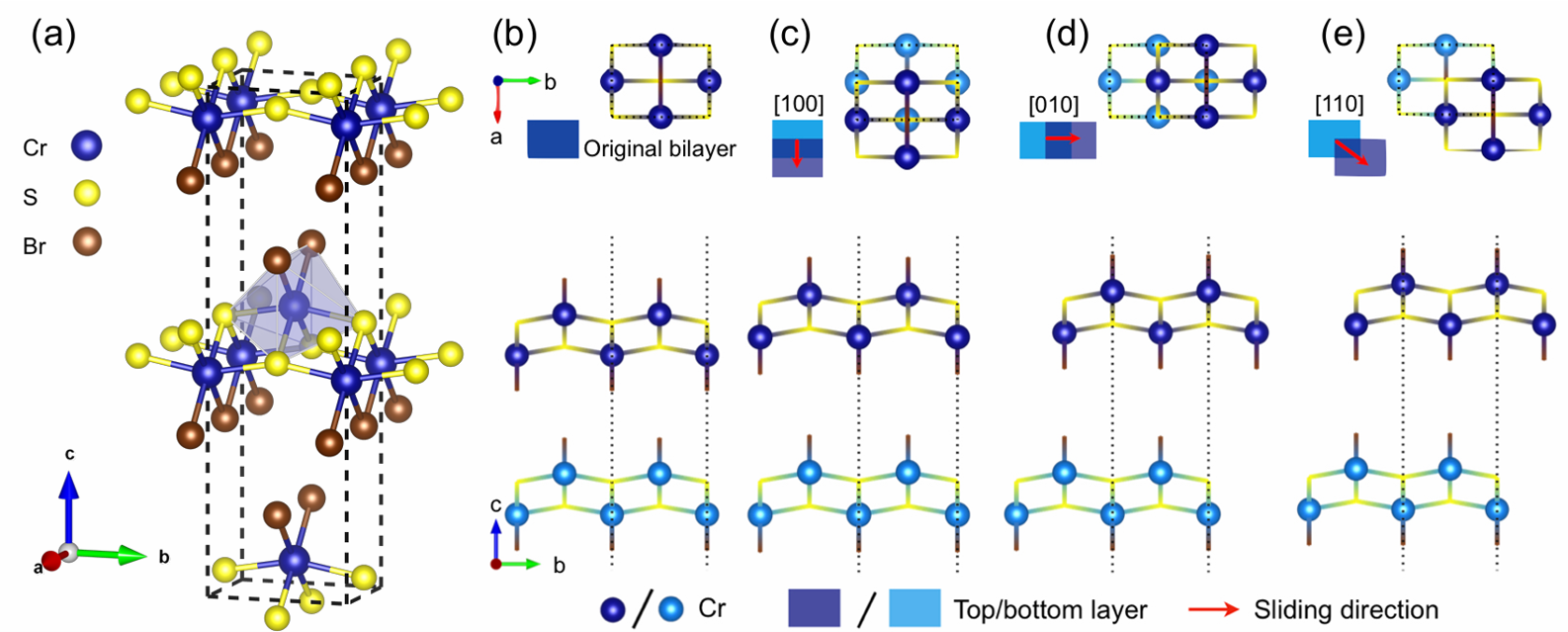}
	\caption{(a) Crystal structure of CrSBr bulk. (b)-(e) Top and side view of the four CrSBr bilayer structures with different stacking modes, which as labeled as AA-0, AA-1, AB-0, and AB-1. Note S and Br atoms are not shown in the bilayer structures for the sake of clarity. The rectangles in light and dark blue represent the bottom and top layers, respectively. The red arrows indicate the sliding directions of one CrSBr layer along the [100], [010], and [110] directions of bilayers.}
	\label{fig:1}
\end{figure*}			
	
\section{Results and discussion}

\subsection{Structures of the stacking patterns}
CrSBr crystallizes in the FeOCl-type structure with the $Pmmn$ (No. 59) space group. This structure comprises CrSBr monolayers interconnected through vdWs interactions along the $c$ axis, as illustrated in Fig.~\ref{fig:1}(a). Within each CrSBr monolayer, the CrS$_4$Br$_2$ octahedra are linked via edge- and corner-sharing along the $a$ and $b$ crystallographic axes, respectively. This configuration produces a rectangular lattice, wherein the lattice constant associated with edge-sharing octahedra is smaller than that of corner-sharing octahedra, resulting in a distinct arrangement of Cr$^{3+}$ cations within the $a$-$b$ plane. The CrS$_4$Br$_2$ heteroleptic octahedral units, each featuring one Cr$^{3+}$ cation coordinated by four S$^{2-}$ and two Br$^{-}$ anions, are stacked and separated by vdWs gaps along the $c$ axis. This structural arrangement arises from the distortion of the CrS$_4$Br$_2$ octahedra.

Prior to investigating the possible bilayer structures of CrSBr, we first fully optimized its bulk crystal structure. The calculated lattice parameters at 0 K are $a$ = 3.523 \AA, $b$ = 4.746 \AA, and $c$ = 7.910 \AA, which closely align with the experimentally measured values ($a$ = 3.511 \AA, $b$ = 4.746 \AA, and $c$ = 7.916 \AA) obtained at 10 K, and the layer spacing is 2.352 \AA which is consistent with the prediction without strain~\cite{a9,PhysRevB.109.214422}. Subsequently, we ``exfoliated'' the bulk structure along the layered direction to derive a bilayer, as illustrated in Fig.~\ref{fig:1}(c). The stacking pattern and space group of the bilayer structure remain the same with those of the bulk, which is labeled as the AA-0 stacking pattern. We then fully relaxed the in-plane lattice constants and atomic positions of the bilayer AA-0 structure while maintaining the out-of-plane lattice parameter fixed. Both FM and AFM spin configurations were considered during the structural relaxations. The calculated total energies indicate that the AFM state is slightly lower in energy (0.112 meV/f.u.) compared to the FM state, as shown in Table~\ref{tab:1}. Furthermore, we tested 15 distinct magnetic configurations and found that the A-type AFM configuration exhibits the lowest energy, as detailed in Table~\textcolor{red}{S2} of the supplementary materials. This finding suggests that the A-type AFM magnetic ordering observed in bulk CrSBr is preserved in its bilayer AA-0 polymorph.

\begin{figure}[htbp!]
	\centering
	\includegraphics[clip,width=1.0\linewidth]{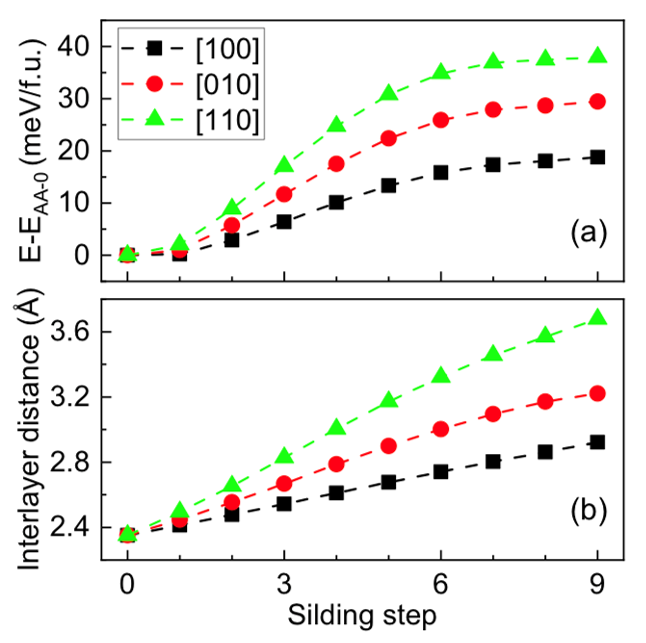}
	\caption{(a) Energy difference when sliding one of the layer along the [100], [010], and [110] directions. (b) The corresponding interlayer distance along the sliding paths.}
	\label{stacking}
\end{figure}	

\begin{table}[htbp!]
\centering
\caption{Lattice constants, space group, interlayer distance, the relative energy difference of stacking structures with respect to AA-0 with A-type AFM configuration, the energy difference between AFM and FM configurations of each CrSBr bilayer stacking pattern.}

\begin{tabular}{cccccc}
\toprule
                                          &             &  AA-0         &    AA-1       &  AB-0         &  AB-1         \\
\midrule
Space group                               &             & $Pmmn$        & $Pmma$        & $Pmma$        & $Pmmm$        \\
\multirow{2}*{Lattice constants (\AA)}    & \textit{a}  & 3.516         & 3.510         & 3.516         & 3.510         \\
			                              & \textit{b}  & 4.737         & 4.735         & 4.730         & 4.729         \\
Interlayer distance (\AA)                 &             & 2.352         & 2.922         & 3.221         & 3.677         \\
$\Delta$E (meV/f.u.)                      &             & 0             & 18.785        & 29.232        & 37.802        \\
E$_{\rm AFM} $ - E$_{\rm FM} $ (meV/f.u.) &             & -0.112        & -0.134        & -0.044        & -0.072        \\
\bottomrule	
\end{tabular}
	\label{tab:1}
\end{table}

Three additional stacking structures were generated by translating one layer of the AA-0 bilayer along three distinct directions: [100], [010], and [110]. The corresponding sliding distances in these directions are $\frac{1}{2}$|\textbf{a}|, $\frac{1}{2}$|\textbf{b}|, and $\frac{1}{2}$|\textbf{a+b}|, respectively. These three generated bilayer structures are tagged as AA-1, AB-0, and AB-1, as illustrated in Fig.~\ref{fig:1}(d)-(f). These bilayer structures exhibit high-symmetry stacking patterns with space groups $Pmma$, $Pmma$, and $Pmmm$. Following a fully relaxation of the FM and AFM states of these three bilayer structures (i.e., optimizing both the in-plane lattice constants and atomic positions), the A-type AFM configuration, which resembles the ground state of the bulk, is favored in all stacked bilayers, as tabulated in Table~\ref{tab:1}. Furthermore, the AA-0 structure demonstrates the lowest total energy among the four bilayer structures, indicating that the ground state of the bilayer structure is the AA-0. We note that in a recent study~\cite{PhysRevB.109.214422}, the AFM ground state of the bilayer CrSBr structure maintains during the sliding of one CrSBr layer along the $a$ axis, which is consistent with our results.

To investigate the energy changes and potential stable/metastable structures during the three sliding processes, we performed cNEB calculations~\cite{cNEB1,cNEB2}. The three cNEB pathways initiate from the AA-0 state and terminate in three distinct configurations: AA-1, AB-0, and AB-1. The AFM spin configuration was employed in these calculations. The results of the cNEB calculations are presented in Fig.~\ref{stacking}(a). It is evident that sliding one CrSBr monolayer along any of the specified directions results in an increase in total energy until the maximum value of each curve is reached, which corresponds to the three generated bilayer structures. Following these maxima, the total energies decrease due to the translation symmetry. This observation indicates that the AA-0 structure represents the ground state of the CrSBr bilayer, while the three generated structures -- AA-1, AB-0, and AB-1 -- are metastable. Notably, no additional metastable structures were identified during the sliding processes. The calculated energy costs for the transitions from AA-0 to AA-1, AB-0, and AB-1 are 19, 29, and 38 meV/f.u., respectively. These values are comparable to those observed in other bilayer systems, such as MoS$_2$ (27 meV/f.u.)~\cite{a54}, InSe (28 meV/f.u.)~\cite{sui2024atomic}, and CrI$_3$ (10 meV/f.u.)~\cite{a55}. Moreover, adopting different stacking patterns significantly alters the vertical distances between adjacent vdWs layers in the AA-0, AA-1, AB-0, and AB-1 structures, as illustrated in Table~\ref{tab:1} and Fig.~\ref{stacking}(b). For instance, the difference of the interlayer distance can reach as much as 1.33 \AA\ between the AA-0 and AB-1 structures. Such substantial variations in interlayer distances are likely to result in corresponding changes in the strength of interlayer interactions, which will discussed in further detail later.

\begin{figure*}[htbp!]
	\centering
	\includegraphics[clip,width=1.0\linewidth]{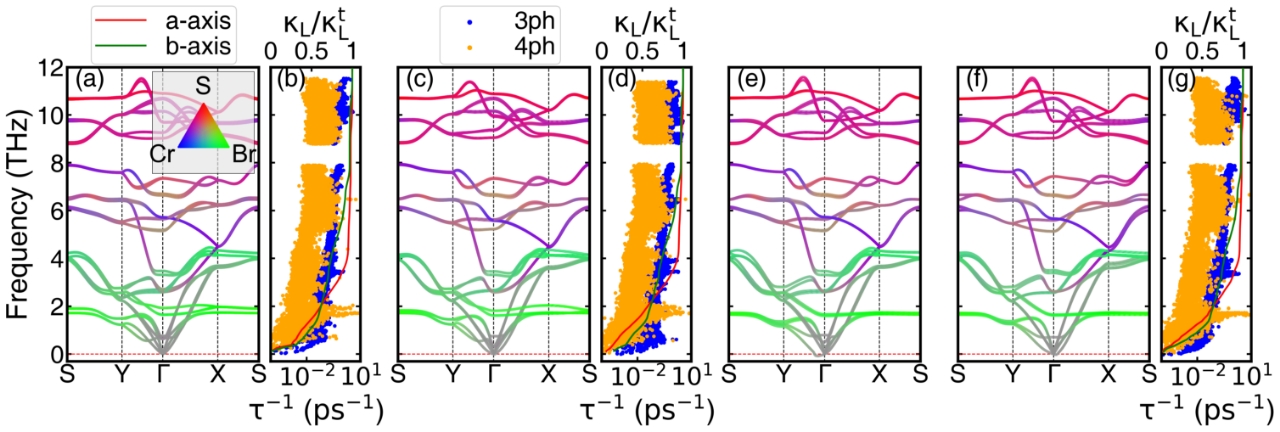}
	\caption{(a), (c), (e), and (f) Phonon dispersion of AA-0, AA-1, AB-0, and AB-1 bilayer structures, respectively. (b), (d), and (g) are the four-phonon scattering rates ($\tau^{-1}$) and relatively accumulated $\kappa_{\rm L}$ of AA-0, AA-1, and AB-0, respectively.}
	\label{phonon}
\end{figure*}

\subsection{Lattice dynamics and lattice thermal conductivity}
The phonon dispersion of these stacking structures at 300 K is shown in Fig.~\ref{phonon}. One acoustic phonon of the AB-0 structure exhibits a subtle imaginary frequency (the maximum value is 0.1 THz) near the $\Gamma$ point along the $\Gamma$-Y direction, which is commonly seen in 2D materials~\cite{FANG2021110688,doi:10.1021/acs.nanolett.2c04497}. Each primitive unit cell of the bilayer structures contains 12 atoms, resulting in a total of 36 phonon modes. Interestingly, all these bilayers exhibit the same phonon mode symmetry, despite they have different space groups. The mode decomposition in the zone center (the $\Gamma$ point) is characterized as 6A$_{\rm g}$ $\oplus$ 6B$_{\rm 1u}$ $\oplus$ 6B$_{\rm 2g}$ $\oplus$ 6B$_{\rm 2u}$ $\oplus$ 6B$_{\rm 3g}$ $\oplus$ 6B$_{\rm 3u}$. Three acoustic modes possess B$_{\rm 1u}$, B$_{\rm 2u}$, and B$_{\rm 3u}$ symmetry. Therefore, the infrared active modes include 5B$_{\rm 1u}$, 5B$_{\rm 2u}$, and 5B$_{\rm 3u}$, while the Raman active modes consist of 6A$_{\rm g}$, 6B$_{\rm 2g}$, and 6B$_{\rm 3g}$. The phonon frequencies and their infrared and Raman activities of all these compounds are tabulated in Table~\textcolor{red}{S3-S6}. The most remarkable features of the phonon dispersion are the flat-phonon bands around 2 THz along the $\Gamma$-X-S direction, which is mainly from the atomic displacement of Br atoms (see Fig.~\textcolor{red}{S1-S4}). Additionally, there are extremely low-frequency optical modes (0.4 THz) at the $\Gamma$ point, as well as an avoid crossing between acoustic and the flat optical phonon modes along the $\Gamma$-Y direction. The frequencies of the flat bands at the $\Gamma$ point gradually decrease from AA-0 (1.64 and 1.92 THz), to AA-1 (1.64 and 1.83 THz), then to AB-0 (1.60 and 1.66 THz), and finally to AB-1 (1.62 and 1.65 THz). The frequencies of the three low-frequency optical phonon modes at the $\Gamma$ point also vary with the stacking pattern: AA-0 (0.62, 0.64, and 0.76 THz), AA-1 (0.34, 0.54, and 0.75 THz), AB-0 (0.31, 0.50, and 0.79 THz), and AB-1 (0.20, 0.21, and 0.70 THz). 

\begin{figure}[htbp!]
	\centering
	\includegraphics[clip,width=1.0\linewidth]{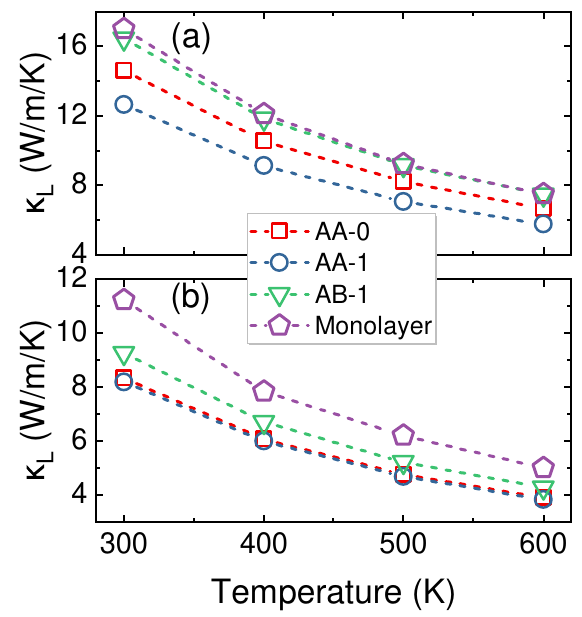}
	\caption{(a) The temperature-dependence $\kappa_{\rm L}$ of CrSBr bilayers along the $a$ axis. (b) The temperature-dependence $\kappa_{\rm L}$ of CrSBr bilayers along the $b$ axis.}
	\label{kappa}
\end{figure}

Due to the presence of a phonon mode with a small imaginary frequency near the $\Gamma$ point in the AB-0 structure, which blocks lattice thermal conductivity ($\kappa_{\rm L}$) calculations, we focused on computing $\kappa_{\rm L}$ for the other three structures. The computed $\kappa_{\rm L}$ values using SCPH method, incorporating three- and four-phonon scattering, denoted as $\kappa_{\rm 3,4ph}^{\rm SCPH}$, are illustrated in Fig.~\ref{kappa}. Firstly, the stacking patterns do have evident effect on $\kappa_{\rm L}$, with AB-1 exhibiting the highest and AA-1 the lowest values. Moreover, the $\kappa_{\rm L}$ of all these structures demonstrates strong anisotropy across all temperatures studied in this work, i.e., $\kappa_{\rm L}$ along the $a$ axis direction is nearly two times larger than that along the $b$ axis. This anisotropy arises primarily from the differences in edge- and corner-sharing CrS$_4$Br$_2$ octahedra along these two axes. It is noteworthy that the $\kappa_{\rm L}$ values of these bilayers at 300 K are considerably lower than those of many other semiconducting few-layers such as MoS$_2$~\cite{doi:10.1021/jp402509w}, WS$_2$~\cite{peimyoo2015thermal}, and BP~\cite{PhysRevB.99.085410}. This reduction in thermal conductivity can be attributed to the high three- and four-phonon scattering rates at the low frequencies ($\sim$ 0.5 and $\sim$ 2 Thz). The analysis of four-phonon scattering is depicted in Fig.~\ref{phonon}. The relative accumulated $\kappa_{\rm L}$ (the ratio between the accumulated $\kappa_{\rm L}$ and the total $\kappa_{\rm L}$, $\kappa_{\rm L}^{\rm t}$) shows that the phonons with frequencies below 2 THz contribute more than 80 \% to $\kappa_{\rm L}^{\rm t}$. In this low-frequency region, strong three- and four-phonon scattering is driven by the avoided crossing between optical and acoustic phonons, as well as the flat bands along the $\Gamma$-X-S-Y direction. The avoid crossing signifies strong phonon-phonon interaction between heat-carrying acoustic and optical phonon modes~\cite{christensen}, while the flat bands provide an extensive scattering phase space~\cite{PhysRevB.91.144304,PhysRevLett.125.245901}. Also, we compared the $\kappa_{\rm L}$ of the bilayers to that of the monolayer. The $\kappa_{\rm L}$ of bilayers are consistently lower than that of the monolayer, with the difference between AB-1 and monolayer being more pronounced in the $b$ direction than the $a$ direction. The temperature dependence of $\kappa_{\rm L}$ of across all these structures remains similar within the temperature range of 300 to 600 K.

\begin{figure}[htbp!]
	\centering
	\includegraphics[clip,width=0.98\linewidth]{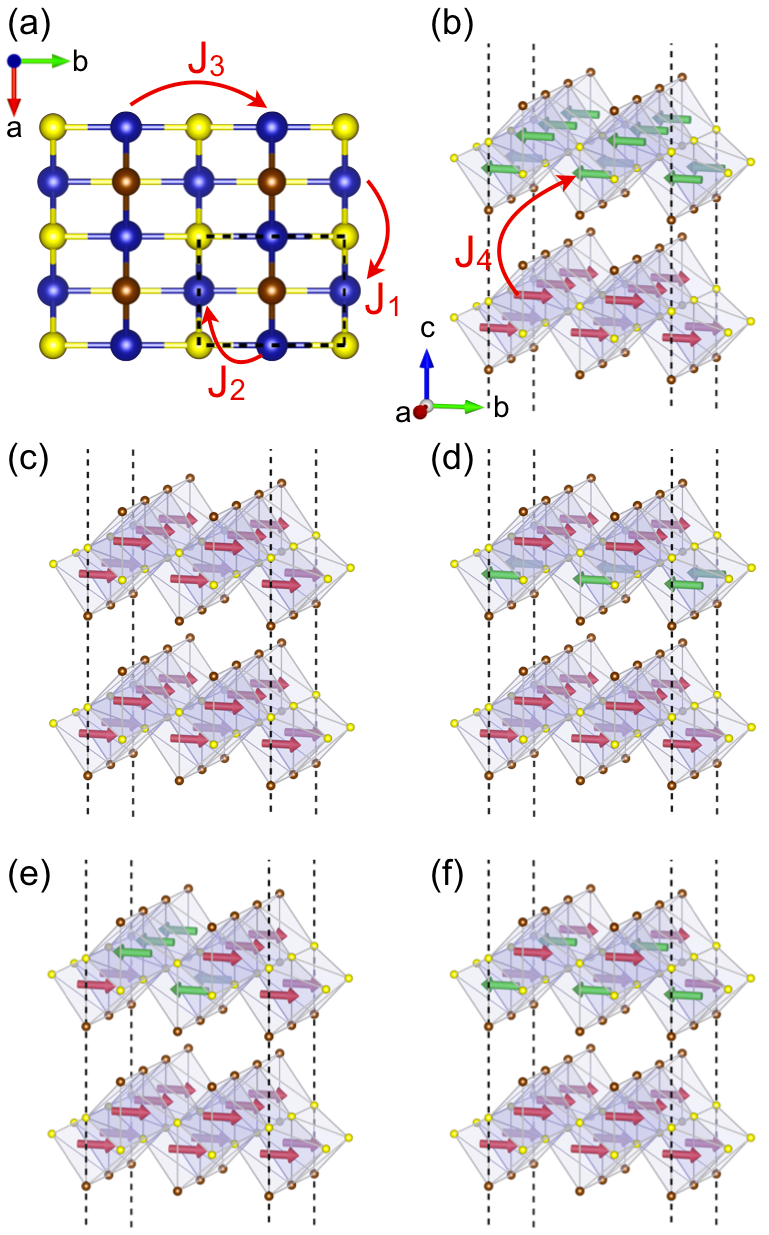}
	\caption{(a)-(b) Intralayer and the nearest interlayer magnetic interactions between Cr ions in CrSBr bilayer structures. (b)-(f) Magnetic configurations used to extract the parameters of the magnetic interactions. The configurations are named as AFM, FM, FIM1, FIM2 and FIM3, respectively.}
	\label{fig:2}
\end{figure}	

\subsection{Magnetic interactions}	
We further studied the impact of stacking patterns on the magnetic properties of CrSBr bilayer structures by analyzing the magnetic coupling parameters between the Cr$^{3+}$ ions. As illustrated in Fig.~\ref{fig:2}(a), the nearest-neighbor (NN), second-NN, and third-NN intralayer magnetic interactions are denoted as $J_1$, $J_2$, and $J_3$, respectively, based on the distance between Cr$^{3+}$ ions. The NN interlayer magnetic interaction is represented as $J_4$, as depicted in Fig.~\ref{fig:2}(b). We then employed the energy mapping method to extract the magnitudes of these interaction parameters~\cite{a39}. For each stacking structure, in addition to the AFM ground state shown in Fig.~\ref{fig:2}(b), we constructed four additional magnetic states using a 2$\times$2$\times$1 supercell. An example of these magnetic configurations for the AA-0 stacking is shown in Fig.\ref{fig:2}(c)-(f), which are labeled as FM, FIM1, FIM2, and FIM3, respectively. The total energy of each magnetic configuration can be expressed as a function of the magnetic interaction parameters: 
$$
	\begin{aligned}
		E_{\rm FM}   & = E_{0} + ( 16J_{1} + 32J_{2} + 16J_{3} + nJ_{4} )S^{2} , \\
		E_{\rm AFM}  & = E_{0} + ( 16J_{1} + 32J_{2} + 16J_{3} - nJ_{4} )S^{2}, \\
		E_{\rm FIM1} & = E_{0} + ( 16J_{1} + 16J_{3} + nJ_{4} )S^{2}, \\
		E_{\rm FIM2} & = E_{0} + ( 16J_{1} + 16J_{2} )S^{2}, \\
		E_{\rm FIM3} & = E_{0} + ( 16J_{2} + 16J_{3} )S^{2},
	\end{aligned}
$$
where $E_0$ is the energy of the system without including the magnetic interactions, $S = \frac{3}{2}$ for the Cr$^{3+}$ cation, and $n$ represents the number of nearest-neighbor interlayer magnetic interactions in the different stacking modes. Specifically, $n = 16$ for the AA-0 stacking, $n = 8$ for the AA-1 and AB-0 stackings, and $n = 4$ for the AB-1 stacking. By performing DFT calculations and utilizing the total energies of the five magnetic configurations for each stacking structure, we obtained the magnetic interaction parameters between the Cr$^{3+}$ cations in the four CrSBr bilayer structures, as detailed in Table~\ref{tab:2}. The negative values of $J_1$, $J_2$, and $J_3$ indicate FM coupling within the CrSBr monolayer, whereas the positive value of $J_4$ indicates AFM coupling between the adjacent layers, aligning with experimental observations~\cite{a45,a46,a47}. Notably, the variations in the magnitudes of $J_1$, $J_2$, and $J_3$ across different stacking structures are relatively minor, which corresponds to the minimal changes in the in-plane lattice constants (less than 0.01 \AA, as shown in Table~\ref{tab:1}). Within a CrSBr layer, two nearest-neighbor Cr$^{3+}$ ions ($t_{\rm 2g}^{\uparrow\uparrow\uparrow}e_{\rm g}$) are connected through S$^{2-}$ and Br$^{-}$ ions at nearly 90$^{\circ}$ angles along the [100] and [110] directions (see Fig.~\ref{fig:2}). This geometric arrangement favors FM superexchange interactions, as the angle between two ions with half-filled $d$ orbitals is approximately 90$^{\circ}$. In the [010] direction, the Cr-S-Cr angle ($\sim$ 160°) deviates from 180$^\circ$, leading to competition between FM and AFM interactions. In this scenario, the FM interaction predominates \cite{g1,g2,g3} In contrast, the calculated interlayer coupling $J_4$ is significantly smaller in magnitude compared to the intralayer magnetic couplings, due to the weak vdWs interlayer interactions and the increased distance between the interlayer Cr$^{3+}$-Cr$^{3+}$ cations. Importantly, the stacking pattern has a substantial influence on the strength of the interlayer AFM coupling $J_4$. Different stacking structures can provide different electron transfer paths through interfacial Br anions (Cr-Br-Br-Cr super-superexchange)~\cite{PhysRevB.109.214422} and alter the interlayer Cr$^{3+}$-Cr$^{3+}$ distances (e.g., the distance between the NN interlayer Cr$^{3+}$-Cr$^{3+}$ cations varies from 6.62 \AA\ in the AA-0 structure to 7.26 \AA\ in the AB-1), leading to difference interlayer Cr-Br-Br-Cr super-superexchange couplings within the CrSBr bilayer structures, as shown in Table~\ref{tab:2} and Fig.~\textcolor{red}{S5} of the Supporting Material. For example, there are two distinct Cr-Br-Br-Cr super-superexchange pathways of equal shortest length between the NN interlayer Cr$^{3+}$-Cr$^{3+}$ cations in the AA-0 structure~\cite{PRB-SSE-Path}. However, this number reduces to one in the AB-1 structure. Such alternations result in differing overlaps between the $p$ orbitals of the NN interlayer Br atoms, which have been proved to play a crucial role in determining interlayer magnetism in CrSBr and some analogs 2D magnetic systems~\cite{PhysRevB.102.020402,PhysRevB.109.214422,a55}, thereby enhancing the NN interlayer AFM coupling strength from 0.03 meV in the AA-0 structure to 0.06 meV in the AB-1 structure. Similarly, the shortest super-superexchange path between the NN interlayer Cr$^{3+}$-Cr$^{3+}$ cations in the AA-1 structure also decreases to one, with the corresponding AFM coupling strength increasing to 0.06 meV. Conversely, in the AB-0 structure, two distinct super-superexchange paths remain between the NN interlayer Cr$^{3+}$-Cr$^{3+}$ cations, yet the AFM coupling strength diminishes to 0.02 meV. Our calculated exchange parameters $J_n$ are consistent with previous calculations that employed the same values for U and J~\cite{a39,PhysRevB.109.214422}. The magnitude of $J_n$ differs from that of the honeycomb lattice compounds Cr$X$ ($X$ = P, As, and Sb), which exhibit three-fold symmetry in the in-plane structure, where the nearest-neighbor exchange parameter $J_1$ is dominant ~\cite{PhysRevApplied.15.064053,PhysRevB.102.024441}.
		
\begin{table}[htbp!]
\scriptsize	
\centering
\caption{Magnetic exchange parameters, $\angle$Br-Br-Cr angles along the super-superexchange paths, distance between NN interlayer Cr-Cr atoms, direct (E$_{\rm g}^{d}$) and indirect band gap (E$_{\rm g}^{i}$), effective mass of electrons ($m^e$) and holes ($m^h$) for the AFM/FM state, and SIA parameters of the bilayer structures with different stacking patterns. The energies of the SIA parameters are referenced to the $a$-axis.}
\setlength{\tabcolsep}{0.7mm}{
\begin{tabular}{cccccc}
\toprule
                                             &            &    AA-0       &  AA-1         &   AB-0        &  AB-1  \\
\midrule
$J_{1} $ (meV)                               &            &   -0.91       &   -0.80       &   -0.96       &  -0.83  \\
$J_{2} $ (meV)                               &            &   -3.30       &   -3.33       &   -3.35       &  -3.37  \\
$J_{3} $ (meV)                               &            &   -2.27       &   -2.29       &   -2.33       &  -2.37  \\
$J_{4} $ (meV)                               &            &    0.03       &    0.06       &    0.02       &   0.06  \\
 $\angle$Br-Br-Cr ($^{\circ}$)               &            & 140.5         & 123.7         & 164.1         & 135.6 \\
 $\angle$Br-Br-Cr ($^{\circ}$)               &            & 96.8          & -             & 106.9         & - \\
 d$_{\rm Cr-Cr}$ (\AA)                       &            & 6.62          & 6.92          & 7.02          & 7.26 \\
\multirow{2}{*}{E$_{\rm g}^{i}$ (eV)}        & AFM        & 1.18          & 1.21          & 1.22          & 1.22 \\
				                             & FM         & 1.03          & 1.10          & 1.17          & 1.18 \\
\multirow{2}{*}{E$_{\rm g}^{d}$ (eV)}        & AFM        & 1.36          & 1.37          & 1.37          & 1.37\\
                                             & FM         & 1.08          & 1.17          & 1.34          & 1.39\\
\multirow{2}{*}{$m_{xx}^e/m_{yy}^e$ ($m_0$)} & AFM        & 6.75/0.26     & 6.05/0.26     & 6.47/0.28     & 6.49/0.27 \\
                                             & FM         & 6.39/0.26     & 7.33/0.26     & 6.15/0.28     & 7.16/0.27 \\
\multirow{2}{*}{$m_{xx}^h/m_{yy}^h$ ($m_0$)} & AFM        & 2.21/0.46     & 2.79/0.57     & 2.96/0.57     & 3.08/0.58 \\
                                             & FM         & 3.25/0.31     & 3.73/0.37     & 3.15/0.46     & 3.26/0.53 \\
\multirow{3}{*}{SIA ($\mu$eV)}&$a$&0     &0     &0      &    0\\
&$b$&-59.69&-62.23&-61.28&-60.82 \\
&$c$&6.42  &8.15  &4.32  &10.25  \\ 
\bottomrule	
\end{tabular}}	
\label{tab:2}
\end{table}
	
\begin{figure}[htbp!]
\centering
\includegraphics[clip,width=1\linewidth]{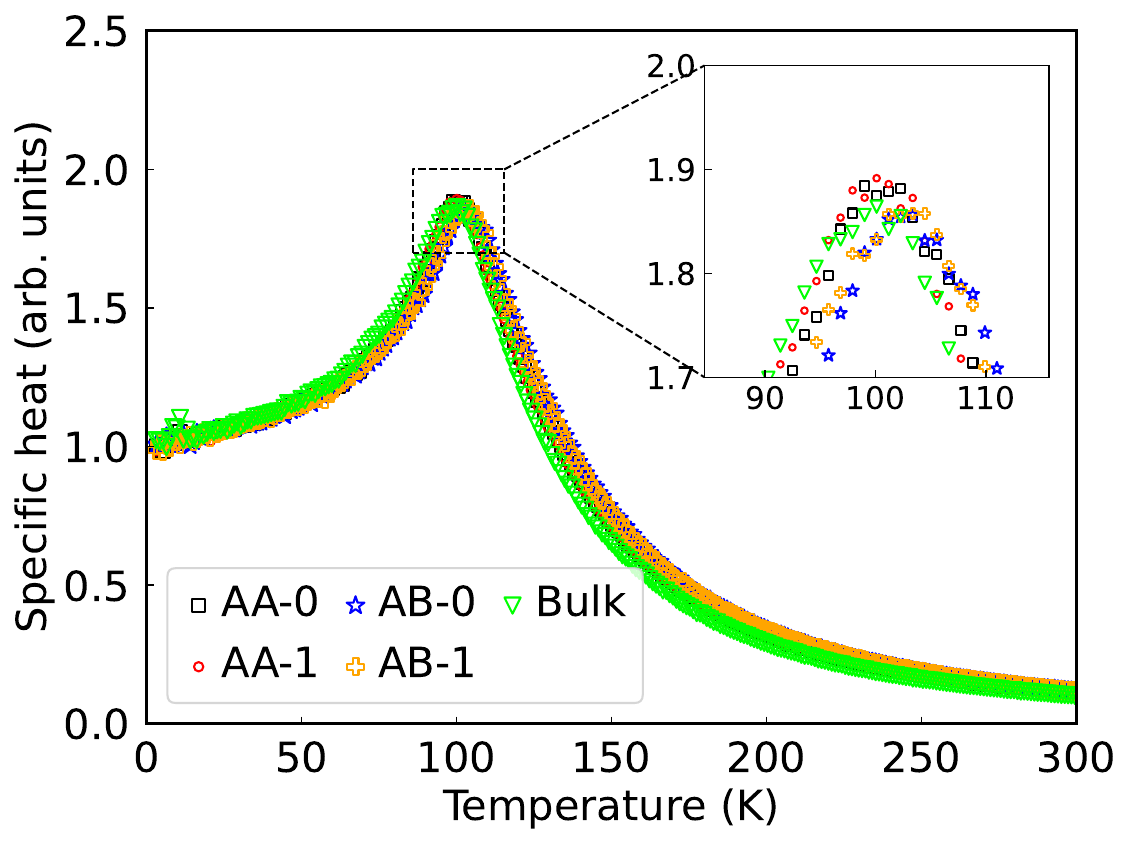}
\caption{Specific heat of CrSBr bilayers and bulk as a function of temperature calculated from Monte Carlo simulation.}
\label{fig:3}
\end{figure}	

\begin{figure*}[htbp!]
	\centering
	\includegraphics[clip,width=1\linewidth]{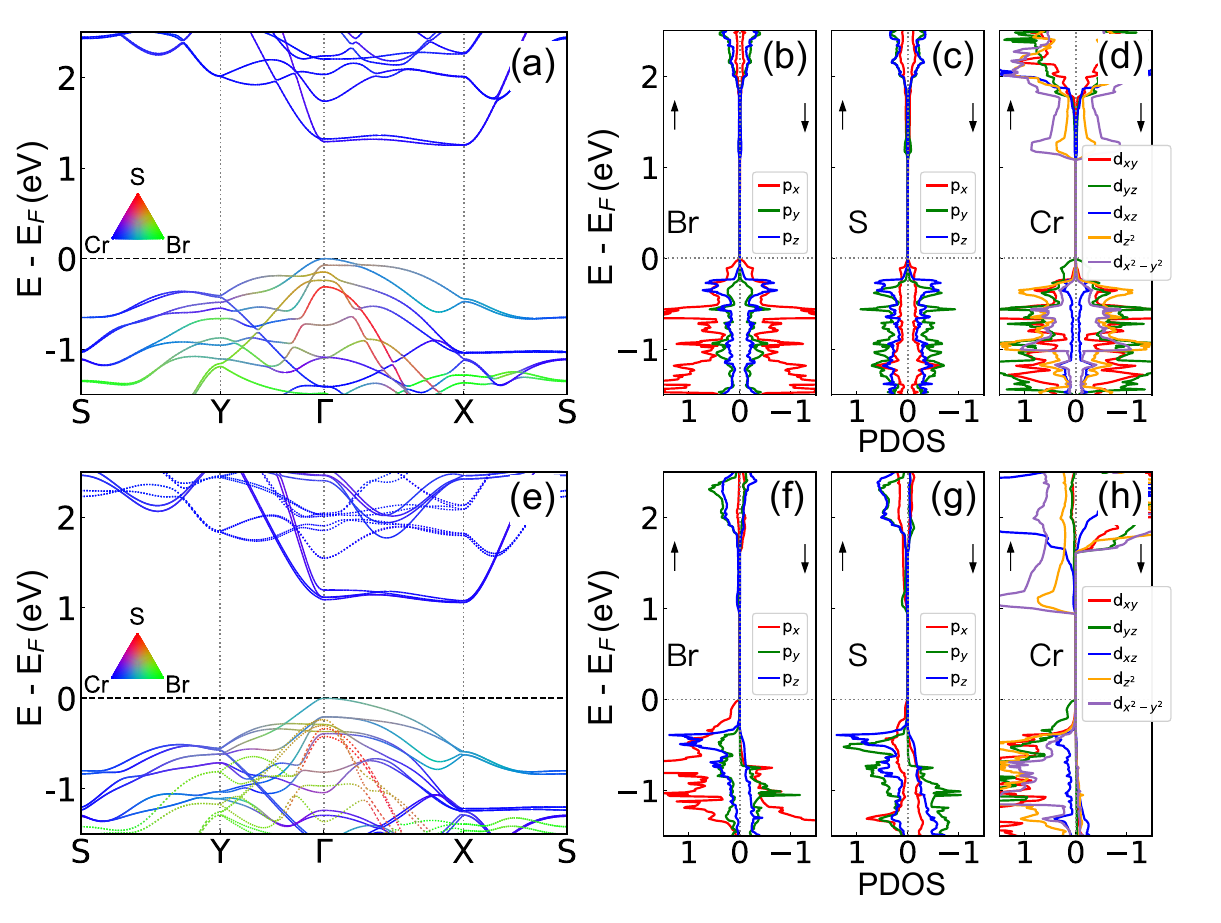}
	\caption{(a) and (e) are band structures of the AA-0 structure with AFM and FM configurations, respectively. Red, blue, and green color represent the contribution of S, Cr, and Br atoms to the bands. The solid and dash lines denote the spin-up and down channels, respectively. (b)-(d) and (f)-(h) are projected density of states (PDOS) of the AA-0 structure with AFM and FM configurations, respectively. The black $\uparrow$ and $\downarrow$ represent spin-up and spin-down channels, respectively.}
	\label{fig:4}
\end{figure*}	
	
Then, the $T_N$ of the four bilayer stacking patterns were calculated with Monte Carlo simulations within a classic Heisenberg spin Hamiltonian model, 
$$
    \begin{aligned}
      H &= \sum_{n=1}^{4}\sum_{i<j} J_n S_i S_j + \sum_{i,\alpha\beta} S_{i\alpha} A_{i,\alpha\beta} S_{i\beta}
    \end{aligned}
$$
where the first and second terms describe the isotropic Heisenberg exchange and single-ion anisotropy, respectively, and $\alpha$, $\beta$ = $a$, $b$, $c$. The isotropic exchange parameters $J_n$ are shown in Table~\ref{tab:2}. Single-ion anisotropy (SIA) is essential for the emergence of 2D magnetism and governs the easy magnetic axis or plane. Our calculated SIA parameters are tabulated in Table~\ref{tab:2}. The off-diagonal elements of the tensor are negligible\cite{PRB-SSE-Path}. Our calculated easy magnetic axis is along the $b$-axis, which is consistent with experimental observations\cite{a25,a26}. Furthermore, these values exhibit insensitivity to stacking due to the weak interlayer vdWs interactions.

 As depicted in Fig.~\ref{fig:3}, the four stacking patterns exhibit nearly identical $T_{\rm N}$ ($\sim$ 102 K), indicating that the stacking has a negligible effect on the magnetic transition temperature in the CrSBr bilayer, despite the remarkable enhancement of $J_4$ in AA-1 and AB-1. For comparison, the $T_{\rm N}$ of the bulk was computed using the same method and found to be close to the bilayers, albeit slightly lower than the experimental value~\cite{a9}. The underestimation of the $T_{\rm N}$ for the bulk and AA-0 bilayer, compared to the experimental or previous calculations, is likely due to the adoption of a smaller Hubbard $U$ in our simulations. Note that larger Hubbard $U$ values were also tested in our simulations; the results show that the intralayer exchange parameters ($J_1$, $J_2$, and $J_3$) are significantly increased with increasing $U$ values, see Table~\textcolor{red}{S7} in the Supplementary Information. However, a larger $U$ leads to a smaller and even positive $J_4$, resulting in ferromagnetism for the CrSBr bulk, which is in contrast with the experimental observation, as observed in a previous study~\cite{rudenko2023dielectric}.

The effect of $J_4$ on $T_N$ is investigated by artificially increasing the magnitude of $J_4$ while keeping the $J_1$, $J_2$, and $J_3$ the unchanged. As depicted in Fig.~\textcolor{red}{S6}, the calculated $T_{\rm N}$ slightly increase when $J_4$ substantially increasing. This result indicates that $T_{\rm N}$ is insensitive to $J_4$ and explains why the $T_{\rm N}$s of these bilayers are very close and agrees with the experimental findings that the $T_{\rm N}$ of CrSBr decreases under the hydrostatic pressure~\cite{https://doi.org/10.1002/apxr.202400052}.

\subsection{Electronic structures}
To investigate the effects of stacking on the electronic structures of bilayer CrSBr, we calculated the atomic-resolved band structures and projected density of states (PDOS) of four bilayer structures with A-type AFM and FM configurations using PBE functional. Given the similarity in band structures of these stacking structures, only the band structure and the PDOS of AA-0 structure are shown in Fig.~\ref{fig:4}. The band structures for the other stacking patterns are illustrated in Fig.~\textcolor{red}{S7}. The band structures of AFM AA-0 bilayer CrSBr exhibit indirect bandgaps of approximately 1.20 eV, with the valence band maximum (VBM) and the conduction band minimum (CBM) at the $\Gamma$ and X points, respectively, as shown in Fig.~\ref{fig:4} and summarized in Table~\ref{tab:2}. This value is comparable to the band gap of the bulk material, and our calculated band gap of 1.21 eV is in good agreement with the experimentally measured optical band gap of 1.50 $\pm$ 0.20 eV when accounting for excitonic effects~\cite{a24,doi:10.1021/acsnano.2c07316,a49}, suggesting that the influence of bilayer stacking on bandgap is minimal. A notable characteristic of the these band structures is the flat valence band laong the $\Gamma$-X direction, contrasted by a highly dispersive band along the $\Gamma$-Y direction, resulting in significant anisotropy. The top of the valence band also exhibits anisotropy, albeit with a smaller difference (effective mass of electron along the $x$ and $y$ directions are $m_{xx}^*$ = 6.75 $m_0$ and $m_{yy}^*$ = 0.12 $m_0$ for A-AFM, $m_{xx}^*$ = 6.39 $m_0$ and $m_{yy}^*$ = 0.14 $m_0$ for FM). The ratio $m_{xx}^e/m_{yy}^e$ is consistent with previous findings in monolayer CrSBr~\cite{doi:10.1021/acsnano.2c07316}.

The dispersive and flat bands along the $\Gamma$-Y and $\Gamma$-X directions arise primarily from the corner- and edge-sharing octahedra along the $a$ and $b$ axis, and the weakly coupling within each layer in combination with a weak interlayer hybridization~\cite{a24,doi:10.1021/acsnano.2c07316}. This is analogous to the band structures of bulk and monolayer CrSBr~\cite{doi:10.1021/acsnano.2c07316}, due to the weak interlayer interactions. The large disparity in effective mass along $a$ and $b$ direction contributes to the pronounced conductivity differences observed in the monolayer~\cite{https://doi.org/10.1002/adma.202109759}. A key distinction between the FM and AFM configurations is the splitting of the CBM due to the Zeeman effect. In the conduction band, two bands at the $\Gamma$ points are degenerate at the X point for the AFM configuration, while they split into four at the $\Gamma$ point, retaining degeneracy at the X point for the FM configuration. Additionally, the differences in band structure among the four stacking structures are subtle for the AFM configuration, showing only minor variations in the band gap (see Table~\ref{tab:2}). In contrast, the differences between stacking structures are pronounced in the FM configuration, with significant changes observed at the both the bottom of the conduction band at the $\Gamma$ point and the valence band maximum at the $\Gamma$ point, as illustrated in Fig.~\textcolor{red}{S7}. The variations in effective mass across different stacking structures are smaller than that between AFM and FM configurations, see Table~\ref{tab:2}. For example, the effective mass $m_{xx}^e$ of AA-0, AA-1, AB-0, and AB-1 are 6.75, 6.05, 6.47, 6.49 $m_0$ for AFM, respectively, while for FM, they are 6.39, 7.33, 6.15, 7.16 $m_0$, respectively. Moreover, the energy difference ($\Delta{\rm E_{VB}}$) between the highest valence band (HVB) and the second highest valence band (HVB-1), as well as the energy difference ($\Delta{\rm E_{CB}}$) between the lowest conduction band (LCB) and the second conduction band (LCB+1), are more sensitive to the magnetic ordering than to stacking pattern, as depicted in Fig.~\textcolor{red}{S8}. Generally, $\Delta{\rm E_{VB}}$ is larger than $\Delta{\rm E_{CB}}$ across all stacking patterns and magnetic ordering. $\Delta{\rm E_{CB}}$ for the AFM configuration is larger than that for the FM configuration for a given stacking structure, although only for the AB-0 and AB-1 configurations does $\Delta{\rm E_{VB}}$ for FM exceed that for AFM.

As shown in Fig.~\ref{fig:4}(b) and (d), the low-energy region of the conduction bands is mainly from $e_g$ ($d_{x^{2}-y^{2}}$ and $d_{z^{2}}$) orbitals of Cr$^{3+}$, while the high-energy region of the valence bands is primarily comprised of contributions from the $t_{2g}$ ($d_{xy}$, $d_{yz}$, $d_{xz}$) orbitals of Cr$^{3+}$, as well as $p_{x}$ the orbitals of Br$^{-}$ and S$^{2-}$, which is in line with the electronic configuration of Cr$^{3+}$ ($t_{2g}^{\uparrow\uparrow\uparrow}e_g$) cation in an octahedral crystal field. The octahedral distortion, mainly driven by the heteroleptic coordinations with S and Br, results in further splitting of the $t_{2g}$ orbitals into distinct $d_{xy}$, $d_{yz}$, and $d_{xz}$~\cite{a30}.

Due to the small energy difference between AFM and FM configurations of these bilayer structures, along with experimental findings that the AFM state of CrSBr can be converted to FM by applying a modest magnetic field ($\sim$ 0.3 T)~\cite{a9,a38}, it is also interesting to study the impact of stacking on the band structures of the FM bilayer structures. The calculated band structures for the FM are shown in Fig.~\ref{fig:4}. Although the band structures of AFM and FM configurations are very similar, their CBM are different. In the FM configuration, the bands near the $\Gamma$ point split into four distinct bands, whereas in AFM, these bands retain the degeneracy. This degeneracy in the AFM bilayer structures arises from the product symmetry of time reversal and spatial inversion~\cite{a49}. The interlayer hybridization of the spin-up or spin-down electrons occupying the bands near the VBN and CBM are suppressed by the interlayer AFM ordering. While in the bilayer structures with the FM configuration, the interlayer hybridization of electrons is enhanced, resulting in the splitting of both the VBM and CBM.

\begin{figure}[htbp!]
	\includegraphics[clip,width=1.0\linewidth]{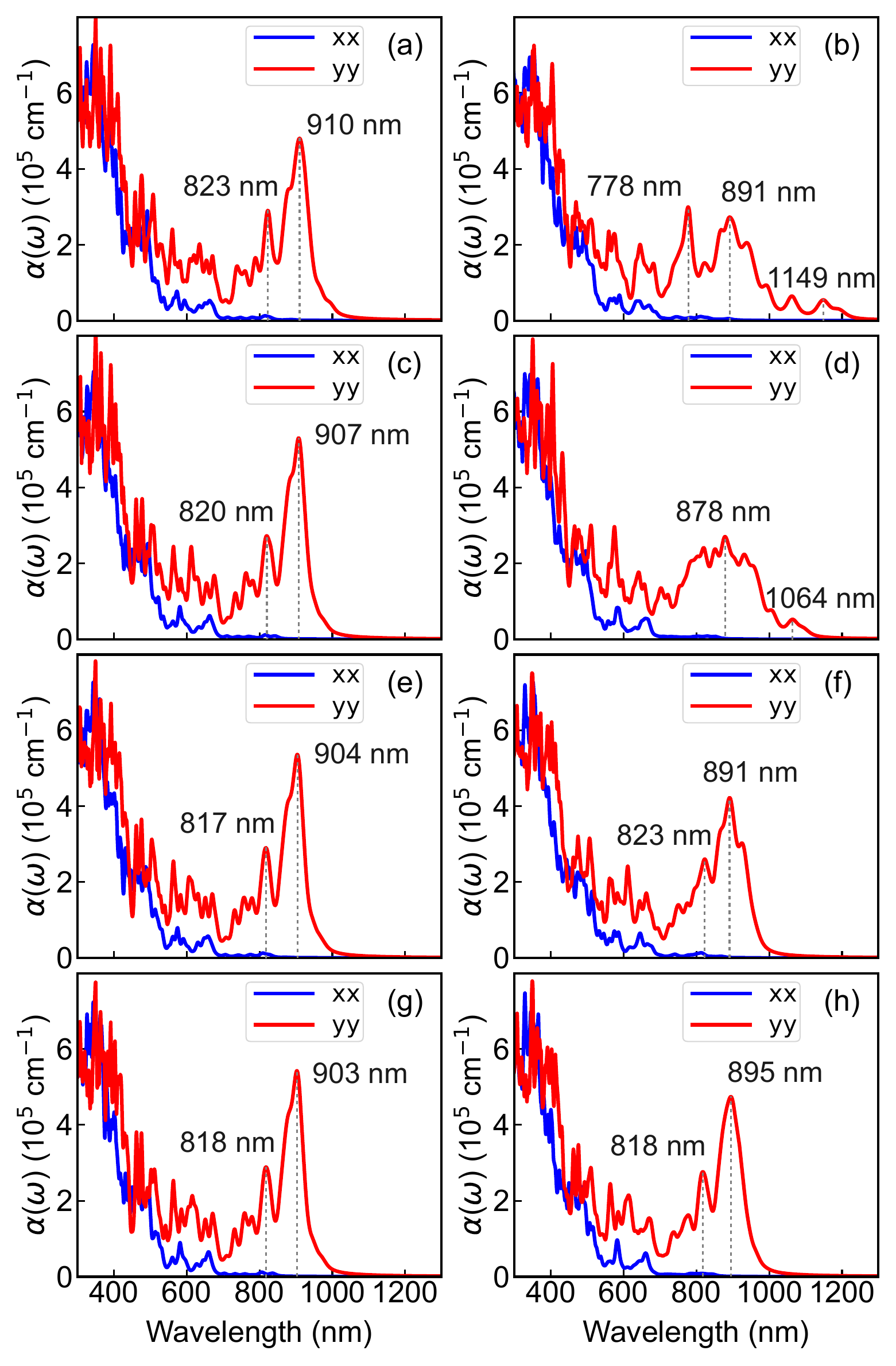}
	\caption{Absorption spectra of the four bilayer structures with AFM and FM configurations. (a), (c), (e), and (g) are absorption spectra of AA-0, AA-1, AB-0, and AB-1 structures with AFM configuration, respectively. (b), (d), (f), and (h) are absorption spectra of AA-0, AA-1, AB-0, and AB-1 with FM configuration, respectively. Vertical dash lines indicate absorption peaks.}
	\label{opitcs}
\end{figure}

\subsection{Optical absorption}
A previous experimental study shows that CrSBr bilayer (AA-0 structure) exhibits strong anistropy in both reflectance and photoluminescence spectra~\cite{a49}. Therefore, the frequency dependent dielectric properties of CrSBr bilayers were computed within the independent particle approximation, with the exciton effect being neglected in this analysis.

The calculated absorption spectra for the four bilayer structures are presented in Fig.~\ref{opitcs}. The optical absorption of these four bilayers in the AFM configuration are remarkably similar, with the primary absorption peak occurring at $\sim$ 910 nm along the $b$ ($\alpha_{yy}$) axis and at $\sim$ 820 nm along the $a$ ($\alpha_{xx}$) axis, due to the similarity in their band structures. The pronounced difference in absorption between $a$ and $b$ axis arises from dipole-allowed interband transitions from the VBM to the CBM along the $\Gamma$-Y direction ($b$ axis), which is forbidden along the $\Gamma$-X direction ($a$ axis)~\cite{a49}. A slight blue-shift of the absorption peak from the AA-0 structure to the other three bilayer structures is due to a modest increase in the band gap for three other stacking bilayer patterns (see Table~\ref{tab:2}), while the second largest peak does not exhibit such a blue-shift, as it involves transitions from the second highest valence band to the LCB, which are influenced by changes in valence band splitting, see Fig.~\textcolor{red}{S7}.

As shown in Fig.~\ref{opitcs}, the absorption spectra of the FM bilayers are quite different from those of the AFM configurations, especially for AA-0 and AA-1 structures. In the bilayers with the FM configuration, only the spin-up channel contributes to the absorption process, resulting a low joint density of state (JDOS). Consequently, the absorption coefficient ($\alpha$) of the FM configurations in the low-energy region is approximately half that of the AFM configuration. Among the FM configurations, AA-1 exhibits the largest valence band splitting and the smallest direct band gap (see Fig.~\textcolor{red}{S7}). As a result, its first absorption peak occurs at the longest wavelength of 1149 nm, corresponding to the excitation from HVB to HCB at the $\Gamma$ point. The Zeeman splitting near the $\Gamma$ point further reduces the JDOS and, leading to a lower absorption coefficient and, consequently, a smaller absorption peak. Interestingly, the low-energy peaks in FM AB-0 and AB-1 are quite narrow, which is different from the broader peaks observed in AA-0 and AA-1, and resembling the characteristic seen in the AFM configurations.
	
\subsection{Tuning magnetic structure by above-gap excitation}	
\begin{figure}[htbp!]
	\includegraphics[clip,width=1.0\linewidth]{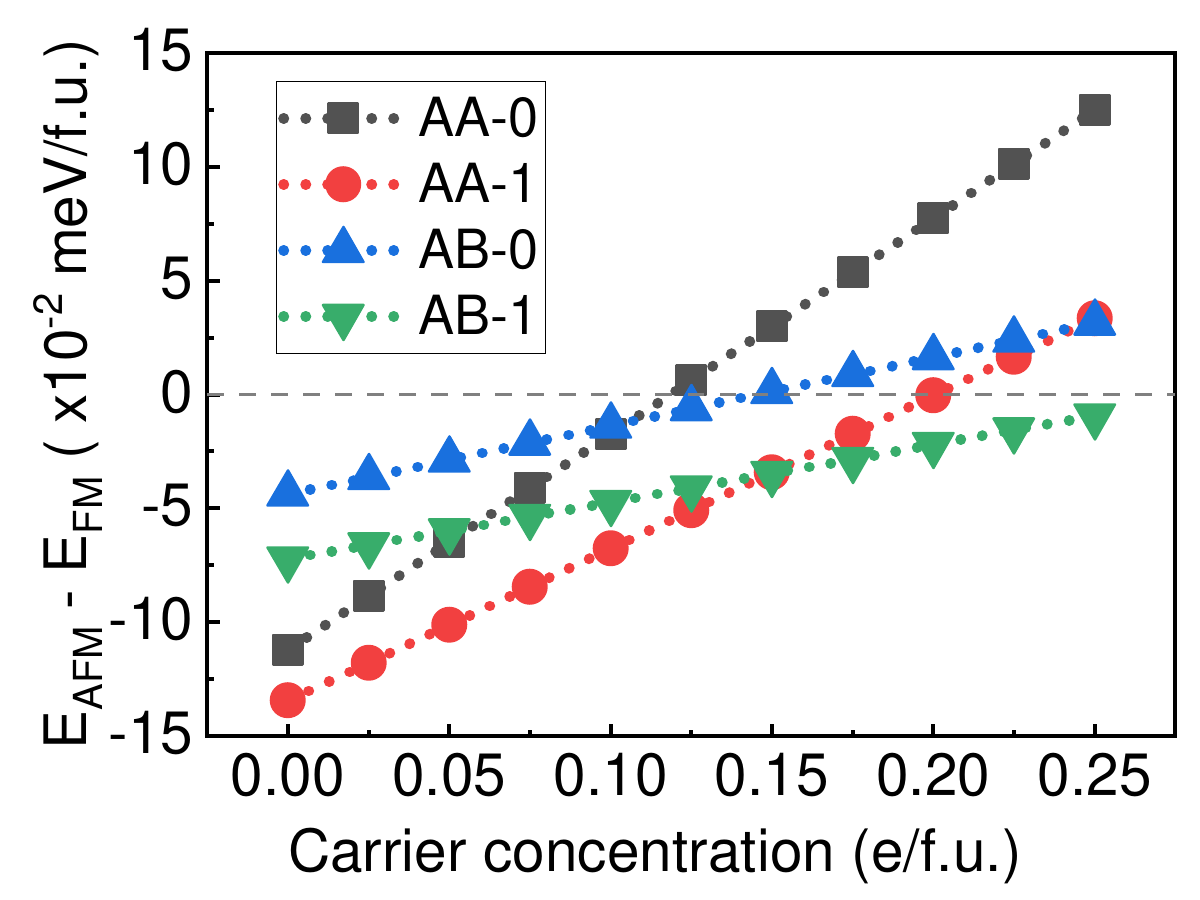}
	\caption{Energy difference between the AFM and FM configurations of each bilayer structure as a function of the photogenerated carrier concentration.}
	\label{fig:6}
\end{figure}

The above-band-gap photo-excitation has been employed to modulate the properties of semiconductors~\cite{10.1063/1.4905505,PhysRevLett.118.227401,doi:10.1021/jacs.4c03296}, particularly in the manipulation of magnetic structures of magnetic semiconductors~\cite{a50,chen2023light}. Given the relatively weak interlayer exchange interaction ($J_4$) in CrSBr bilayers, the influence of photo-generated carriers on the magnetic structure is expected to be significant. Therefore, the effects of above-gap excitation on the magnetic properties of these stacking patterns is simulated by moving electrons from the valence bands to the conduction bands. This is accomplished using a constrained DFT method \cite{PhysRevLett.116.247401}, which involves modifying the occupation numbers of the valence band maximum (VBM) and the conduction band minima (CBM). All the bilayer crystal structures were fully relaxed while maintaining a fixed occupation number for each carrier concentration, with the out-of-plane lattice parameter also held constant. As illustrated in Fig.~\ref{fig:6}, the energy differences between AFM and FM configurations for each of the four stacking patterns decrease linearly with an increase in photo-generated carrier concentration. When the concentration of photo-generated carriers surpasses a critical threshold, the FM structure becomes energetically more favorable than the AFM. As illustrated in Fig.~\textcolor{red}{S9} of the supplementary material, the interlayer magnetic coupling parameter $J_4$ is calculated for the four bilayer structures under varying concentrations of photoexcited electrons. The sign of the $J_4$ value transitions from positive to negative, indicating a shift in the magnetic state from FM to AFM. However, the critical concentration of photo-generated carriers varies among the stacking patterns. For example, the AB-1 configuration requires more than 0.25 e/f.u. of carriers to transition to the FM configuration, which corresponds to a carrier concentration of 6.0 × 10$^{14}$ cm$^{-2}$, while the AA-0 structure necessitates only 0.12 e/f.u. (5.0 × 10$^{14}$ cm$^{-2}$) of photo-generated carriers. The critical carrier concentrations for the AA-1 and AB-0 configurations lie between those of AA-0 and AB-1 structures. Therefore, the magnetic phase transition from AFM to FM configuration induced by photo-excitation can be modulated through bilayer stacking. It is noteworthy that the critical carrier concentrations for these structures are significantly lower than those reported for MnPS$_3$~\cite{a50}.

\section{Conclusions}
In summary, we conducted a systematic investigation into the structures, lattice dynamics, electronic structures, magnetism, and optical properties of CrSBr bilayers with four distinct stacking structures using first-principles calculations. Our results indicate that while the most stable CrSBr bilayer retains the same stacking pattern as its bulk counterpart, the other three structures can be realized by sliding one layer along the [100], [010], and [110] directions, with energy costs comparable to that of MoS$_2$ bilayer. Similar to the bulk material, all stacking structures favor an AFM configuration and exhibit semiconducting behavior, characterized by indirect band gaps of approximately 1.2 eV as determined by the PBE+U method. These bilayers display distinctive features, including flat phonon bands along the $\Gamma$-X-S direction in the low-frequency region and avoided crossing between acoustic and optical phonon modes. These features contribute to relatively low lattice thermal conductivities and strong anisotropy of lattice thermal conductivity in these bilayers due to the enhanced phonon-phonon scattering. The interlayer AFM coupling strength in these bilayers is significantly weaker than the intralayer FM coupling and can be effectively modulated by the stacking pattern, as the shortest interlayer superexchange path is sensitive to these stacking patterns. The band structures and optical absorption characteristic of these bilayers are independent of the stacking patterns in the AFM configuration, while they are sensitive to the FM configuration. Moreover, the magnetic structure can be tuned from AFM to FM configuration under visible light illumination. Thus, stacking engineering, in conjunction with visible light illumination, provides a viable approach to modulating the properties of CrSBr bilayers. Our findings offer valuable insights into the effects of stacking on the electronic, magnetic, and optical properties of CrSBr bilayers, highlighting alternative avenues for property control.

\section{ACKNOWLEDGMENTS}
This work was supported by National Natural Science Foundation of China (Grant No. 12034002). Y.Y. acknowledges the support by National Natural Science Foundation of China (grant no. 12304115), Fundamental Research Funds for the Central Universities (Grant No. FRF-TP-24-039A) and 2023 Fund for Fostering Young Scholars of the School of Mathematics and Physics, USTB (Grant No. FRF-BR-23-01B), J.H. acknowledges the support of the Fundamental Research Funds for the Central Universities China (USTB) and National Natural Science Foundation of China (Grant No. 12374024). The computing resource was supported by USTB MatCom of Beijing Advanced Innovation Center for Materials Genome Engineering.

\bibliography{ref}

\providecommand{\noopsort}[1]{}\providecommand{\singleletter}[1]{#1}%
\begin{thebibliography}{104}%
\makeatletter
\providecommand \@ifxundefined [1]{%
 \@ifx{#1\undefined}
}%
\providecommand \@ifnum [1]{%
 \ifnum #1\expandafter \@firstoftwo
 \else \expandafter \@secondoftwo
 \fi
}%
\providecommand \@ifx [1]{%
 \ifx #1\expandafter \@firstoftwo
 \else \expandafter \@secondoftwo
 \fi
}%
\providecommand \natexlab [1]{#1}%
\providecommand \enquote  [1]{``#1''}%
\providecommand \bibnamefont  [1]{#1}%
\providecommand \bibfnamefont [1]{#1}%
\providecommand \citenamefont [1]{#1}%
\providecommand \href@noop [0]{\@secondoftwo}%
\providecommand \href [0]{\begingroup \@sanitize@url \@href}%
\providecommand \@href[1]{\@@startlink{#1}\@@href}%
\providecommand \@@href[1]{\endgroup#1\@@endlink}%
\providecommand \@sanitize@url [0]{\catcode `\\12\catcode `\$12\catcode
  `\&12\catcode `\#12\catcode `\^12\catcode `\_12\catcode `\%12\relax}%
\providecommand \@@startlink[1]{}%
\providecommand \@@endlink[0]{}%
\providecommand \url  [0]{\begingroup\@sanitize@url \@url }%
\providecommand \@url [1]{\endgroup\@href {#1}{\urlprefix }}%
\providecommand \urlprefix  [0]{URL }%
\providecommand \Eprint [0]{\href }%
\providecommand \doibase [0]{https://doi.org/}%
\providecommand \selectlanguage [0]{\@gobble}%
\providecommand \bibinfo  [0]{\@secondoftwo}%
\providecommand \bibfield  [0]{\@secondoftwo}%
\providecommand \translation [1]{[#1]}%
\providecommand \BibitemOpen [0]{}%
\providecommand \bibitemStop [0]{}%
\providecommand \bibitemNoStop [0]{.\EOS\space}%
\providecommand \EOS [0]{\spacefactor3000\relax}%
\providecommand \BibitemShut  [1]{\csname bibitem#1\endcsname}%
\let\auto@bib@innerbib\@empty
\bibitem [{\citenamefont {Geim}\ and\ \citenamefont {Grigorieva}(2013)}]{a1}%
  \BibitemOpen
  \bibfield  {author} {\bibinfo {author} {\bibfnamefont {A.~K.}\ \bibnamefont
  {Geim}}\ and\ \bibinfo {author} {\bibfnamefont {I.~V.}\ \bibnamefont
  {Grigorieva}},\ }\bibfield  {title} {\bibinfo {title} {Van der
  $\mathrm{W}$aals heterostructures},\ }\href
  {https://doi.org/10.1038/nature12385} {\bibfield  {journal} {\bibinfo
  {journal} {Nature}\ }\textbf {\bibinfo {volume} {499}},\ \bibinfo {pages}
  {419} (\bibinfo {year} {2013})}\BibitemShut {NoStop}%
\bibitem [{\citenamefont {Novoselov}\ \emph {et~al.}(2016)\citenamefont
  {Novoselov}, \citenamefont {Mishchenko}, \citenamefont {Carvalho},\ and\
  \citenamefont {Neto}}]{a2}%
  \BibitemOpen
  \bibfield  {author} {\bibinfo {author} {\bibfnamefont {K.~S.}\ \bibnamefont
  {Novoselov}}, \bibinfo {author} {\bibfnamefont {A.}~\bibnamefont
  {Mishchenko}}, \bibinfo {author} {\bibfnamefont {A.}~\bibnamefont
  {Carvalho}},\ and\ \bibinfo {author} {\bibfnamefont {A.~H.~C.}\ \bibnamefont
  {Neto}},\ }\bibfield  {title} {\bibinfo {title} {2$\mathrm{D}$ materials and
  van der $\mathrm{W}$aals heterostructures},\ }\href
  {https://doi.org/10.1126/science.aac9439} {\bibfield  {journal} {\bibinfo
  {journal} {Science}\ }\textbf {\bibinfo {volume} {353}},\ \bibinfo {pages}
  {aac9439} (\bibinfo {year} {2016})}\BibitemShut {NoStop}%
\bibitem [{\citenamefont {Cao}\ \emph {et~al.}(2018)\citenamefont {Cao},
  \citenamefont {Fatemi}, \citenamefont {Fang}, \citenamefont {Watanabe},
  \citenamefont {Taniguchi}, \citenamefont {Kaxiras},\ and\ \citenamefont
  {Jarillo-Herrero}}]{a3}%
  \BibitemOpen
  \bibfield  {author} {\bibinfo {author} {\bibfnamefont {Y.}~\bibnamefont
  {Cao}}, \bibinfo {author} {\bibfnamefont {V.}~\bibnamefont {Fatemi}},
  \bibinfo {author} {\bibfnamefont {S.}~\bibnamefont {Fang}}, \bibinfo {author}
  {\bibfnamefont {K.}~\bibnamefont {Watanabe}}, \bibinfo {author}
  {\bibfnamefont {T.}~\bibnamefont {Taniguchi}}, \bibinfo {author}
  {\bibfnamefont {E.}~\bibnamefont {Kaxiras}},\ and\ \bibinfo {author}
  {\bibfnamefont {P.}~\bibnamefont {Jarillo-Herrero}},\ }\bibfield  {title}
  {\bibinfo {title} {Unconventional superconductivity in magic-angle graphene
  superlattices},\ }\href {https://doi.org/10.1038/nature26160} {\bibfield
  {journal} {\bibinfo  {journal} {Nature}\ }\textbf {\bibinfo {volume} {556}},\
  \bibinfo {pages} {43} (\bibinfo {year} {2018})}\BibitemShut {NoStop}%
\bibitem [{\citenamefont {Yang}\ \emph {et~al.}(2022)\citenamefont {Yang},
  \citenamefont {Bao}, \citenamefont {Lou}, \citenamefont {Li}, \citenamefont
  {Jiang}, \citenamefont {Wang}, \citenamefont {Sun}, \citenamefont {Liu},
  \citenamefont {Guo}, \citenamefont {Ramakrishnan}, \citenamefont {Kotla},
  \citenamefont {Tolkiehn}, \citenamefont {Paulmann}, \citenamefont {Cao},
  \citenamefont {Nie}, \citenamefont {Li}, \citenamefont {Liu}, \citenamefont
  {van Smaalen}, \citenamefont {Lin},\ and\ \citenamefont {Xu}}]{a4}%
  \BibitemOpen
  \bibfield  {author} {\bibinfo {author} {\bibfnamefont {X.}~\bibnamefont
  {Yang}}, \bibinfo {author} {\bibfnamefont {J.-K.}\ \bibnamefont {Bao}},
  \bibinfo {author} {\bibfnamefont {Z.}~\bibnamefont {Lou}}, \bibinfo {author}
  {\bibfnamefont {P.}~\bibnamefont {Li}}, \bibinfo {author} {\bibfnamefont
  {C.}~\bibnamefont {Jiang}}, \bibinfo {author} {\bibfnamefont
  {J.}~\bibnamefont {Wang}}, \bibinfo {author} {\bibfnamefont {T.}~\bibnamefont
  {Sun}}, \bibinfo {author} {\bibfnamefont {Y.}~\bibnamefont {Liu}}, \bibinfo
  {author} {\bibfnamefont {W.}~\bibnamefont {Guo}}, \bibinfo {author}
  {\bibfnamefont {S.}~\bibnamefont {Ramakrishnan}}, \bibinfo {author}
  {\bibfnamefont {S.~R.}\ \bibnamefont {Kotla}}, \bibinfo {author}
  {\bibfnamefont {M.}~\bibnamefont {Tolkiehn}}, \bibinfo {author}
  {\bibfnamefont {C.}~\bibnamefont {Paulmann}}, \bibinfo {author}
  {\bibfnamefont {G.-H.}\ \bibnamefont {Cao}}, \bibinfo {author} {\bibfnamefont
  {Y.}~\bibnamefont {Nie}}, \bibinfo {author} {\bibfnamefont {W.}~\bibnamefont
  {Li}}, \bibinfo {author} {\bibfnamefont {Y.}~\bibnamefont {Liu}}, \bibinfo
  {author} {\bibfnamefont {S.}~\bibnamefont {van Smaalen}}, \bibinfo {author}
  {\bibfnamefont {X.}~\bibnamefont {Lin}},\ and\ \bibinfo {author}
  {\bibfnamefont {Z.-A.}\ \bibnamefont {Xu}},\ }\bibfield  {title} {\bibinfo
  {title} {Commensurate stacking phase transitions in an intercalated
  transition metal dichalcogenide},\ }\href
  {https://doi.org/10.1002/adma.202108550} {\bibfield  {journal} {\bibinfo
  {journal} {Adv. Mater.}\ }\textbf {\bibinfo {volume} {34}},\ \bibinfo {pages}
  {2108550} (\bibinfo {year} {2022})}\BibitemShut {NoStop}%
\bibitem [{\citenamefont {Hart}\ \emph {et~al.}(2023)\citenamefont {Hart},
  \citenamefont {Bhatt}, \citenamefont {Zhu}, \citenamefont {Han},
  \citenamefont {Bianco}, \citenamefont {Li}, \citenamefont {Hynek},
  \citenamefont {Schneeloch}, \citenamefont {Tao}, \citenamefont {Louca},
  \citenamefont {Guo}, \citenamefont {Zhu}, \citenamefont {Jornada},
  \citenamefont {Reed}, \citenamefont {Kourkoutis},\ and\ \citenamefont
  {Cha}}]{a5}%
  \BibitemOpen
  \bibfield  {author} {\bibinfo {author} {\bibfnamefont {J.~L.}\ \bibnamefont
  {Hart}}, \bibinfo {author} {\bibfnamefont {L.}~\bibnamefont {Bhatt}},
  \bibinfo {author} {\bibfnamefont {Y.}~\bibnamefont {Zhu}}, \bibinfo {author}
  {\bibfnamefont {M.-G.}\ \bibnamefont {Han}}, \bibinfo {author} {\bibfnamefont
  {E.}~\bibnamefont {Bianco}}, \bibinfo {author} {\bibfnamefont
  {S.}~\bibnamefont {Li}}, \bibinfo {author} {\bibfnamefont {D.~J.}\
  \bibnamefont {Hynek}}, \bibinfo {author} {\bibfnamefont {J.~A.}\ \bibnamefont
  {Schneeloch}}, \bibinfo {author} {\bibfnamefont {Y.}~\bibnamefont {Tao}},
  \bibinfo {author} {\bibfnamefont {D.}~\bibnamefont {Louca}}, \bibinfo
  {author} {\bibfnamefont {P.}~\bibnamefont {Guo}}, \bibinfo {author}
  {\bibfnamefont {Y.}~\bibnamefont {Zhu}}, \bibinfo {author} {\bibfnamefont
  {F.}~\bibnamefont {Jornada}}, \bibinfo {author} {\bibfnamefont {E.~J.}\
  \bibnamefont {Reed}}, \bibinfo {author} {\bibfnamefont {L.~F.}\ \bibnamefont
  {Kourkoutis}},\ and\ \bibinfo {author} {\bibfnamefont {J.~J.}\ \bibnamefont
  {Cha}},\ }\bibfield  {title} {\bibinfo {title} {Emergent layer stacking
  arrangements in c-axis confined $\mathrm{MoTe_2}$},\ }\href
  {https://doi.org/10.1038/s41467-023-40528-y} {\bibfield  {journal} {\bibinfo
  {journal} {Nat. Commun.}\ }\textbf {\bibinfo {volume} {14}},\ \bibinfo
  {pages} {4803} (\bibinfo {year} {2023})}\BibitemShut {NoStop}%
\bibitem [{\citenamefont {Wu}\ \emph {et~al.}(2023)\citenamefont {Wu},
  \citenamefont {Tong}, \citenamefont {Deng}, \citenamefont {Luo},
  \citenamefont {Tian}, \citenamefont {Qin},\ and\ \citenamefont {Zhang}}]{a6}%
  \BibitemOpen
  \bibfield  {author} {\bibinfo {author} {\bibfnamefont {Y.}~\bibnamefont
  {Wu}}, \bibinfo {author} {\bibfnamefont {J.}~\bibnamefont {Tong}}, \bibinfo
  {author} {\bibfnamefont {L.}~\bibnamefont {Deng}}, \bibinfo {author}
  {\bibfnamefont {F.}~\bibnamefont {Luo}}, \bibinfo {author} {\bibfnamefont
  {F.}~\bibnamefont {Tian}}, \bibinfo {author} {\bibfnamefont {G.}~\bibnamefont
  {Qin}},\ and\ \bibinfo {author} {\bibfnamefont {X.}~\bibnamefont {Zhang}},\
  }\bibfield  {title} {\bibinfo {title} {Coexisting ferroelectric and
  ferrovalley polarizations in bilayer stacked magnetic semiconductors},\
  }\href {https://doi.org/10.1021/acs.nanolett.3c01948} {\bibfield  {journal}
  {\bibinfo  {journal} {Nano Lett.}\ }\textbf {\bibinfo {volume} {23}},\
  \bibinfo {pages} {6226} (\bibinfo {year} {2023})}\BibitemShut {NoStop}%
\bibitem [{\citenamefont {He}\ \emph {et~al.}(2014)\citenamefont {He},
  \citenamefont {Hummer},\ and\ \citenamefont
  {Franchini}}]{PhysRevB.89.075409}%
  \BibitemOpen
  \bibfield  {author} {\bibinfo {author} {\bibfnamefont {J.}~\bibnamefont
  {He}}, \bibinfo {author} {\bibfnamefont {K.}~\bibnamefont {Hummer}},\ and\
  \bibinfo {author} {\bibfnamefont {C.}~\bibnamefont {Franchini}},\ }\bibfield
  {title} {\bibinfo {title} {Stacking effects on the electronic and optical
  properties of bilayer transition metal dichalcogenides $\mathrm{MoS_2}$,
  $\mathrm{MoSe_2}$, $\mathrm{WS_2}$, and $\mathrm{WSe_2}$},\ }\href
  {https://doi.org/10.1103/PhysRevB.89.075409} {\bibfield  {journal} {\bibinfo
  {journal} {Phys. Rev. B}\ }\textbf {\bibinfo {volume} {89}},\ \bibinfo
  {pages} {075409} (\bibinfo {year} {2014})}\BibitemShut {NoStop}%
\bibitem [{\citenamefont {Huang}\ \emph {et~al.}(2017)\citenamefont {Huang},
  \citenamefont {Clark}, \citenamefont {Navarro-Moratalla}, \citenamefont
  {Klein}, \citenamefont {Cheng}, \citenamefont {Seyler}, \citenamefont
  {Zhong}, \citenamefont {Schmidgall}, \citenamefont {McGuire}, \citenamefont
  {Cobden}, \citenamefont {Yao}, \citenamefont {Xiao}, \citenamefont
  {Jarillo-Herrero},\ and\ \citenamefont {Xu}}]{a17}%
  \BibitemOpen
  \bibfield  {author} {\bibinfo {author} {\bibfnamefont {B.}~\bibnamefont
  {Huang}}, \bibinfo {author} {\bibfnamefont {G.}~\bibnamefont {Clark}},
  \bibinfo {author} {\bibfnamefont {E.}~\bibnamefont {Navarro-Moratalla}},
  \bibinfo {author} {\bibfnamefont {D.~R.}\ \bibnamefont {Klein}}, \bibinfo
  {author} {\bibfnamefont {R.}~\bibnamefont {Cheng}}, \bibinfo {author}
  {\bibfnamefont {K.~L.}\ \bibnamefont {Seyler}}, \bibinfo {author}
  {\bibfnamefont {D.}~\bibnamefont {Zhong}}, \bibinfo {author} {\bibfnamefont
  {E.}~\bibnamefont {Schmidgall}}, \bibinfo {author} {\bibfnamefont {M.~A.}\
  \bibnamefont {McGuire}}, \bibinfo {author} {\bibfnamefont {D.~H.}\
  \bibnamefont {Cobden}}, \bibinfo {author} {\bibfnamefont {W.}~\bibnamefont
  {Yao}}, \bibinfo {author} {\bibfnamefont {D.}~\bibnamefont {Xiao}}, \bibinfo
  {author} {\bibfnamefont {P.}~\bibnamefont {Jarillo-Herrero}},\ and\ \bibinfo
  {author} {\bibfnamefont {X.}~\bibnamefont {Xu}},\ }\bibfield  {title}
  {\bibinfo {title} {Layer-dependent ferromagnetism in a van der
  $\mathrm{W}$aals crystal down to the monolayer limit},\ }\href
  {https://doi.org/10.1038/nature22391} {\bibfield  {journal} {\bibinfo
  {journal} {Nature}\ }\textbf {\bibinfo {volume} {546}},\ \bibinfo {pages}
  {270} (\bibinfo {year} {2017})}\BibitemShut {NoStop}%
\bibitem [{\citenamefont {Xu}\ \emph {et~al.}(2022{\natexlab{a}})\citenamefont
  {Xu}, \citenamefont {Ray}, \citenamefont {Shao}, \citenamefont {Jiang},
  \citenamefont {Lee}, \citenamefont {Weber}, \citenamefont {Goldberger},
  \citenamefont {Watanabe}, \citenamefont {Taniguchi}, \citenamefont {Muller},
  \citenamefont {Mak},\ and\ \citenamefont {Shan}}]{a28}%
  \BibitemOpen
  \bibfield  {author} {\bibinfo {author} {\bibfnamefont {Y.}~\bibnamefont
  {Xu}}, \bibinfo {author} {\bibfnamefont {A.}~\bibnamefont {Ray}}, \bibinfo
  {author} {\bibfnamefont {Y.-T.}\ \bibnamefont {Shao}}, \bibinfo {author}
  {\bibfnamefont {S.}~\bibnamefont {Jiang}}, \bibinfo {author} {\bibfnamefont
  {K.}~\bibnamefont {Lee}}, \bibinfo {author} {\bibfnamefont {D.}~\bibnamefont
  {Weber}}, \bibinfo {author} {\bibfnamefont {J.~E.}\ \bibnamefont
  {Goldberger}}, \bibinfo {author} {\bibfnamefont {K.}~\bibnamefont
  {Watanabe}}, \bibinfo {author} {\bibfnamefont {T.}~\bibnamefont {Taniguchi}},
  \bibinfo {author} {\bibfnamefont {D.~A.}\ \bibnamefont {Muller}}, \bibinfo
  {author} {\bibfnamefont {K.~F.}\ \bibnamefont {Mak}},\ and\ \bibinfo {author}
  {\bibfnamefont {J.}~\bibnamefont {Shan}},\ }\bibfield  {title} {\bibinfo
  {title} {Coexisting ferromagnetic-antiferromagnetic state in twisted bilayer
  $\mathrm{CrI_3}$},\ }\href {https://doi.org/10.1038/s41565-021-01014-y}
  {\bibfield  {journal} {\bibinfo  {journal} {Nature Nanotech.}\ }\textbf
  {\bibinfo {volume} {17}},\ \bibinfo {pages} {143} (\bibinfo {year}
  {2022}{\natexlab{a}})}\BibitemShut {NoStop}%
\bibitem [{\citenamefont {Tong}\ \emph {et~al.}(2018)\citenamefont {Tong},
  \citenamefont {Liu}, \citenamefont {Xiao},\ and\ \citenamefont {Yao}}]{a29}%
  \BibitemOpen
  \bibfield  {author} {\bibinfo {author} {\bibfnamefont {Q.}~\bibnamefont
  {Tong}}, \bibinfo {author} {\bibfnamefont {F.}~\bibnamefont {Liu}}, \bibinfo
  {author} {\bibfnamefont {J.}~\bibnamefont {Xiao}},\ and\ \bibinfo {author}
  {\bibfnamefont {W.}~\bibnamefont {Yao}},\ }\bibfield  {title} {\bibinfo
  {title} {Skyrmions in the $\mathrm{Moir\acute{e}}$ of van der
  $\mathrm{Waals}$ 2$\mathrm{D}$ magnets},\ }\href
  {https://doi.org/10.1021/acs.nanolett.8b03315} {\bibfield  {journal}
  {\bibinfo  {journal} {Nano Lett.}\ }\textbf {\bibinfo {volume} {18}},\
  \bibinfo {pages} {7194} (\bibinfo {year} {2018})}\BibitemShut {NoStop}%
\bibitem [{\citenamefont {Jiang}\ \emph {et~al.}(2019)\citenamefont {Jiang},
  \citenamefont {Wang}, \citenamefont {Chen}, \citenamefont {Zhong},
  \citenamefont {Yuan}, \citenamefont {Lu},\ and\ \citenamefont {Ji}}]{a55}%
  \BibitemOpen
  \bibfield  {author} {\bibinfo {author} {\bibfnamefont {P.}~\bibnamefont
  {Jiang}}, \bibinfo {author} {\bibfnamefont {C.}~\bibnamefont {Wang}},
  \bibinfo {author} {\bibfnamefont {D.}~\bibnamefont {Chen}}, \bibinfo {author}
  {\bibfnamefont {Z.}~\bibnamefont {Zhong}}, \bibinfo {author} {\bibfnamefont
  {Z.}~\bibnamefont {Yuan}}, \bibinfo {author} {\bibfnamefont {Z.-Y.}\
  \bibnamefont {Lu}},\ and\ \bibinfo {author} {\bibfnamefont {W.}~\bibnamefont
  {Ji}},\ }\bibfield  {title} {\bibinfo {title} {Stacking tunable interlayer
  magnetism in bilayer $\mathrm{CrI_3}$},\ }\href
  {https://doi.org/10.1103/PhysRevB.99.144401} {\bibfield  {journal} {\bibinfo
  {journal} {Phys. Rev. B}\ }\textbf {\bibinfo {volume} {99}},\ \bibinfo
  {pages} {144401} (\bibinfo {year} {2019})}\BibitemShut {NoStop}%
\bibitem [{\citenamefont {Sivadas}\ \emph {et~al.}(2018)\citenamefont
  {Sivadas}, \citenamefont {Okamoto}, \citenamefont {Xu}, \citenamefont
  {Fennie},\ and\ \citenamefont {Xiao}}]{doi:10.1021/acs.nanolett.8b03321}%
  \BibitemOpen
  \bibfield  {author} {\bibinfo {author} {\bibfnamefont {N.}~\bibnamefont
  {Sivadas}}, \bibinfo {author} {\bibfnamefont {S.}~\bibnamefont {Okamoto}},
  \bibinfo {author} {\bibfnamefont {X.}~\bibnamefont {Xu}}, \bibinfo {author}
  {\bibfnamefont {C.~J.}\ \bibnamefont {Fennie}},\ and\ \bibinfo {author}
  {\bibfnamefont {D.}~\bibnamefont {Xiao}},\ }\bibfield  {title} {\bibinfo
  {title} {Stacking-dependent magnetism in bilayer $\mathrm{CrI_3}$},\ }\href
  {https://doi.org/10.1021/acs.nanolett.8b03321} {\bibfield  {journal}
  {\bibinfo  {journal} {Nano Lett.}\ }\textbf {\bibinfo {volume} {18}},\
  \bibinfo {pages} {7658} (\bibinfo {year} {2018})}\BibitemShut {NoStop}%
\bibitem [{\citenamefont {Wu}\ \emph {et~al.}(2019)\citenamefont {Wu},
  \citenamefont {Zhou}, \citenamefont {Cai}, \citenamefont {Cheung},
  \citenamefont {Liu}, \citenamefont {Huang}, \citenamefont {Lin},
  \citenamefont {Han}, \citenamefont {An}, \citenamefont {Wang} \emph
  {et~al.}}]{wu2019intrinsic}%
  \BibitemOpen
  \bibfield  {author} {\bibinfo {author} {\bibfnamefont {Z.}~\bibnamefont
  {Wu}}, \bibinfo {author} {\bibfnamefont {B.~T.}\ \bibnamefont {Zhou}},
  \bibinfo {author} {\bibfnamefont {X.}~\bibnamefont {Cai}}, \bibinfo {author}
  {\bibfnamefont {P.}~\bibnamefont {Cheung}}, \bibinfo {author} {\bibfnamefont
  {G.-B.}\ \bibnamefont {Liu}}, \bibinfo {author} {\bibfnamefont
  {M.}~\bibnamefont {Huang}}, \bibinfo {author} {\bibfnamefont
  {J.}~\bibnamefont {Lin}}, \bibinfo {author} {\bibfnamefont {T.}~\bibnamefont
  {Han}}, \bibinfo {author} {\bibfnamefont {L.}~\bibnamefont {An}}, \bibinfo
  {author} {\bibfnamefont {Y.}~\bibnamefont {Wang}}, \emph {et~al.},\
  }\bibfield  {title} {\bibinfo {title} {Intrinsic valley $\mathrm{Hall}$
  transport in atomically thin $\mathrm{MoS_2}$},\ }\href
  {https://doi.org/doi.org/10.1038/s41467-019-08629-9} {\bibfield  {journal}
  {\bibinfo  {journal} {Nat. Commun.}\ }\textbf {\bibinfo {volume} {10}},\
  \bibinfo {pages} {611} (\bibinfo {year} {2019})}\BibitemShut {NoStop}%
\bibitem [{\citenamefont {Fox}\ \emph {et~al.}(2024)\citenamefont {Fox},
  \citenamefont {Mao}, \citenamefont {Zhang}, \citenamefont {Wang},\ and\
  \citenamefont {Xiao}}]{doi:10.1021/acs.chemrev.3c00618}%
  \BibitemOpen
  \bibfield  {author} {\bibinfo {author} {\bibfnamefont {C.}~\bibnamefont
  {Fox}}, \bibinfo {author} {\bibfnamefont {Y.}~\bibnamefont {Mao}}, \bibinfo
  {author} {\bibfnamefont {X.}~\bibnamefont {Zhang}}, \bibinfo {author}
  {\bibfnamefont {Y.}~\bibnamefont {Wang}},\ and\ \bibinfo {author}
  {\bibfnamefont {J.}~\bibnamefont {Xiao}},\ }\bibfield  {title} {\bibinfo
  {title} {Stacking order engineering of two-dimensional materials and device
  applications},\ }\href {https://doi.org/10.1021/acs.chemrev.3c00618}
  {\bibfield  {journal} {\bibinfo  {journal} {Chem. Rev.}\ }\textbf {\bibinfo
  {volume} {124}},\ \bibinfo {pages} {1862} (\bibinfo {year}
  {2024})}\BibitemShut {NoStop}%
\bibitem [{\citenamefont {Yasuda}\ \emph {et~al.}(2021)\citenamefont {Yasuda},
  \citenamefont {Wang}, \citenamefont {Watanabe}, \citenamefont {Taniguchi},\
  and\ \citenamefont {Jarillo-Herrero}}]{doi:10.1126/science.abd3230}%
  \BibitemOpen
  \bibfield  {author} {\bibinfo {author} {\bibfnamefont {K.}~\bibnamefont
  {Yasuda}}, \bibinfo {author} {\bibfnamefont {X.}~\bibnamefont {Wang}},
  \bibinfo {author} {\bibfnamefont {K.}~\bibnamefont {Watanabe}}, \bibinfo
  {author} {\bibfnamefont {T.}~\bibnamefont {Taniguchi}},\ and\ \bibinfo
  {author} {\bibfnamefont {P.}~\bibnamefont {Jarillo-Herrero}},\ }\bibfield
  {title} {\bibinfo {title} {Stacking-engineered ferroelectricity in bilayer
  boron nitride},\ }\href {https://doi.org/10.1126/science.abd3230} {\bibfield
  {journal} {\bibinfo  {journal} {Science}\ }\textbf {\bibinfo {volume}
  {372}},\ \bibinfo {pages} {1458} (\bibinfo {year} {2021})}\BibitemShut
  {NoStop}%
\bibitem [{\citenamefont {Li}\ \emph {et~al.}(2023)\citenamefont {Li},
  \citenamefont {Shi}, \citenamefont {Marian}, \citenamefont {Soriano},
  \citenamefont {Cusati}, \citenamefont {Iannaccone}, \citenamefont {Fiori},
  \citenamefont {Guo}, \citenamefont {Zhao},\ and\ \citenamefont
  {Wu}}]{mobilityWS2}%
  \BibitemOpen
  \bibfield  {author} {\bibinfo {author} {\bibfnamefont {X.}~\bibnamefont
  {Li}}, \bibinfo {author} {\bibfnamefont {X.}~\bibnamefont {Shi}}, \bibinfo
  {author} {\bibfnamefont {D.}~\bibnamefont {Marian}}, \bibinfo {author}
  {\bibfnamefont {D.}~\bibnamefont {Soriano}}, \bibinfo {author} {\bibfnamefont
  {T.}~\bibnamefont {Cusati}}, \bibinfo {author} {\bibfnamefont
  {G.}~\bibnamefont {Iannaccone}}, \bibinfo {author} {\bibfnamefont
  {G.}~\bibnamefont {Fiori}}, \bibinfo {author} {\bibfnamefont
  {Q.}~\bibnamefont {Guo}}, \bibinfo {author} {\bibfnamefont {W.}~\bibnamefont
  {Zhao}},\ and\ \bibinfo {author} {\bibfnamefont {Y.}~\bibnamefont {Wu}},\
  }\bibfield  {title} {\bibinfo {title} {Rhombohedral-stacked bilayer
  transition metal dichalcogenides for high-performance atomically thin cmos
  devices},\ }\href {https://doi.org/10.1126/sciadv.ade5706} {\bibfield
  {journal} {\bibinfo  {journal} {Sci. Adv.}\ }\textbf {\bibinfo {volume}
  {9}},\ \bibinfo {pages} {eade5706} (\bibinfo {year} {2023})}\BibitemShut
  {NoStop}%
\bibitem [{\citenamefont {Cantos-Prieto}\ \emph {et~al.}(2021)\citenamefont
  {Cantos-Prieto}, \citenamefont {Falin}, \citenamefont {Alliati},
  \citenamefont {Qian}, \citenamefont {Zhang}, \citenamefont {Tao},
  \citenamefont {Barnett}, \citenamefont {Santos}, \citenamefont {Li},\ and\
  \citenamefont {Navarro-Moratalla}}]{doi:10.1021/acs.nanolett.0c04794}%
  \BibitemOpen
  \bibfield  {author} {\bibinfo {author} {\bibfnamefont {F.}~\bibnamefont
  {Cantos-Prieto}}, \bibinfo {author} {\bibfnamefont {A.}~\bibnamefont
  {Falin}}, \bibinfo {author} {\bibfnamefont {M.}~\bibnamefont {Alliati}},
  \bibinfo {author} {\bibfnamefont {D.}~\bibnamefont {Qian}}, \bibinfo {author}
  {\bibfnamefont {R.}~\bibnamefont {Zhang}}, \bibinfo {author} {\bibfnamefont
  {T.}~\bibnamefont {Tao}}, \bibinfo {author} {\bibfnamefont {M.~R.}\
  \bibnamefont {Barnett}}, \bibinfo {author} {\bibfnamefont {E.~J.~G.}\
  \bibnamefont {Santos}}, \bibinfo {author} {\bibfnamefont {L.~H.}\
  \bibnamefont {Li}},\ and\ \bibinfo {author} {\bibfnamefont {E.}~\bibnamefont
  {Navarro-Moratalla}},\ }\bibfield  {title} {\bibinfo {title} {Layer-dependent
  mechanical properties and enhanced plasticity in the van der $\mathrm{W}$aals
  chromium trihalide magnets},\ }\href
  {https://doi.org/10.1021/acs.nanolett.0c04794} {\bibfield  {journal}
  {\bibinfo  {journal} {Nano Lett.}\ }\textbf {\bibinfo {volume} {21}},\
  \bibinfo {pages} {3379} (\bibinfo {year} {2021})}\BibitemShut {NoStop}%
\bibitem [{\citenamefont {Yang}\ \emph {et~al.}(2023)\citenamefont {Yang},
  \citenamefont {Xu}, \citenamefont {Han}, \citenamefont {Gu}, \citenamefont
  {Guzman}, \citenamefont {Song}, \citenamefont {Lin}, \citenamefont {Gao},
  \citenamefont {Zhou}, \citenamefont {Yang}, \citenamefont {Chen},\ and\
  \citenamefont {Ye}}]{doi:10.1021/jacs.3c10777}%
  \BibitemOpen
  \bibfield  {author} {\bibinfo {author} {\bibfnamefont {S.}~\bibnamefont
  {Yang}}, \bibinfo {author} {\bibfnamefont {X.}~\bibnamefont {Xu}}, \bibinfo
  {author} {\bibfnamefont {B.}~\bibnamefont {Han}}, \bibinfo {author}
  {\bibfnamefont {P.}~\bibnamefont {Gu}}, \bibinfo {author} {\bibfnamefont
  {R.}~\bibnamefont {Guzman}}, \bibinfo {author} {\bibfnamefont
  {Y.}~\bibnamefont {Song}}, \bibinfo {author} {\bibfnamefont {Z.}~\bibnamefont
  {Lin}}, \bibinfo {author} {\bibfnamefont {P.}~\bibnamefont {Gao}}, \bibinfo
  {author} {\bibfnamefont {W.}~\bibnamefont {Zhou}}, \bibinfo {author}
  {\bibfnamefont {J.}~\bibnamefont {Yang}}, \bibinfo {author} {\bibfnamefont
  {Z.}~\bibnamefont {Chen}},\ and\ \bibinfo {author} {\bibfnamefont
  {Y.}~\bibnamefont {Ye}},\ }\bibfield  {title} {\bibinfo {title} {Controlling
  the 2$\mathrm{D}$ magnetism of $\mathrm{CrBr_3}$ by van der $\mathrm{W}$aals
  stacking engineering},\ }\href {https://doi.org/10.1021/jacs.3c10777}
  {\bibfield  {journal} {\bibinfo  {journal} {J. Am. Chem. Soc.}\ }\textbf
  {\bibinfo {volume} {145}},\ \bibinfo {pages} {28184} (\bibinfo {year}
  {2023})}\BibitemShut {NoStop}%
\bibitem [{\citenamefont {Park}(2016)}]{a7}%
  \BibitemOpen
  \bibfield  {author} {\bibinfo {author} {\bibfnamefont {J.-G.}\ \bibnamefont
  {Park}},\ }\bibfield  {title} {\bibinfo {title} {Opportunities and challenges
  of 2$\mathrm{D}$ magnetic van der $\mathrm{W}$aals materials: magnetic
  graphene?},\ }\href {https://doi.org/10.1088/0953-8984/28/30/301001}
  {\bibfield  {journal} {\bibinfo  {journal} {J. Phys.: Condens. Matter}\
  }\textbf {\bibinfo {volume} {28}},\ \bibinfo {pages} {301001} (\bibinfo
  {year} {2016})}\BibitemShut {NoStop}%
\bibitem [{\citenamefont {Samarth}(2017)}]{a8}%
  \BibitemOpen
  \bibfield  {author} {\bibinfo {author} {\bibfnamefont {N.}~\bibnamefont
  {Samarth}},\ }\bibfield  {title} {\bibinfo {title} {Magnetism in flatland},\
  }\href {https://doi.org/10.1038/546216a} {\bibfield  {journal} {\bibinfo
  {journal} {Nature}\ }\textbf {\bibinfo {volume} {546}},\ \bibinfo {pages}
  {216} (\bibinfo {year} {2017})}\BibitemShut {NoStop}%
\bibitem [{\citenamefont {Wang}\ \emph
  {et~al.}(2020{\natexlab{a}})\citenamefont {Wang}, \citenamefont {Huang},
  \citenamefont {Cheung}, \citenamefont {Chen}, \citenamefont {Tan},
  \citenamefont {Huang}, \citenamefont {Zhao}, \citenamefont {Zhao},
  \citenamefont {Wu}, \citenamefont {Feng}, \citenamefont {Wu},\ and\
  \citenamefont {Chang}}]{https://doi.org/10.1002/andp.201900452}%
  \BibitemOpen
  \bibfield  {author} {\bibinfo {author} {\bibfnamefont {M.-C.}\ \bibnamefont
  {Wang}}, \bibinfo {author} {\bibfnamefont {C.-C.}\ \bibnamefont {Huang}},
  \bibinfo {author} {\bibfnamefont {C.-H.}\ \bibnamefont {Cheung}}, \bibinfo
  {author} {\bibfnamefont {C.-Y.}\ \bibnamefont {Chen}}, \bibinfo {author}
  {\bibfnamefont {S.~G.}\ \bibnamefont {Tan}}, \bibinfo {author} {\bibfnamefont
  {T.-W.}\ \bibnamefont {Huang}}, \bibinfo {author} {\bibfnamefont
  {Y.}~\bibnamefont {Zhao}}, \bibinfo {author} {\bibfnamefont {Y.}~\bibnamefont
  {Zhao}}, \bibinfo {author} {\bibfnamefont {G.}~\bibnamefont {Wu}}, \bibinfo
  {author} {\bibfnamefont {Y.-P.}\ \bibnamefont {Feng}}, \bibinfo {author}
  {\bibfnamefont {H.-C.}\ \bibnamefont {Wu}},\ and\ \bibinfo {author}
  {\bibfnamefont {C.-R.}\ \bibnamefont {Chang}},\ }\bibfield  {title} {\bibinfo
  {title} {Prospects and opportunities of 2$\mathrm{D}$ van der
  $\mathrm{W}$aals magnetic systems},\ }\href
  {https://doi.org/https://doi.org/10.1002/andp.201900452} {\bibfield
  {journal} {\bibinfo  {journal} {Ann. Phys.}\ }\textbf {\bibinfo {volume}
  {532}},\ \bibinfo {pages} {1900452} (\bibinfo {year}
  {2020}{\natexlab{a}})}\BibitemShut {NoStop}%
\bibitem [{\citenamefont {López-Paz}\ \emph {et~al.}(2022)\citenamefont
  {López-Paz}, \citenamefont {Guguchia}, \citenamefont {Pomjakushin},
  \citenamefont {Witteveen}, \citenamefont {Cervellino}, \citenamefont
  {Luetkens}, \citenamefont {Casati}, \citenamefont {Morpurgo},\ and\
  \citenamefont {von Rohr}}]{a9}%
  \BibitemOpen
  \bibfield  {author} {\bibinfo {author} {\bibfnamefont {S.~A.}\ \bibnamefont
  {López-Paz}}, \bibinfo {author} {\bibfnamefont {Z.}~\bibnamefont
  {Guguchia}}, \bibinfo {author} {\bibfnamefont {V.~Y.}\ \bibnamefont
  {Pomjakushin}}, \bibinfo {author} {\bibfnamefont {C.}~\bibnamefont
  {Witteveen}}, \bibinfo {author} {\bibfnamefont {A.}~\bibnamefont
  {Cervellino}}, \bibinfo {author} {\bibfnamefont {H.}~\bibnamefont
  {Luetkens}}, \bibinfo {author} {\bibfnamefont {N.}~\bibnamefont {Casati}},
  \bibinfo {author} {\bibfnamefont {A.~F.}\ \bibnamefont {Morpurgo}},\ and\
  \bibinfo {author} {\bibfnamefont {F.~O.}\ \bibnamefont {von Rohr}},\
  }\bibfield  {title} {\bibinfo {title} {Dynamic magnetic crossover at the
  origin of the hidden-order in van der $\mathrm{W}$aals antiferromagnet
  $\mathrm{CrSBr}$},\ }\href {https://doi.org/10.1038/s41467-022-32290-4}
  {\bibfield  {journal} {\bibinfo  {journal} {Nat. Commun.}\ }\textbf {\bibinfo
  {volume} {13}},\ \bibinfo {pages} {4745} (\bibinfo {year}
  {2022})}\BibitemShut {NoStop}%
\bibitem [{\citenamefont {Klein}\ \emph {et~al.}(2018)\citenamefont {Klein},
  \citenamefont {MacNeill}, \citenamefont {Lado}, \citenamefont {Soriano},
  \citenamefont {Navarro-Moratalla}, \citenamefont {Watanabe}, \citenamefont
  {Taniguchi}, \citenamefont {Manni}, \citenamefont {Canfield}, \citenamefont
  {Fernández-Rossier},\ and\ \citenamefont {Jarillo-Herrero}}]{a10}%
  \BibitemOpen
  \bibfield  {author} {\bibinfo {author} {\bibfnamefont {D.~R.}\ \bibnamefont
  {Klein}}, \bibinfo {author} {\bibfnamefont {D.}~\bibnamefont {MacNeill}},
  \bibinfo {author} {\bibfnamefont {J.~L.}\ \bibnamefont {Lado}}, \bibinfo
  {author} {\bibfnamefont {D.}~\bibnamefont {Soriano}}, \bibinfo {author}
  {\bibfnamefont {E.}~\bibnamefont {Navarro-Moratalla}}, \bibinfo {author}
  {\bibfnamefont {K.}~\bibnamefont {Watanabe}}, \bibinfo {author}
  {\bibfnamefont {T.}~\bibnamefont {Taniguchi}}, \bibinfo {author}
  {\bibfnamefont {S.}~\bibnamefont {Manni}}, \bibinfo {author} {\bibfnamefont
  {P.~C.}\ \bibnamefont {Canfield}}, \bibinfo {author} {\bibfnamefont
  {J.}~\bibnamefont {Fernández-Rossier}},\ and\ \bibinfo {author}
  {\bibfnamefont {P.}~\bibnamefont {Jarillo-Herrero}},\ }\bibfield  {title}
  {\bibinfo {title} {Probing magnetism in 2$\mathrm{D}$ van der
  $\mathrm{W}$aals crystalline insulators via electron tunneling},\ }\href
  {https://doi.org/10.1126/science.aar3617} {\bibfield  {journal} {\bibinfo
  {journal} {Science}\ }\textbf {\bibinfo {volume} {360}},\ \bibinfo {pages}
  {1218} (\bibinfo {year} {2018})}\BibitemShut {NoStop}%
\bibitem [{\citenamefont {Telford}\ \emph {et~al.}(2022)\citenamefont
  {Telford}, \citenamefont {Dismukes}, \citenamefont {Dudley}, \citenamefont
  {Wiscons}, \citenamefont {Lee}, \citenamefont {Chica}, \citenamefont
  {Ziebel}, \citenamefont {Han}, \citenamefont {Yu}, \citenamefont {Shabani},
  \citenamefont {Scheie}, \citenamefont {Watanabe}, \citenamefont {Taniguchi},
  \citenamefont {Xiao}, \citenamefont {Zhu}, \citenamefont {Pasupathy},
  \citenamefont {Nuckolls}, \citenamefont {Zhu}, \citenamefont {Dean},\ and\
  \citenamefont {Roy}}]{a11}%
  \BibitemOpen
  \bibfield  {author} {\bibinfo {author} {\bibfnamefont {E.~J.}\ \bibnamefont
  {Telford}}, \bibinfo {author} {\bibfnamefont {A.~H.}\ \bibnamefont
  {Dismukes}}, \bibinfo {author} {\bibfnamefont {R.~L.}\ \bibnamefont
  {Dudley}}, \bibinfo {author} {\bibfnamefont {R.~A.}\ \bibnamefont {Wiscons}},
  \bibinfo {author} {\bibfnamefont {K.}~\bibnamefont {Lee}}, \bibinfo {author}
  {\bibfnamefont {D.~G.}\ \bibnamefont {Chica}}, \bibinfo {author}
  {\bibfnamefont {M.~E.}\ \bibnamefont {Ziebel}}, \bibinfo {author}
  {\bibfnamefont {M.-G.}\ \bibnamefont {Han}}, \bibinfo {author} {\bibfnamefont
  {J.}~\bibnamefont {Yu}}, \bibinfo {author} {\bibfnamefont {S.}~\bibnamefont
  {Shabani}}, \bibinfo {author} {\bibfnamefont {A.}~\bibnamefont {Scheie}},
  \bibinfo {author} {\bibfnamefont {K.}~\bibnamefont {Watanabe}}, \bibinfo
  {author} {\bibfnamefont {T.}~\bibnamefont {Taniguchi}}, \bibinfo {author}
  {\bibfnamefont {D.}~\bibnamefont {Xiao}}, \bibinfo {author} {\bibfnamefont
  {Y.}~\bibnamefont {Zhu}}, \bibinfo {author} {\bibfnamefont {A.~N.}\
  \bibnamefont {Pasupathy}}, \bibinfo {author} {\bibfnamefont {C.}~\bibnamefont
  {Nuckolls}}, \bibinfo {author} {\bibfnamefont {X.}~\bibnamefont {Zhu}},
  \bibinfo {author} {\bibfnamefont {C.~R.}\ \bibnamefont {Dean}},\ and\
  \bibinfo {author} {\bibfnamefont {X.}~\bibnamefont {Roy}},\ }\bibfield
  {title} {\bibinfo {title} {Coupling between magnetic order and charge
  transport in a two-dimensional magnetic semiconductor},\ }\href
  {https://doi.org/10.1038/s41563-022-01245-x} {\bibfield  {journal} {\bibinfo
  {journal} {Nat. Mater.}\ }\textbf {\bibinfo {volume} {21}},\ \bibinfo {pages}
  {754} (\bibinfo {year} {2022})}\BibitemShut {NoStop}%
\bibitem [{\citenamefont {Herrero}\ \emph {et~al.}(2016)\citenamefont
  {Herrero}, \citenamefont {Gómez-Rodríguez}, \citenamefont {Mallet},
  \citenamefont {Moaied}, \citenamefont {Palacios}, \citenamefont {Salgado},
  \citenamefont {Ugeda}, \citenamefont {Veuillen}, \citenamefont {Yndurain},\
  and\ \citenamefont {Brihuega}}]{a12}%
  \BibitemOpen
  \bibfield  {author} {\bibinfo {author} {\bibfnamefont {H.~G.}\ \bibnamefont
  {Herrero}}, \bibinfo {author} {\bibfnamefont {J.~M.}\ \bibnamefont
  {Gómez-Rodríguez}}, \bibinfo {author} {\bibfnamefont {P.}~\bibnamefont
  {Mallet}}, \bibinfo {author} {\bibfnamefont {M.}~\bibnamefont {Moaied}},
  \bibinfo {author} {\bibfnamefont {J.~J.}\ \bibnamefont {Palacios}}, \bibinfo
  {author} {\bibfnamefont {C.}~\bibnamefont {Salgado}}, \bibinfo {author}
  {\bibfnamefont {M.~M.}\ \bibnamefont {Ugeda}}, \bibinfo {author}
  {\bibfnamefont {J.-Y.}\ \bibnamefont {Veuillen}}, \bibinfo {author}
  {\bibfnamefont {F.}~\bibnamefont {Yndurain}},\ and\ \bibinfo {author}
  {\bibfnamefont {I.}~\bibnamefont {Brihuega}},\ }\bibfield  {title} {\bibinfo
  {title} {Atomic-scale control of graphene magnetism by using hydrogen
  atoms},\ }\href {https://doi.org/10.1126/science.aad8038} {\bibfield
  {journal} {\bibinfo  {journal} {Science}\ }\textbf {\bibinfo {volume}
  {352}},\ \bibinfo {pages} {437} (\bibinfo {year} {2016})}\BibitemShut
  {NoStop}%
\bibitem [{\citenamefont {Nair}\ \emph {et~al.}(2012)\citenamefont {Nair},
  \citenamefont {Sepioni}, \citenamefont {Tsai}, \citenamefont {Lehtinen},
  \citenamefont {Keinonen}, \citenamefont {Krasheninnikov}, \citenamefont
  {Thomson}, \citenamefont {Geim},\ and\ \citenamefont {Grigorieva}}]{a13}%
  \BibitemOpen
  \bibfield  {author} {\bibinfo {author} {\bibfnamefont {R.~R.}\ \bibnamefont
  {Nair}}, \bibinfo {author} {\bibfnamefont {M.}~\bibnamefont {Sepioni}},
  \bibinfo {author} {\bibfnamefont {I.-L.}\ \bibnamefont {Tsai}}, \bibinfo
  {author} {\bibfnamefont {O.}~\bibnamefont {Lehtinen}}, \bibinfo {author}
  {\bibfnamefont {J.}~\bibnamefont {Keinonen}}, \bibinfo {author}
  {\bibfnamefont {A.~V.}\ \bibnamefont {Krasheninnikov}}, \bibinfo {author}
  {\bibfnamefont {T.}~\bibnamefont {Thomson}}, \bibinfo {author} {\bibfnamefont
  {A.~K.}\ \bibnamefont {Geim}},\ and\ \bibinfo {author} {\bibfnamefont
  {I.~V.}\ \bibnamefont {Grigorieva}},\ }\bibfield  {title} {\bibinfo {title}
  {Spin-half paramagnetism in graphene induced by point defects},\ }\href
  {https://doi.org/10.1038/nphys2183} {\bibfield  {journal} {\bibinfo
  {journal} {Nature Phys.}\ }\textbf {\bibinfo {volume} {8}},\ \bibinfo {pages}
  {199} (\bibinfo {year} {2012})}\BibitemShut {NoStop}%
\bibitem [{\citenamefont {Červenka}\ \emph {et~al.}(2009)\citenamefont
  {Červenka}, \citenamefont {Katsnelson},\ and\ \citenamefont {Flipse}}]{a14}%
  \BibitemOpen
  \bibfield  {author} {\bibinfo {author} {\bibfnamefont {J.}~\bibnamefont
  {Červenka}}, \bibinfo {author} {\bibfnamefont {M.~I.}\ \bibnamefont
  {Katsnelson}},\ and\ \bibinfo {author} {\bibfnamefont {C.~F.~J.}\
  \bibnamefont {Flipse}},\ }\bibfield  {title} {\bibinfo {title}
  {Room-temperature ferromagnetism in graphite driven by two-dimensional
  networks of point defects},\ }\href {https://doi.org/10.1038/nphys1399}
  {\bibfield  {journal} {\bibinfo  {journal} {Nature Phys.}\ }\textbf {\bibinfo
  {volume} {5}},\ \bibinfo {pages} {840} (\bibinfo {year} {2009})}\BibitemShut
  {NoStop}%
\bibitem [{\citenamefont {Mermin}\ and\ \citenamefont {Wagner}(1966)}]{a15}%
  \BibitemOpen
  \bibfield  {author} {\bibinfo {author} {\bibfnamefont {N.~D.}\ \bibnamefont
  {Mermin}}\ and\ \bibinfo {author} {\bibfnamefont {H.}~\bibnamefont
  {Wagner}},\ }\bibfield  {title} {\bibinfo {title} {Absence of ferromagnetism
  or antiferromagnetism in one- or two-dimensional isotropic heisenberg
  models},\ }\href {https://doi.org/10.1103/PhysRevLett.17.1133} {\bibfield
  {journal} {\bibinfo  {journal} {Phys. Rev. Lett.}\ }\textbf {\bibinfo
  {volume} {17}},\ \bibinfo {pages} {1133} (\bibinfo {year}
  {1966})}\BibitemShut {NoStop}%
\bibitem [{\citenamefont {Ghazaryan}\ \emph {et~al.}(2018)\citenamefont
  {Ghazaryan}, \citenamefont {Greenaway}, \citenamefont {Wang}, \citenamefont
  {Guarochico-Moreira}, \citenamefont {Vera-Marun}, \citenamefont {Yin},
  \citenamefont {Liao}, \citenamefont {Morozov}, \citenamefont {Kristanovski},
  \citenamefont {Lichtenstein} \emph {et~al.}}]{ghazaryan2018magnon}%
  \BibitemOpen
  \bibfield  {author} {\bibinfo {author} {\bibfnamefont {D.}~\bibnamefont
  {Ghazaryan}}, \bibinfo {author} {\bibfnamefont {M.~T.}\ \bibnamefont
  {Greenaway}}, \bibinfo {author} {\bibfnamefont {Z.}~\bibnamefont {Wang}},
  \bibinfo {author} {\bibfnamefont {V.~H.}\ \bibnamefont {Guarochico-Moreira}},
  \bibinfo {author} {\bibfnamefont {I.~J.}\ \bibnamefont {Vera-Marun}},
  \bibinfo {author} {\bibfnamefont {J.}~\bibnamefont {Yin}}, \bibinfo {author}
  {\bibfnamefont {Y.}~\bibnamefont {Liao}}, \bibinfo {author} {\bibfnamefont
  {S.~V.}\ \bibnamefont {Morozov}}, \bibinfo {author} {\bibfnamefont
  {O.}~\bibnamefont {Kristanovski}}, \bibinfo {author} {\bibfnamefont {A.~I.}\
  \bibnamefont {Lichtenstein}}, \emph {et~al.},\ }\bibfield  {title} {\bibinfo
  {title} {Magnon-assisted tunnelling in van der $\mathrm{W}$aals
  heterostructures based on $\mathrm{CrBr_3}$},\ }\href
  {https://doi.org/10.1038/s41928-018-0087-z} {\bibfield  {journal} {\bibinfo
  {journal} {Nat. Electron.}\ }\textbf {\bibinfo {volume} {1}},\ \bibinfo
  {pages} {344} (\bibinfo {year} {2018})}\BibitemShut {NoStop}%
\bibitem [{\citenamefont {Cai}\ \emph {et~al.}(2019)\citenamefont {Cai},
  \citenamefont {Song}, \citenamefont {Wilson}, \citenamefont {Clark},
  \citenamefont {He}, \citenamefont {Zhang}, \citenamefont {Taniguchi},
  \citenamefont {Watanabe}, \citenamefont {Yao}, \citenamefont {Xiao},
  \citenamefont {McGuire}, \citenamefont {Cobden},\ and\ \citenamefont
  {Xu}}]{doi:10.1021/acs.nanolett.9b01317}%
  \BibitemOpen
  \bibfield  {author} {\bibinfo {author} {\bibfnamefont {X.}~\bibnamefont
  {Cai}}, \bibinfo {author} {\bibfnamefont {T.}~\bibnamefont {Song}}, \bibinfo
  {author} {\bibfnamefont {N.~P.}\ \bibnamefont {Wilson}}, \bibinfo {author}
  {\bibfnamefont {G.}~\bibnamefont {Clark}}, \bibinfo {author} {\bibfnamefont
  {M.}~\bibnamefont {He}}, \bibinfo {author} {\bibfnamefont {X.}~\bibnamefont
  {Zhang}}, \bibinfo {author} {\bibfnamefont {T.}~\bibnamefont {Taniguchi}},
  \bibinfo {author} {\bibfnamefont {K.}~\bibnamefont {Watanabe}}, \bibinfo
  {author} {\bibfnamefont {W.}~\bibnamefont {Yao}}, \bibinfo {author}
  {\bibfnamefont {D.}~\bibnamefont {Xiao}}, \bibinfo {author} {\bibfnamefont
  {M.~A.}\ \bibnamefont {McGuire}}, \bibinfo {author} {\bibfnamefont {D.~H.}\
  \bibnamefont {Cobden}},\ and\ \bibinfo {author} {\bibfnamefont
  {X.}~\bibnamefont {Xu}},\ }\bibfield  {title} {\bibinfo {title} {Atomically
  thin $\mathrm{CrCl_3}$: $\mathrm{An}$ in-plane layered antiferromagnetic
  insulator},\ }\href {https://doi.org/10.1021/acs.nanolett.9b01317} {\bibfield
   {journal} {\bibinfo  {journal} {Nano Lett.}\ }\textbf {\bibinfo {volume}
  {19}},\ \bibinfo {pages} {3993} (\bibinfo {year} {2019})}\BibitemShut
  {NoStop}%
\bibitem [{\citenamefont {Gong}\ \emph {et~al.}(2017)\citenamefont {Gong},
  \citenamefont {Li}, \citenamefont {Li}, \citenamefont {Ji}, \citenamefont
  {Stern}, \citenamefont {Xia}, \citenamefont {Cao}, \citenamefont {Bao},
  \citenamefont {Wang}, \citenamefont {Wang}, \citenamefont {Qiu},
  \citenamefont {Cava}, \citenamefont {Louie}, \citenamefont {Xia},\ and\
  \citenamefont {Zhang}}]{a16}%
  \BibitemOpen
  \bibfield  {author} {\bibinfo {author} {\bibfnamefont {C.}~\bibnamefont
  {Gong}}, \bibinfo {author} {\bibfnamefont {L.}~\bibnamefont {Li}}, \bibinfo
  {author} {\bibfnamefont {Z.}~\bibnamefont {Li}}, \bibinfo {author}
  {\bibfnamefont {H.}~\bibnamefont {Ji}}, \bibinfo {author} {\bibfnamefont
  {A.}~\bibnamefont {Stern}}, \bibinfo {author} {\bibfnamefont
  {Y.}~\bibnamefont {Xia}}, \bibinfo {author} {\bibfnamefont {T.}~\bibnamefont
  {Cao}}, \bibinfo {author} {\bibfnamefont {W.}~\bibnamefont {Bao}}, \bibinfo
  {author} {\bibfnamefont {C.}~\bibnamefont {Wang}}, \bibinfo {author}
  {\bibfnamefont {Y.}~\bibnamefont {Wang}}, \bibinfo {author} {\bibfnamefont
  {Z.~Q.}\ \bibnamefont {Qiu}}, \bibinfo {author} {\bibfnamefont {R.~J.}\
  \bibnamefont {Cava}}, \bibinfo {author} {\bibfnamefont {S.~G.}\ \bibnamefont
  {Louie}}, \bibinfo {author} {\bibfnamefont {J.}~\bibnamefont {Xia}},\ and\
  \bibinfo {author} {\bibfnamefont {X.}~\bibnamefont {Zhang}},\ }\bibfield
  {title} {\bibinfo {title} {Discovery of intrinsic ferromagnetism in
  two-dimensional van der $\mathrm{W}$aals crystals},\ }\href
  {https://doi.org/10.1038/nature22060} {\bibfield  {journal} {\bibinfo
  {journal} {Nature}\ }\textbf {\bibinfo {volume} {546}},\ \bibinfo {pages}
  {265} (\bibinfo {year} {2017})}\BibitemShut {NoStop}%
\bibitem [{\citenamefont {Kuo}\ \emph {et~al.}(2016)\citenamefont {Kuo},
  \citenamefont {Neumann}, \citenamefont {Balamurugan}, \citenamefont {Park},
  \citenamefont {Kang}, \citenamefont {Shiu}, \citenamefont {Kang},
  \citenamefont {Hong}, \citenamefont {Han}, \citenamefont {Noh},\ and\
  \citenamefont {Park}}]{10.1038/srep20904}%
  \BibitemOpen
  \bibfield  {author} {\bibinfo {author} {\bibfnamefont {C.-T.}\ \bibnamefont
  {Kuo}}, \bibinfo {author} {\bibfnamefont {M.}~\bibnamefont {Neumann}},
  \bibinfo {author} {\bibfnamefont {K.}~\bibnamefont {Balamurugan}}, \bibinfo
  {author} {\bibfnamefont {H.~J.}\ \bibnamefont {Park}}, \bibinfo {author}
  {\bibfnamefont {S.}~\bibnamefont {Kang}}, \bibinfo {author} {\bibfnamefont
  {H.~W.}\ \bibnamefont {Shiu}}, \bibinfo {author} {\bibfnamefont {J.~H.}\
  \bibnamefont {Kang}}, \bibinfo {author} {\bibfnamefont {B.~H.}\ \bibnamefont
  {Hong}}, \bibinfo {author} {\bibfnamefont {M.}~\bibnamefont {Han}}, \bibinfo
  {author} {\bibfnamefont {T.~W.}\ \bibnamefont {Noh}},\ and\ \bibinfo {author}
  {\bibfnamefont {J.-G.}\ \bibnamefont {Park}},\ }\bibfield  {title} {\bibinfo
  {title} {Exfoliation and raman spectroscopic fingerprint of few-layer
  $\mathrm{NiPS_3}$ van der $\mathrm{W}$aals crystals},\ }\href
  {https://doi.org/10.1038/srep20904} {\bibfield  {journal} {\bibinfo
  {journal} {Sci. Rep.}\ }\textbf {\bibinfo {volume} {6}},\ \bibinfo {pages}
  {20904} (\bibinfo {year} {2016})}\BibitemShut {NoStop}%
\bibitem [{\citenamefont {Lee}\ \emph {et~al.}(2016)\citenamefont {Lee},
  \citenamefont {Lee}, \citenamefont {Ryoo}, \citenamefont {Kang},
  \citenamefont {Kim}, \citenamefont {Kim}, \citenamefont {Park}, \citenamefont
  {Park},\ and\ \citenamefont {Cheong}}]{10.1021/acs.nanolett.6b03052}%
  \BibitemOpen
  \bibfield  {author} {\bibinfo {author} {\bibfnamefont {J.-U.}\ \bibnamefont
  {Lee}}, \bibinfo {author} {\bibfnamefont {S.}~\bibnamefont {Lee}}, \bibinfo
  {author} {\bibfnamefont {J.~H.}\ \bibnamefont {Ryoo}}, \bibinfo {author}
  {\bibfnamefont {S.}~\bibnamefont {Kang}}, \bibinfo {author} {\bibfnamefont
  {T.~Y.}\ \bibnamefont {Kim}}, \bibinfo {author} {\bibfnamefont
  {P.}~\bibnamefont {Kim}}, \bibinfo {author} {\bibfnamefont {C.-H.}\
  \bibnamefont {Park}}, \bibinfo {author} {\bibfnamefont {J.-G.}\ \bibnamefont
  {Park}},\ and\ \bibinfo {author} {\bibfnamefont {H.}~\bibnamefont {Cheong}},\
  }\bibfield  {title} {\bibinfo {title} {Ising-type magnetic ordering in
  atomically thin $\mathrm{FePS_3}$},\ }\href
  {https://doi.org/10.1021/acs.nanolett.6b03052} {\bibfield  {journal}
  {\bibinfo  {journal} {Nano Lett.}\ }\textbf {\bibinfo {volume} {16}},\
  \bibinfo {pages} {7433} (\bibinfo {year} {2016})}\BibitemShut {NoStop}%
\bibitem [{\citenamefont {Li}\ \emph {et~al.}(2021)\citenamefont {Li},
  \citenamefont {Wan}, \citenamefont {Wang}, \citenamefont {Chen},
  \citenamefont {Huang}, \citenamefont {Cheng}, \citenamefont {Qian},
  \citenamefont {Li}, \citenamefont {Zhang}, \citenamefont {Sun}, \citenamefont
  {Zhao}, \citenamefont {Ma}, \citenamefont {Wu}, \citenamefont {Wei},
  \citenamefont {Liu}, \citenamefont {Liao}, \citenamefont {Ye}, \citenamefont
  {Huang}, \citenamefont {Xu}, \citenamefont {Duan}, \citenamefont {Ji},\ and\
  \citenamefont {Duan}}]{10.1038/s41563-021-00927-2}%
  \BibitemOpen
  \bibfield  {author} {\bibinfo {author} {\bibfnamefont {B.}~\bibnamefont
  {Li}}, \bibinfo {author} {\bibfnamefont {Z.}~\bibnamefont {Wan}}, \bibinfo
  {author} {\bibfnamefont {C.}~\bibnamefont {Wang}}, \bibinfo {author}
  {\bibfnamefont {P.}~\bibnamefont {Chen}}, \bibinfo {author} {\bibfnamefont
  {B.}~\bibnamefont {Huang}}, \bibinfo {author} {\bibfnamefont
  {X.}~\bibnamefont {Cheng}}, \bibinfo {author} {\bibfnamefont
  {Q.}~\bibnamefont {Qian}}, \bibinfo {author} {\bibfnamefont {J.}~\bibnamefont
  {Li}}, \bibinfo {author} {\bibfnamefont {Z.}~\bibnamefont {Zhang}}, \bibinfo
  {author} {\bibfnamefont {G.}~\bibnamefont {Sun}}, \bibinfo {author}
  {\bibfnamefont {B.}~\bibnamefont {Zhao}}, \bibinfo {author} {\bibfnamefont
  {H.}~\bibnamefont {Ma}}, \bibinfo {author} {\bibfnamefont {R.}~\bibnamefont
  {Wu}}, \bibinfo {author} {\bibfnamefont {Z.}~\bibnamefont {Wei}}, \bibinfo
  {author} {\bibfnamefont {Y.}~\bibnamefont {Liu}}, \bibinfo {author}
  {\bibfnamefont {L.}~\bibnamefont {Liao}}, \bibinfo {author} {\bibfnamefont
  {Y.}~\bibnamefont {Ye}}, \bibinfo {author} {\bibfnamefont {Y.}~\bibnamefont
  {Huang}}, \bibinfo {author} {\bibfnamefont {X.}~\bibnamefont {Xu}}, \bibinfo
  {author} {\bibfnamefont {X.}~\bibnamefont {Duan}}, \bibinfo {author}
  {\bibfnamefont {W.}~\bibnamefont {Ji}},\ and\ \bibinfo {author}
  {\bibfnamefont {X.}~\bibnamefont {Duan}},\ }\bibfield  {title} {\bibinfo
  {title} {van der $\mathrm{W}$aals epitaxial growth of air-stable
  $\mathrm{CrSe_2}$ nanosheets with thickness-tunable magnetic order},\ }\href
  {https://doi.org/10.1038/s41563-021-00927-2} {\bibfield  {journal} {\bibinfo
  {journal} {Nat. Mater.}\ }\textbf {\bibinfo {volume} {20}},\ \bibinfo {pages}
  {818} (\bibinfo {year} {2021})}\BibitemShut {NoStop}%
\bibitem [{\citenamefont {Zhang}\ \emph {et~al.}(2021)\citenamefont {Zhang},
  \citenamefont {Lu}, \citenamefont {Liu}, \citenamefont {Niu}, \citenamefont
  {Sun}, \citenamefont {Cook}, \citenamefont {Vaninger}, \citenamefont
  {Miceli}, \citenamefont {Singh}, \citenamefont {Lian}, \citenamefont {Chang},
  \citenamefont {He}, \citenamefont {Du}, \citenamefont {He}, \citenamefont
  {Zhang}, \citenamefont {Bian},\ and\ \citenamefont
  {Xu}}]{WOS:000656508100006}%
  \BibitemOpen
  \bibfield  {author} {\bibinfo {author} {\bibfnamefont {X.}~\bibnamefont
  {Zhang}}, \bibinfo {author} {\bibfnamefont {Q.}~\bibnamefont {Lu}}, \bibinfo
  {author} {\bibfnamefont {W.}~\bibnamefont {Liu}}, \bibinfo {author}
  {\bibfnamefont {W.}~\bibnamefont {Niu}}, \bibinfo {author} {\bibfnamefont
  {J.}~\bibnamefont {Sun}}, \bibinfo {author} {\bibfnamefont {J.}~\bibnamefont
  {Cook}}, \bibinfo {author} {\bibfnamefont {M.}~\bibnamefont {Vaninger}},
  \bibinfo {author} {\bibfnamefont {P.~F.}\ \bibnamefont {Miceli}}, \bibinfo
  {author} {\bibfnamefont {D.~J.}\ \bibnamefont {Singh}}, \bibinfo {author}
  {\bibfnamefont {S.-W.}\ \bibnamefont {Lian}}, \bibinfo {author}
  {\bibfnamefont {T.-R.}\ \bibnamefont {Chang}}, \bibinfo {author}
  {\bibfnamefont {X.}~\bibnamefont {He}}, \bibinfo {author} {\bibfnamefont
  {J.}~\bibnamefont {Du}}, \bibinfo {author} {\bibfnamefont {L.}~\bibnamefont
  {He}}, \bibinfo {author} {\bibfnamefont {R.}~\bibnamefont {Zhang}}, \bibinfo
  {author} {\bibfnamefont {G.}~\bibnamefont {Bian}},\ and\ \bibinfo {author}
  {\bibfnamefont {Y.}~\bibnamefont {Xu}},\ }\bibfield  {title} {\bibinfo
  {title} {Room-temperature intrinsic ferromagnetism in epitaxial
  $\mathrm{CrTe_2}$ ultrathin films},\ }\href
  {https://doi.org/10.1038/s41467-021-22777-x} {\bibfield  {journal} {\bibinfo
  {journal} {Nat. Commun.}\ }\textbf {\bibinfo {volume} {12}},\ \bibinfo
  {pages} {2492} (\bibinfo {year} {2021})}\BibitemShut {NoStop}%
\bibitem [{\citenamefont {Meng}\ \emph {et~al.}(2021)\citenamefont {Meng},
  \citenamefont {Zhou}, \citenamefont {Xu}, \citenamefont {Yang}, \citenamefont
  {Si}, \citenamefont {Liu}, \citenamefont {Wang}, \citenamefont {Jiang},
  \citenamefont {Li}, \citenamefont {Qin}, \citenamefont {Zhang}, \citenamefont
  {Wang}, \citenamefont {Liu}, \citenamefont {Tang}, \citenamefont {Ye},
  \citenamefont {Zhou}, \citenamefont {Bao}, \citenamefont {Gao},\ and\
  \citenamefont {Gong}}]{WOS:000617499600011}%
  \BibitemOpen
  \bibfield  {author} {\bibinfo {author} {\bibfnamefont {L.}~\bibnamefont
  {Meng}}, \bibinfo {author} {\bibfnamefont {Z.}~\bibnamefont {Zhou}}, \bibinfo
  {author} {\bibfnamefont {M.}~\bibnamefont {Xu}}, \bibinfo {author}
  {\bibfnamefont {S.}~\bibnamefont {Yang}}, \bibinfo {author} {\bibfnamefont
  {K.}~\bibnamefont {Si}}, \bibinfo {author} {\bibfnamefont {L.}~\bibnamefont
  {Liu}}, \bibinfo {author} {\bibfnamefont {X.}~\bibnamefont {Wang}}, \bibinfo
  {author} {\bibfnamefont {H.}~\bibnamefont {Jiang}}, \bibinfo {author}
  {\bibfnamefont {B.}~\bibnamefont {Li}}, \bibinfo {author} {\bibfnamefont
  {P.}~\bibnamefont {Qin}}, \bibinfo {author} {\bibfnamefont {P.}~\bibnamefont
  {Zhang}}, \bibinfo {author} {\bibfnamefont {J.}~\bibnamefont {Wang}},
  \bibinfo {author} {\bibfnamefont {Z.}~\bibnamefont {Liu}}, \bibinfo {author}
  {\bibfnamefont {P.}~\bibnamefont {Tang}}, \bibinfo {author} {\bibfnamefont
  {Y.}~\bibnamefont {Ye}}, \bibinfo {author} {\bibfnamefont {W.}~\bibnamefont
  {Zhou}}, \bibinfo {author} {\bibfnamefont {L.}~\bibnamefont {Bao}}, \bibinfo
  {author} {\bibfnamefont {H.-J.}\ \bibnamefont {Gao}},\ and\ \bibinfo {author}
  {\bibfnamefont {Y.}~\bibnamefont {Gong}},\ }\bibfield  {title} {\bibinfo
  {title} {Anomalous thickness dependence of $\mathrm{Curie}$ temperature in
  air-stable two-dimensional ferromagnetic $\mathrm{1T-CrTe_2}$ grown by
  chemical vapor deposition},\ }\href
  {https://doi.org/10.1038/s41467-021-21072-z} {\bibfield  {journal} {\bibinfo
  {journal} {Nat. Commun.}\ }\textbf {\bibinfo {volume} {12}},\ \bibinfo
  {pages} {809} (\bibinfo {year} {2021})}\BibitemShut {NoStop}%
\bibitem [{\citenamefont {Xian}\ \emph {et~al.}(2022)\citenamefont {Xian},
  \citenamefont {Wang}, \citenamefont {Nie}, \citenamefont {Li}, \citenamefont
  {Han}, \citenamefont {Lin}, \citenamefont {Zhang}, \citenamefont {Liu},
  \citenamefont {Zhang}, \citenamefont {Miao}, \citenamefont {Yi},
  \citenamefont {Wu}, \citenamefont {Chen}, \citenamefont {Han}, \citenamefont
  {Xia}, \citenamefont {Ji},\ and\ \citenamefont {Fu}}]{WOS:000741852200021}%
  \BibitemOpen
  \bibfield  {author} {\bibinfo {author} {\bibfnamefont {J.-J.}\ \bibnamefont
  {Xian}}, \bibinfo {author} {\bibfnamefont {C.}~\bibnamefont {Wang}}, \bibinfo
  {author} {\bibfnamefont {J.-H.}\ \bibnamefont {Nie}}, \bibinfo {author}
  {\bibfnamefont {R.}~\bibnamefont {Li}}, \bibinfo {author} {\bibfnamefont
  {M.}~\bibnamefont {Han}}, \bibinfo {author} {\bibfnamefont {J.}~\bibnamefont
  {Lin}}, \bibinfo {author} {\bibfnamefont {W.-H.}\ \bibnamefont {Zhang}},
  \bibinfo {author} {\bibfnamefont {Z.-Y.}\ \bibnamefont {Liu}}, \bibinfo
  {author} {\bibfnamefont {Z.-M.}\ \bibnamefont {Zhang}}, \bibinfo {author}
  {\bibfnamefont {M.-P.}\ \bibnamefont {Miao}}, \bibinfo {author}
  {\bibfnamefont {Y.}~\bibnamefont {Yi}}, \bibinfo {author} {\bibfnamefont
  {S.}~\bibnamefont {Wu}}, \bibinfo {author} {\bibfnamefont {X.}~\bibnamefont
  {Chen}}, \bibinfo {author} {\bibfnamefont {J.}~\bibnamefont {Han}}, \bibinfo
  {author} {\bibfnamefont {Z.}~\bibnamefont {Xia}}, \bibinfo {author}
  {\bibfnamefont {W.}~\bibnamefont {Ji}},\ and\ \bibinfo {author}
  {\bibfnamefont {Y.-S.}\ \bibnamefont {Fu}},\ }\bibfield  {title} {\bibinfo
  {title} {Spin mapping of intralayer antiferromagnetism and field-induced spin
  reorientation in monolayer $\mathrm{CrTe_2}$},\ }\href
  {https://doi.org/10.1038/s41467-021-27834-z} {\bibfield  {journal} {\bibinfo
  {journal} {Nat. Commun.}\ }\textbf {\bibinfo {volume} {13}},\ \bibinfo
  {pages} {257} (\bibinfo {year} {2022})}\BibitemShut {NoStop}%
\bibitem [{\citenamefont {Xu}\ \emph {et~al.}(2018)\citenamefont {Xu},
  \citenamefont {Feng}, \citenamefont {Xiang},\ and\ \citenamefont
  {Bellaiche}}]{a18}%
  \BibitemOpen
  \bibfield  {author} {\bibinfo {author} {\bibfnamefont {C.}~\bibnamefont
  {Xu}}, \bibinfo {author} {\bibfnamefont {J.}~\bibnamefont {Feng}}, \bibinfo
  {author} {\bibfnamefont {H.}~\bibnamefont {Xiang}},\ and\ \bibinfo {author}
  {\bibfnamefont {L.}~\bibnamefont {Bellaiche}},\ }\bibfield  {title} {\bibinfo
  {title} {Interplay between $\mathrm{Kitaev}$ interaction and single ion
  anisotropy in ferromagnetic $\mathrm{CrI_3}$ and $\mathrm{CrGeTe_3}$
  monolayers},\ }\href {https://doi.org/10.1038/s41524-018-0115-6} {\bibfield
  {journal} {\bibinfo  {journal} {npj Comput. Mater.}\ }\textbf {\bibinfo
  {volume} {4}},\ \bibinfo {pages} {57} (\bibinfo {year} {2018})}\BibitemShut
  {NoStop}%
\bibitem [{\citenamefont {Ni}\ \emph {et~al.}(2021)\citenamefont {Ni},
  \citenamefont {Haglund}, \citenamefont {Wang}, \citenamefont {Xu},
  \citenamefont {Bernhard}, \citenamefont {Mandrus}, \citenamefont {Qian},
  \citenamefont {Mele}, \citenamefont {Kane},\ and\ \citenamefont {Wu}}]{a19}%
  \BibitemOpen
  \bibfield  {author} {\bibinfo {author} {\bibfnamefont {Z.}~\bibnamefont
  {Ni}}, \bibinfo {author} {\bibfnamefont {A.~V.}\ \bibnamefont {Haglund}},
  \bibinfo {author} {\bibfnamefont {H.}~\bibnamefont {Wang}}, \bibinfo {author}
  {\bibfnamefont {B.}~\bibnamefont {Xu}}, \bibinfo {author} {\bibfnamefont
  {C.}~\bibnamefont {Bernhard}}, \bibinfo {author} {\bibfnamefont {D.~G.}\
  \bibnamefont {Mandrus}}, \bibinfo {author} {\bibfnamefont {X.}~\bibnamefont
  {Qian}}, \bibinfo {author} {\bibfnamefont {E.~J.}\ \bibnamefont {Mele}},
  \bibinfo {author} {\bibfnamefont {C.~L.}\ \bibnamefont {Kane}},\ and\
  \bibinfo {author} {\bibfnamefont {L.}~\bibnamefont {Wu}},\ }\bibfield
  {title} {\bibinfo {title} {Imaging the $\mathrm{N\acute{e}el}$ vector
  switching in the monolayer antiferromagnet $\mathrm{MnPSe_3}$ with
  strain-controlled $\mathrm{Ising}$ order},\ }\href
  {https://doi.org/10.1038/s41565-021-00885-5} {\bibfield  {journal} {\bibinfo
  {journal} {Nature Nanotech.}\ }\textbf {\bibinfo {volume} {16}},\ \bibinfo
  {pages} {782} (\bibinfo {year} {2021})}\BibitemShut {NoStop}%
\bibitem [{\citenamefont {Gong}\ and\ \citenamefont {Zhang}(2019)}]{a20}%
  \BibitemOpen
  \bibfield  {author} {\bibinfo {author} {\bibfnamefont {C.}~\bibnamefont
  {Gong}}\ and\ \bibinfo {author} {\bibfnamefont {X.}~\bibnamefont {Zhang}},\
  }\bibfield  {title} {\bibinfo {title} {Two-dimensional magnetic crystals and
  emergent heterostructure devices},\ }\href
  {https://doi.org/10.1126/science.aav4450} {\bibfield  {journal} {\bibinfo
  {journal} {Science}\ }\textbf {\bibinfo {volume} {363}},\ \bibinfo {pages}
  {eaav4450} (\bibinfo {year} {2019})}\BibitemShut {NoStop}%
\bibitem [{\citenamefont {Lee}\ \emph {et~al.}(2021)\citenamefont {Lee},
  \citenamefont {Dismukes}, \citenamefont {Telford}, \citenamefont {Wiscons},
  \citenamefont {Wang}, \citenamefont {Xu}, \citenamefont {Nuckolls},
  \citenamefont {Dean}, \citenamefont {Roy},\ and\ \citenamefont {Zhu}}]{a25}%
  \BibitemOpen
  \bibfield  {author} {\bibinfo {author} {\bibfnamefont {K.}~\bibnamefont
  {Lee}}, \bibinfo {author} {\bibfnamefont {A.~H.}\ \bibnamefont {Dismukes}},
  \bibinfo {author} {\bibfnamefont {E.~J.}\ \bibnamefont {Telford}}, \bibinfo
  {author} {\bibfnamefont {R.~A.}\ \bibnamefont {Wiscons}}, \bibinfo {author}
  {\bibfnamefont {J.}~\bibnamefont {Wang}}, \bibinfo {author} {\bibfnamefont
  {X.}~\bibnamefont {Xu}}, \bibinfo {author} {\bibfnamefont {C.}~\bibnamefont
  {Nuckolls}}, \bibinfo {author} {\bibfnamefont {C.~R.}\ \bibnamefont {Dean}},
  \bibinfo {author} {\bibfnamefont {X.}~\bibnamefont {Roy}},\ and\ \bibinfo
  {author} {\bibfnamefont {X.}~\bibnamefont {Zhu}},\ }\bibfield  {title}
  {\bibinfo {title} {Magnetic order and symmetry in the 2$\mathrm{D}$
  semiconductor $\mathrm{CrSBr}$},\ }\href
  {https://doi.org/10.1021/acs.nanolett.1c00219} {\bibfield  {journal}
  {\bibinfo  {journal} {Nano Lett.}\ }\textbf {\bibinfo {volume} {21}},\
  \bibinfo {pages} {3511} (\bibinfo {year} {2021})}\BibitemShut {NoStop}%
\bibitem [{\citenamefont {Klein}\ \emph {et~al.}(2023)\citenamefont {Klein},
  \citenamefont {Pingault}, \citenamefont {Florian}, \citenamefont
  {Heißenbüttel}, \citenamefont {Steinhoff}, \citenamefont {Song},
  \citenamefont {Torres}, \citenamefont {Dirnberger}, \citenamefont {Curtis},
  \citenamefont {Weile}, \citenamefont {Penn}, \citenamefont {Deilmann},
  \citenamefont {Dana}, \citenamefont {Bushati}, \citenamefont {Quan},
  \citenamefont {Luxa}, \citenamefont {Sofer}, \citenamefont {Alù},
  \citenamefont {Menon}, \citenamefont {Wurstbauer}, \citenamefont {Rohlfing},
  \citenamefont {Narang}, \citenamefont {Lončar},\ and\ \citenamefont
  {Ross}}]{doi:10.1021/acsnano.2c07316}%
  \BibitemOpen
  \bibfield  {author} {\bibinfo {author} {\bibfnamefont {J.}~\bibnamefont
  {Klein}}, \bibinfo {author} {\bibfnamefont {B.}~\bibnamefont {Pingault}},
  \bibinfo {author} {\bibfnamefont {M.}~\bibnamefont {Florian}}, \bibinfo
  {author} {\bibfnamefont {M.-C.}\ \bibnamefont {Heißenbüttel}}, \bibinfo
  {author} {\bibfnamefont {A.}~\bibnamefont {Steinhoff}}, \bibinfo {author}
  {\bibfnamefont {Z.}~\bibnamefont {Song}}, \bibinfo {author} {\bibfnamefont
  {K.}~\bibnamefont {Torres}}, \bibinfo {author} {\bibfnamefont
  {F.}~\bibnamefont {Dirnberger}}, \bibinfo {author} {\bibfnamefont {J.~B.}\
  \bibnamefont {Curtis}}, \bibinfo {author} {\bibfnamefont {M.}~\bibnamefont
  {Weile}}, \bibinfo {author} {\bibfnamefont {A.}~\bibnamefont {Penn}},
  \bibinfo {author} {\bibfnamefont {T.}~\bibnamefont {Deilmann}}, \bibinfo
  {author} {\bibfnamefont {R.}~\bibnamefont {Dana}}, \bibinfo {author}
  {\bibfnamefont {R.}~\bibnamefont {Bushati}}, \bibinfo {author} {\bibfnamefont
  {J.}~\bibnamefont {Quan}}, \bibinfo {author} {\bibfnamefont {J.}~\bibnamefont
  {Luxa}}, \bibinfo {author} {\bibfnamefont {Z.}~\bibnamefont {Sofer}},
  \bibinfo {author} {\bibfnamefont {A.}~\bibnamefont {Alù}}, \bibinfo {author}
  {\bibfnamefont {V.~M.}\ \bibnamefont {Menon}}, \bibinfo {author}
  {\bibfnamefont {U.}~\bibnamefont {Wurstbauer}}, \bibinfo {author}
  {\bibfnamefont {M.}~\bibnamefont {Rohlfing}}, \bibinfo {author}
  {\bibfnamefont {P.}~\bibnamefont {Narang}}, \bibinfo {author} {\bibfnamefont
  {M.}~\bibnamefont {Lončar}},\ and\ \bibinfo {author} {\bibfnamefont {F.~M.}\
  \bibnamefont {Ross}},\ }\bibfield  {title} {\bibinfo {title} {The bulk van
  der $\mathrm{W}$aals layered magnet $\mathrm{CrSBr}$ is a $\mathrm{Quasi-1D}$
  material},\ }\href {https://doi.org/10.1021/acsnano.2c07316} {\bibfield
  {journal} {\bibinfo  {journal} {ACS Nano}\ }\textbf {\bibinfo {volume}
  {17}},\ \bibinfo {pages} {5316} (\bibinfo {year} {2023})}\BibitemShut
  {NoStop}%
\bibitem [{\citenamefont {Wang}\ \emph
  {et~al.}(2020{\natexlab{b}})\citenamefont {Wang}, \citenamefont {Qi},\ and\
  \citenamefont {Qian}}]{a30}%
  \BibitemOpen
  \bibfield  {author} {\bibinfo {author} {\bibfnamefont {H.}~\bibnamefont
  {Wang}}, \bibinfo {author} {\bibfnamefont {J.}~\bibnamefont {Qi}},\ and\
  \bibinfo {author} {\bibfnamefont {X.}~\bibnamefont {Qian}},\ }\bibfield
  {title} {\bibinfo {title} {Electrically tunable high $\mathrm{Curie}$
  temperature two-dimensional ferromagnetism in van der $\mathrm{Waals}$
  layered crystals},\ }\href {https://doi.org/10.1063/5.0014865} {\bibfield
  {journal} {\bibinfo  {journal} {Appl. Phys. Lett.}\ }\textbf {\bibinfo
  {volume} {117}},\ \bibinfo {pages} {083102} (\bibinfo {year}
  {2020}{\natexlab{b}})}\BibitemShut {NoStop}%
\bibitem [{\citenamefont {Guo}\ \emph {et~al.}(2018)\citenamefont {Guo},
  \citenamefont {Zhang}, \citenamefont {Yuan}, \citenamefont {Wang},\ and\
  \citenamefont {Wang}}]{a46}%
  \BibitemOpen
  \bibfield  {author} {\bibinfo {author} {\bibfnamefont {Y.}~\bibnamefont
  {Guo}}, \bibinfo {author} {\bibfnamefont {Y.}~\bibnamefont {Zhang}}, \bibinfo
  {author} {\bibfnamefont {S.}~\bibnamefont {Yuan}}, \bibinfo {author}
  {\bibfnamefont {B.}~\bibnamefont {Wang}},\ and\ \bibinfo {author}
  {\bibfnamefont {J.}~\bibnamefont {Wang}},\ }\bibfield  {title} {\bibinfo
  {title} {Chromium sulfide halide monolayers: intrinsic ferromagnetic
  semiconductors with large spin polarization and high carrier mobility},\
  }\href {https://doi.org/10.1039/C8NR06368K} {\bibfield  {journal} {\bibinfo
  {journal} {Nanoscale}\ }\textbf {\bibinfo {volume} {10}},\ \bibinfo {pages}
  {18036} (\bibinfo {year} {2018})}\BibitemShut {NoStop}%
\bibitem [{\citenamefont {Cong~Wang}(2019)}]{tcmono}%
  \BibitemOpen
  \bibfield  {author} {\bibinfo {author} {\bibfnamefont {L.~Z. N.-H. T. Z.-Y.
  L. W.~J.}\ \bibnamefont {Cong~Wang}, \bibfnamefont {Xieyu~Zhou}},\ }\bibfield
   {title} {\bibinfo {title} {A family of high-temperature ferromagnetic
  monolayers with locked spin-dichroism-mobility anisotropy: $\mathrm{MnNX}$
  and $\mathrm{CrCX}$ ($\mathrm{X=Cl, Br, I; C=S, Se, Te}$)},\ }\href
  {http://www.sciengine.com/publisher/Science China Press/journal/Science
  Bulletin/64/5/10.1016/j.scib.2019.02.011, doi =} {\bibfield  {journal}
  {\bibinfo  {journal} {Sci. Bull.}\ }\textbf {\bibinfo {volume} {64}},\
  \bibinfo {pages} {293} (\bibinfo {year} {2019})}\BibitemShut {NoStop}%
\bibitem [{\citenamefont {Wang}\ \emph
  {et~al.}(2023{\natexlab{a}})\citenamefont {Wang}, \citenamefont {Wu},
  \citenamefont {Bai}, \citenamefont {Shi}, \citenamefont {Zhang},
  \citenamefont {Zhang},\ and\ \citenamefont {Liu}}]{a23}%
  \BibitemOpen
  \bibfield  {author} {\bibinfo {author} {\bibfnamefont {B.}~\bibnamefont
  {Wang}}, \bibinfo {author} {\bibfnamefont {Y.}~\bibnamefont {Wu}}, \bibinfo
  {author} {\bibfnamefont {Y.}~\bibnamefont {Bai}}, \bibinfo {author}
  {\bibfnamefont {P.}~\bibnamefont {Shi}}, \bibinfo {author} {\bibfnamefont
  {G.}~\bibnamefont {Zhang}}, \bibinfo {author} {\bibfnamefont
  {Y.}~\bibnamefont {Zhang}},\ and\ \bibinfo {author} {\bibfnamefont
  {C.}~\bibnamefont {Liu}},\ }\bibfield  {title} {\bibinfo {title} {Origin and
  regulation of triaxial magnetic anisotropy in the ferromagnetic semiconductor
  $\mathrm{CrSBr}$ monolayer},\ }\href {https://doi.org/10.1039/D3NR02518G}
  {\bibfield  {journal} {\bibinfo  {journal} {Nanoscale}\ }\textbf {\bibinfo
  {volume} {15}},\ \bibinfo {pages} {13402} (\bibinfo {year}
  {2023}{\natexlab{a}})}\BibitemShut {NoStop}%
\bibitem [{\citenamefont {Telford}\ \emph {et~al.}(2020)\citenamefont
  {Telford}, \citenamefont {Dismukes}, \citenamefont {Lee}, \citenamefont
  {Cheng}, \citenamefont {Wieteska}, \citenamefont {Bartholomew}, \citenamefont
  {Chen}, \citenamefont {Xu}, \citenamefont {Pasupathy}, \citenamefont {Zhu},
  \citenamefont {Dean},\ and\ \citenamefont {Roy}}]{a24}%
  \BibitemOpen
  \bibfield  {author} {\bibinfo {author} {\bibfnamefont {E.~J.}\ \bibnamefont
  {Telford}}, \bibinfo {author} {\bibfnamefont {A.~H.}\ \bibnamefont
  {Dismukes}}, \bibinfo {author} {\bibfnamefont {K.}~\bibnamefont {Lee}},
  \bibinfo {author} {\bibfnamefont {M.}~\bibnamefont {Cheng}}, \bibinfo
  {author} {\bibfnamefont {A.}~\bibnamefont {Wieteska}}, \bibinfo {author}
  {\bibfnamefont {A.~K.}\ \bibnamefont {Bartholomew}}, \bibinfo {author}
  {\bibfnamefont {Y.-S.}\ \bibnamefont {Chen}}, \bibinfo {author}
  {\bibfnamefont {X.}~\bibnamefont {Xu}}, \bibinfo {author} {\bibfnamefont
  {A.~N.}\ \bibnamefont {Pasupathy}}, \bibinfo {author} {\bibfnamefont
  {X.}~\bibnamefont {Zhu}}, \bibinfo {author} {\bibfnamefont {C.~R.}\
  \bibnamefont {Dean}},\ and\ \bibinfo {author} {\bibfnamefont
  {X.}~\bibnamefont {Roy}},\ }\bibfield  {title} {\bibinfo {title} {Layered
  antiferromagnetism induces large negative magnetoresistance in the van der
  $\mathrm{Waals}$ semiconductor $\mathrm{CrSBr}$},\ }\href
  {https://doi.org/10.1002/adma.202003240} {\bibfield  {journal} {\bibinfo
  {journal} {Adv. Mater.}\ }\textbf {\bibinfo {volume} {32}},\ \bibinfo {pages}
  {2003240} (\bibinfo {year} {2020})}\BibitemShut {NoStop}%
\bibitem [{\citenamefont {Long}\ \emph {et~al.}(2023)\citenamefont {Long},
  \citenamefont {Ghorbani-Asl}, \citenamefont {Mosina}, \citenamefont {Li},
  \citenamefont {Lin}, \citenamefont {Ganss}, \citenamefont {Hübner},
  \citenamefont {Sofer}, \citenamefont {Dirnberger}, \citenamefont {Kamra},
  \citenamefont {Krasheninnikov}, \citenamefont {Prucnal}, \citenamefont
  {Helm},\ and\ \citenamefont {Zhou}}]{a26}%
  \BibitemOpen
  \bibfield  {author} {\bibinfo {author} {\bibfnamefont {F.}~\bibnamefont
  {Long}}, \bibinfo {author} {\bibfnamefont {M.}~\bibnamefont {Ghorbani-Asl}},
  \bibinfo {author} {\bibfnamefont {K.}~\bibnamefont {Mosina}}, \bibinfo
  {author} {\bibfnamefont {Y.}~\bibnamefont {Li}}, \bibinfo {author}
  {\bibfnamefont {K.}~\bibnamefont {Lin}}, \bibinfo {author} {\bibfnamefont
  {F.}~\bibnamefont {Ganss}}, \bibinfo {author} {\bibfnamefont
  {R.}~\bibnamefont {Hübner}}, \bibinfo {author} {\bibfnamefont
  {Z.}~\bibnamefont {Sofer}}, \bibinfo {author} {\bibfnamefont
  {F.}~\bibnamefont {Dirnberger}}, \bibinfo {author} {\bibfnamefont
  {A.}~\bibnamefont {Kamra}}, \bibinfo {author} {\bibfnamefont {A.~V.}\
  \bibnamefont {Krasheninnikov}}, \bibinfo {author} {\bibfnamefont
  {S.}~\bibnamefont {Prucnal}}, \bibinfo {author} {\bibfnamefont
  {M.}~\bibnamefont {Helm}},\ and\ \bibinfo {author} {\bibfnamefont
  {S.}~\bibnamefont {Zhou}},\ }\bibfield  {title} {\bibinfo {title}
  {Ferromagnetic interlayer coupling in $\mathrm{CrSBr}$ crystals irradiated by
  ions},\ }\href {https://doi.org/10.1021/acs.nanolett.3c01920} {\bibfield
  {journal} {\bibinfo  {journal} {Nano Lett.}\ }\textbf {\bibinfo {volume}
  {23}},\ \bibinfo {pages} {8468} (\bibinfo {year} {2023})}\BibitemShut
  {NoStop}%
\bibitem [{\citenamefont {Ye}\ \emph {et~al.}(2022)\citenamefont {Ye},
  \citenamefont {Wang}, \citenamefont {Wu}, \citenamefont {Liu}, \citenamefont
  {Zhou}, \citenamefont {Wang}, \citenamefont {Söll}, \citenamefont {Sofer},
  \citenamefont {Yue}, \citenamefont {Liu}, \citenamefont {Tian}, \citenamefont
  {Xiong}, \citenamefont {Ji},\ and\ \citenamefont {Wang}}]{a27}%
  \BibitemOpen
  \bibfield  {author} {\bibinfo {author} {\bibfnamefont {C.}~\bibnamefont
  {Ye}}, \bibinfo {author} {\bibfnamefont {C.}~\bibnamefont {Wang}}, \bibinfo
  {author} {\bibfnamefont {Q.}~\bibnamefont {Wu}}, \bibinfo {author}
  {\bibfnamefont {S.}~\bibnamefont {Liu}}, \bibinfo {author} {\bibfnamefont
  {J.}~\bibnamefont {Zhou}}, \bibinfo {author} {\bibfnamefont {G.}~\bibnamefont
  {Wang}}, \bibinfo {author} {\bibfnamefont {A.}~\bibnamefont {Söll}},
  \bibinfo {author} {\bibfnamefont {Z.}~\bibnamefont {Sofer}}, \bibinfo
  {author} {\bibfnamefont {M.}~\bibnamefont {Yue}}, \bibinfo {author}
  {\bibfnamefont {X.}~\bibnamefont {Liu}}, \bibinfo {author} {\bibfnamefont
  {M.}~\bibnamefont {Tian}}, \bibinfo {author} {\bibfnamefont {Q.}~\bibnamefont
  {Xiong}}, \bibinfo {author} {\bibfnamefont {W.}~\bibnamefont {Ji}},\ and\
  \bibinfo {author} {\bibfnamefont {X.~R.}\ \bibnamefont {Wang}},\ }\bibfield
  {title} {\bibinfo {title} {Layer-dependent interlayer antiferromagnetic spin
  reorientation in air-stable semiconductor $\mathrm{CrSBr}$},\ }\href
  {https://doi.org/10.1021/acsnano.2c01151} {\bibfield  {journal} {\bibinfo
  {journal} {ACS Nano}\ }\textbf {\bibinfo {volume} {16}},\ \bibinfo {pages}
  {11876} (\bibinfo {year} {2022})}\BibitemShut {NoStop}%
\bibitem [{\citenamefont {Wilson}\ \emph {et~al.}(2021)\citenamefont {Wilson},
  \citenamefont {Lee}, \citenamefont {Cenker}, \citenamefont {Xie},
  \citenamefont {Dismukes}, \citenamefont {Telford}, \citenamefont {Fonseca},
  \citenamefont {Sivakumar}, \citenamefont {Dean}, \citenamefont {Cao},
  \citenamefont {Roy}, \citenamefont {Xu},\ and\ \citenamefont {Zhu}}]{a49}%
  \BibitemOpen
  \bibfield  {author} {\bibinfo {author} {\bibfnamefont {N.~P.}\ \bibnamefont
  {Wilson}}, \bibinfo {author} {\bibfnamefont {K.}~\bibnamefont {Lee}},
  \bibinfo {author} {\bibfnamefont {J.}~\bibnamefont {Cenker}}, \bibinfo
  {author} {\bibfnamefont {K.}~\bibnamefont {Xie}}, \bibinfo {author}
  {\bibfnamefont {A.~H.}\ \bibnamefont {Dismukes}}, \bibinfo {author}
  {\bibfnamefont {E.~J.}\ \bibnamefont {Telford}}, \bibinfo {author}
  {\bibfnamefont {J.}~\bibnamefont {Fonseca}}, \bibinfo {author} {\bibfnamefont
  {S.}~\bibnamefont {Sivakumar}}, \bibinfo {author} {\bibfnamefont
  {C.}~\bibnamefont {Dean}}, \bibinfo {author} {\bibfnamefont {T.}~\bibnamefont
  {Cao}}, \bibinfo {author} {\bibfnamefont {X.}~\bibnamefont {Roy}}, \bibinfo
  {author} {\bibfnamefont {X.}~\bibnamefont {Xu}},\ and\ \bibinfo {author}
  {\bibfnamefont {X.}~\bibnamefont {Zhu}},\ }\bibfield  {title} {\bibinfo
  {title} {Interlayer electronic coupling on demand in a 2$\mathrm{D}$ magnetic
  semiconductor},\ }\href {https://doi.org/10.1038/s41563-021-01070-8}
  {\bibfield  {journal} {\bibinfo  {journal} {Nat. Mater.}\ }\textbf {\bibinfo
  {volume} {20}},\ \bibinfo {pages} {1657} (\bibinfo {year}
  {2021})}\BibitemShut {NoStop}%
\bibitem [{\citenamefont {Avsar}(2022)}]{avsar2022highly}%
  \BibitemOpen
  \bibfield  {author} {\bibinfo {author} {\bibfnamefont {A.}~\bibnamefont
  {Avsar}},\ }\bibfield  {title} {\bibinfo {title} {Highly anisotropic van der
  $\mathrm{W}$aals magnetism},\ }\href
  {https://doi.org/doi.org/10.1038/s41563-022-01299-x} {\bibfield  {journal}
  {\bibinfo  {journal} {Nat. Mater.}\ }\textbf {\bibinfo {volume} {21}},\
  \bibinfo {pages} {731} (\bibinfo {year} {2022})}\BibitemShut {NoStop}%
\bibitem [{\citenamefont {Kresse}\ and\ \citenamefont
  {Furthmüller}(1996{\natexlab{a}})}]{a31}%
  \BibitemOpen
  \bibfield  {author} {\bibinfo {author} {\bibfnamefont {G.}~\bibnamefont
  {Kresse}}\ and\ \bibinfo {author} {\bibfnamefont {J.}~\bibnamefont
  {Furthmüller}},\ }\bibfield  {title} {\bibinfo {title} {Efficiency of
  ab-initio total energy calculations for metals and semiconductors using a
  plane-wave basis set},\ }\href {https://doi.org/10.1016/0927-0256(96)00008-0}
  {\bibfield  {journal} {\bibinfo  {journal} {Comput. Mater. Sci.}\ }\textbf
  {\bibinfo {volume} {6}},\ \bibinfo {pages} {15} (\bibinfo {year}
  {1996}{\natexlab{a}})}\BibitemShut {NoStop}%
\bibitem [{\citenamefont {Kresse}\ and\ \citenamefont
  {Furthmüller}(1996{\natexlab{b}})}]{a32}%
  \BibitemOpen
  \bibfield  {author} {\bibinfo {author} {\bibfnamefont {G.}~\bibnamefont
  {Kresse}}\ and\ \bibinfo {author} {\bibfnamefont {J.}~\bibnamefont
  {Furthmüller}},\ }\bibfield  {title} {\bibinfo {title} {Efficient iterative
  schemes for ab initio total-energy calculations using a plane-wave basis
  set},\ }\href {https://doi.org/10.1103/PhysRevB.54.11169} {\bibfield
  {journal} {\bibinfo  {journal} {Phys. Rev. B}\ }\textbf {\bibinfo {volume}
  {54}},\ \bibinfo {pages} {11169} (\bibinfo {year}
  {1996}{\natexlab{b}})}\BibitemShut {NoStop}%
\bibitem [{\citenamefont {Wang}\ \emph {et~al.}(2021)\citenamefont {Wang},
  \citenamefont {Xu}, \citenamefont {Liu}, \citenamefont {Tang},\ and\
  \citenamefont {Geng}}]{a33}%
  \BibitemOpen
  \bibfield  {author} {\bibinfo {author} {\bibfnamefont {V.}~\bibnamefont
  {Wang}}, \bibinfo {author} {\bibfnamefont {N.}~\bibnamefont {Xu}}, \bibinfo
  {author} {\bibfnamefont {J.-C.}\ \bibnamefont {Liu}}, \bibinfo {author}
  {\bibfnamefont {G.}~\bibnamefont {Tang}},\ and\ \bibinfo {author}
  {\bibfnamefont {W.-T.}\ \bibnamefont {Geng}},\ }\bibfield  {title} {\bibinfo
  {title} {$\mathrm{VASPKIT: A}$ user-friendly interface facilitating
  high-throughput computing and analysis using $\mathrm{VASP}$ code},\ }\href
  {https://doi.org/10.1016/j.cpc.2021.108033} {\bibfield  {journal} {\bibinfo
  {journal} {Comput. Phys. Commun.}\ }\textbf {\bibinfo {volume} {267}},\
  \bibinfo {pages} {108033} (\bibinfo {year} {2021})}\BibitemShut {NoStop}%
\bibitem [{\citenamefont {Blöchl}(1994)}]{a34}%
  \BibitemOpen
  \bibfield  {author} {\bibinfo {author} {\bibfnamefont {P.~E.}\ \bibnamefont
  {Blöchl}},\ }\bibfield  {title} {\bibinfo {title} {Projector augmented-wave
  method},\ }\href {https://doi.org/10.1103/PhysRevB.50.17953} {\bibfield
  {journal} {\bibinfo  {journal} {Phys. Rev. B}\ }\textbf {\bibinfo {volume}
  {50}},\ \bibinfo {pages} {17953} (\bibinfo {year} {1994})}\BibitemShut
  {NoStop}%
\bibitem [{\citenamefont {Kresse}\ and\ \citenamefont {Joubert}(1999)}]{PAW2}%
  \BibitemOpen
  \bibfield  {author} {\bibinfo {author} {\bibfnamefont {G.}~\bibnamefont
  {Kresse}}\ and\ \bibinfo {author} {\bibfnamefont {D.}~\bibnamefont
  {Joubert}},\ }\bibfield  {title} {\bibinfo {title} {From ultrasoft
  pseudopotentials to the projector augmented-wave method},\ }\href
  {https://doi.org/10.1103/PhysRevB.59.1758} {\bibfield  {journal} {\bibinfo
  {journal} {Phys. Rev. B}\ }\textbf {\bibinfo {volume} {59}},\ \bibinfo
  {pages} {1758} (\bibinfo {year} {1999})}\BibitemShut {NoStop}%
\bibitem [{\citenamefont {Klimeš}\ \emph {et~al.}(2011)\citenamefont
  {Klimeš}, \citenamefont {Bowler},\ and\ \citenamefont {Michaelides}}]{a36}%
  \BibitemOpen
  \bibfield  {author} {\bibinfo {author} {\bibfnamefont {J.}~\bibnamefont
  {Klimeš}}, \bibinfo {author} {\bibfnamefont {D.~R.}\ \bibnamefont
  {Bowler}},\ and\ \bibinfo {author} {\bibfnamefont {A.}~\bibnamefont
  {Michaelides}},\ }\bibfield  {title} {\bibinfo {title} {van der
  $\mathrm{W}$aals density functionals applied to solids},\ }\href
  {https://doi.org/10.1103/PhysRevB.83.195131} {\bibfield  {journal} {\bibinfo
  {journal} {Phys. Rev. B}\ }\textbf {\bibinfo {volume} {83}},\ \bibinfo
  {pages} {195131} (\bibinfo {year} {2011})}\BibitemShut {NoStop}%
\bibitem [{\citenamefont {Rohrbach}\ \emph {et~al.}(2003)\citenamefont
  {Rohrbach}, \citenamefont {Hafner},\ and\ \citenamefont {Kresse}}]{a37}%
  \BibitemOpen
  \bibfield  {author} {\bibinfo {author} {\bibfnamefont {A.~C.~W.}\
  \bibnamefont {Rohrbach}}, \bibinfo {author} {\bibfnamefont {J.}~\bibnamefont
  {Hafner}},\ and\ \bibinfo {author} {\bibfnamefont {G.}~\bibnamefont
  {Kresse}},\ }\bibfield  {title} {\bibinfo {title} {Electronic correlation
  effects in transition-metal sulfides},\ }\href
  {https://doi.org/10.1088/0953-8984/15/6/325} {\bibfield  {journal} {\bibinfo
  {journal} {J. Phys.: Condens. Matter}\ }\textbf {\bibinfo {volume} {15}},\
  \bibinfo {pages} {979} (\bibinfo {year} {2003})}\BibitemShut {NoStop}%
\bibitem [{\citenamefont {Dudarev}\ \emph {et~al.}(1998)\citenamefont
  {Dudarev}, \citenamefont {Botton}, \citenamefont {Savrasov}, \citenamefont
  {Humphreys},\ and\ \citenamefont {Sutton}}]{a38}%
  \BibitemOpen
  \bibfield  {author} {\bibinfo {author} {\bibfnamefont {S.~L.}\ \bibnamefont
  {Dudarev}}, \bibinfo {author} {\bibfnamefont {G.~A.}\ \bibnamefont {Botton}},
  \bibinfo {author} {\bibfnamefont {S.~Y.}\ \bibnamefont {Savrasov}}, \bibinfo
  {author} {\bibfnamefont {C.~J.}\ \bibnamefont {Humphreys}},\ and\ \bibinfo
  {author} {\bibfnamefont {A.~P.}\ \bibnamefont {Sutton}},\ }\bibfield  {title}
  {\bibinfo {title} {Electron-energy-loss spectra and the structural stability
  of nickel oxide: $\mathrm{An LSDA+U}$ study},\ }\href
  {https://doi.org/10.1103/PhysRevB.57.1505} {\bibfield  {journal} {\bibinfo
  {journal} {Phys. Rev. B}\ }\textbf {\bibinfo {volume} {57}},\ \bibinfo
  {pages} {1505} (\bibinfo {year} {1998})}\BibitemShut {NoStop}%
\bibitem [{\citenamefont {Rudenko}\ \emph
  {et~al.}(2023{\natexlab{a}})\citenamefont {Rudenko}, \citenamefont
  {Rösner},\ and\ \citenamefont {Katsnelson}}]{a39}%
  \BibitemOpen
  \bibfield  {author} {\bibinfo {author} {\bibfnamefont {A.~N.}\ \bibnamefont
  {Rudenko}}, \bibinfo {author} {\bibfnamefont {M.}~\bibnamefont {Rösner}},\
  and\ \bibinfo {author} {\bibfnamefont {M.~I.}\ \bibnamefont {Katsnelson}},\
  }\bibfield  {title} {\bibinfo {title} {Dielectric tunability of magnetic
  properties in orthorhombic ferromagnetic monolayer $\mathrm{CrSBr}$},\ }\href
  {https://doi.org/10.1038/s41524-023-01050-3} {\bibfield  {journal} {\bibinfo
  {journal} {npj Comput. Mater.}\ }\textbf {\bibinfo {volume} {9}},\ \bibinfo
  {pages} {83} (\bibinfo {year} {2023}{\natexlab{a}})}\BibitemShut {NoStop}%
\bibitem [{\citenamefont {Esteras}\ \emph {et~al.}(2022)\citenamefont
  {Esteras}, \citenamefont {Rybakov}, \citenamefont {Ruiz},\ and\ \citenamefont
  {Baldoví}}]{a51}%
  \BibitemOpen
  \bibfield  {author} {\bibinfo {author} {\bibfnamefont {D.~L.}\ \bibnamefont
  {Esteras}}, \bibinfo {author} {\bibfnamefont {A.}~\bibnamefont {Rybakov}},
  \bibinfo {author} {\bibfnamefont {A.~M.}\ \bibnamefont {Ruiz}},\ and\
  \bibinfo {author} {\bibfnamefont {J.~J.}\ \bibnamefont {Baldoví}},\
  }\bibfield  {title} {\bibinfo {title} {Magnon straintronics in the
  2$\mathrm{D}$ van der $\mathrm{W}$aals ferromagnet $\mathrm{CrSBr}$ from
  first-principles},\ }\href {https://doi.org/10.1021/acs.nanolett.2c02863}
  {\bibfield  {journal} {\bibinfo  {journal} {Nano Lett.}\ }\textbf {\bibinfo
  {volume} {22}},\ \bibinfo {pages} {8771} (\bibinfo {year}
  {2022})}\BibitemShut {NoStop}%
\bibitem [{\citenamefont {Henkelman}\ \emph {et~al.}(2000)\citenamefont
  {Henkelman}, \citenamefont {Uberuaga},\ and\ \citenamefont
  {Jónsson}}]{cNEB1}%
  \BibitemOpen
  \bibfield  {author} {\bibinfo {author} {\bibfnamefont {G.}~\bibnamefont
  {Henkelman}}, \bibinfo {author} {\bibfnamefont {B.~P.}\ \bibnamefont
  {Uberuaga}},\ and\ \bibinfo {author} {\bibfnamefont {H.}~\bibnamefont
  {Jónsson}},\ }\bibfield  {title} {\bibinfo {title} {A climbing image nudged
  elastic band method for finding saddle points and minimum energy paths},\
  }\href {https://doi.org/10.1063/1.1329672} {\bibfield  {journal} {\bibinfo
  {journal} {J. Chem. Phys.}\ }\textbf {\bibinfo {volume} {113}},\ \bibinfo
  {pages} {9901} (\bibinfo {year} {2000})}\BibitemShut {NoStop}%
\bibitem [{\citenamefont {Henkelman}\ and\ \citenamefont
  {Jónsson}(2000)}]{cNEB2}%
  \BibitemOpen
  \bibfield  {author} {\bibinfo {author} {\bibfnamefont {G.}~\bibnamefont
  {Henkelman}}\ and\ \bibinfo {author} {\bibfnamefont {H.}~\bibnamefont
  {Jónsson}},\ }\bibfield  {title} {\bibinfo {title} {Improved tangent
  estimate in the nudged elastic band method for finding minimum energy paths
  and saddle points},\ }\href {https://doi.org/10.1063/1.1323224} {\bibfield
  {journal} {\bibinfo  {journal} {J. Chem. Phys.}\ }\textbf {\bibinfo {volume}
  {113}},\ \bibinfo {pages} {9978} (\bibinfo {year} {2000})}\BibitemShut
  {NoStop}%
\bibitem [{\citenamefont {Togo}\ and\ \citenamefont
  {Tanaka}(2015)}]{TOGO20151}%
  \BibitemOpen
  \bibfield  {author} {\bibinfo {author} {\bibfnamefont {A.}~\bibnamefont
  {Togo}}\ and\ \bibinfo {author} {\bibfnamefont {I.}~\bibnamefont {Tanaka}},\
  }\bibfield  {title} {\bibinfo {title} {First principles phonon calculations
  in materials science},\ }\href
  {https://doi.org/https://doi.org/10.1016/j.scriptamat.2015.07.021} {\bibfield
   {journal} {\bibinfo  {journal} {Scr. Metall.}\ }\textbf {\bibinfo {volume}
  {108}},\ \bibinfo {pages} {1 } (\bibinfo {year} {2015})}\BibitemShut
  {NoStop}%
\bibitem [{\citenamefont {Zhou}\ \emph {et~al.}(2014)\citenamefont {Zhou},
  \citenamefont {Nielson}, \citenamefont {Xia},\ and\ \citenamefont
  {Ozoli\ifmmode \mbox{\c{n}}\else \c{n}\fi{}\ifmmode~\check{s}\else
  \v{s}\fi{}}}]{PhysRevLett.113.185501}%
  \BibitemOpen
  \bibfield  {author} {\bibinfo {author} {\bibfnamefont {F.}~\bibnamefont
  {Zhou}}, \bibinfo {author} {\bibfnamefont {W.}~\bibnamefont {Nielson}},
  \bibinfo {author} {\bibfnamefont {Y.}~\bibnamefont {Xia}},\ and\ \bibinfo
  {author} {\bibfnamefont {V.}~\bibnamefont {Ozoli\ifmmode \mbox{\c{n}}\else
  \c{n}\fi{}\ifmmode~\check{s}\else \v{s}\fi{}}},\ }\bibfield  {title}
  {\bibinfo {title} {Lattice anharmonicity and thermal conductivity from
  compressive sensing of first-principles calculations},\ }\href
  {https://doi.org/10.1103/PhysRevLett.113.185501} {\bibfield  {journal}
  {\bibinfo  {journal} {Phys. Rev. Lett.}\ }\textbf {\bibinfo {volume} {113}},\
  \bibinfo {pages} {185501} (\bibinfo {year} {2014})}\BibitemShut {NoStop}%
\bibitem [{\citenamefont {Zhou}\ \emph
  {et~al.}(2019{\natexlab{a}})\citenamefont {Zhou}, \citenamefont {Nielson},
  \citenamefont {Xia},\ and\ \citenamefont {Ozoli\ifmmode \mbox{\c{n}}\else
  \c{n}\fi{}\ifmmode~\check{s}\else \v{s}\fi{}}}]{PhysRevB.100.184308}%
  \BibitemOpen
  \bibfield  {author} {\bibinfo {author} {\bibfnamefont {F.}~\bibnamefont
  {Zhou}}, \bibinfo {author} {\bibfnamefont {W.}~\bibnamefont {Nielson}},
  \bibinfo {author} {\bibfnamefont {Y.}~\bibnamefont {Xia}},\ and\ \bibinfo
  {author} {\bibfnamefont {V.}~\bibnamefont {Ozoli\ifmmode \mbox{\c{n}}\else
  \c{n}\fi{}\ifmmode~\check{s}\else \v{s}\fi{}}},\ }\bibfield  {title}
  {\bibinfo {title} {Compressive sensing lattice $\mathrm{dynamics. I.
  General}$ formalism},\ }\href {https://doi.org/10.1103/PhysRevB.100.184308}
  {\bibfield  {journal} {\bibinfo  {journal} {Phys. Rev. B}\ }\textbf {\bibinfo
  {volume} {100}},\ \bibinfo {pages} {184308} (\bibinfo {year}
  {2019}{\natexlab{a}})}\BibitemShut {NoStop}%
\bibitem [{\citenamefont {Zhou}\ \emph
  {et~al.}(2019{\natexlab{b}})\citenamefont {Zhou}, \citenamefont {Sadigh},
  \citenamefont {\AA{}berg}, \citenamefont {Xia},\ and\ \citenamefont
  {Ozoli\ifmmode \mbox{\c{n}}\else \c{n}\fi{}\ifmmode~\check{s}\else
  \v{s}\fi{}}}]{PhysRevB.100.184309}%
  \BibitemOpen
  \bibfield  {author} {\bibinfo {author} {\bibfnamefont {F.}~\bibnamefont
  {Zhou}}, \bibinfo {author} {\bibfnamefont {B.}~\bibnamefont {Sadigh}},
  \bibinfo {author} {\bibfnamefont {D.}~\bibnamefont {\AA{}berg}}, \bibinfo
  {author} {\bibfnamefont {Y.}~\bibnamefont {Xia}},\ and\ \bibinfo {author}
  {\bibfnamefont {V.}~\bibnamefont {Ozoli\ifmmode \mbox{\c{n}}\else
  \c{n}\fi{}\ifmmode~\check{s}\else \v{s}\fi{}}},\ }\bibfield  {title}
  {\bibinfo {title} {Compressive sensing lattice $\mathrm{dynamics. II.
  Efficient}$ phonon calculations and long-range interactions},\ }\href
  {https://doi.org/10.1103/PhysRevB.100.184309} {\bibfield  {journal} {\bibinfo
   {journal} {Phys. Rev. B}\ }\textbf {\bibinfo {volume} {100}},\ \bibinfo
  {pages} {184309} (\bibinfo {year} {2019}{\natexlab{b}})}\BibitemShut
  {NoStop}%
\bibitem [{\citenamefont {Hooton}(1955)}]{doi:10.1080/14786440408520575}%
  \BibitemOpen
  \bibfield  {author} {\bibinfo {author} {\bibfnamefont {D.}~\bibnamefont
  {Hooton}},\ }\bibfield  {title} {\bibinfo {title} {$\mathrm{LI.}$
  $\mathrm{A}$ new treatment of anharmonicity in lattice
  thermodynamics$\mathrm{: I}$},\ }\href
  {https://doi.org/10.1080/14786440408520575} {\bibfield  {journal} {\bibinfo
  {journal} {Philos. Mag. J. Sci.}\ }\textbf {\bibinfo {volume} {46}},\
  \bibinfo {pages} {422} (\bibinfo {year} {1955})}\BibitemShut {NoStop}%
\bibitem [{\citenamefont {Pawley}\ \emph {et~al.}(1966)\citenamefont {Pawley},
  \citenamefont {Cochran}, \citenamefont {Cowley},\ and\ \citenamefont
  {Dolling}}]{PhysRevLett.17.753}%
  \BibitemOpen
  \bibfield  {author} {\bibinfo {author} {\bibfnamefont {G.~S.}\ \bibnamefont
  {Pawley}}, \bibinfo {author} {\bibfnamefont {W.}~\bibnamefont {Cochran}},
  \bibinfo {author} {\bibfnamefont {R.~A.}\ \bibnamefont {Cowley}},\ and\
  \bibinfo {author} {\bibfnamefont {G.}~\bibnamefont {Dolling}},\ }\bibfield
  {title} {\bibinfo {title} {Diatomic ferroelectrics},\ }\href
  {https://doi.org/10.1103/PhysRevLett.17.753} {\bibfield  {journal} {\bibinfo
  {journal} {Phys. Rev. Lett.}\ }\textbf {\bibinfo {volume} {17}},\ \bibinfo
  {pages} {753} (\bibinfo {year} {1966})}\BibitemShut {NoStop}%
\bibitem [{\citenamefont {Werthamer}(1970)}]{PhysRevB.1.572}%
  \BibitemOpen
  \bibfield  {author} {\bibinfo {author} {\bibfnamefont {N.~R.}\ \bibnamefont
  {Werthamer}},\ }\bibfield  {title} {\bibinfo {title} {Self-consistent phonon
  formulation of anharmonic lattice dynamics},\ }\href
  {https://doi.org/10.1103/PhysRevB.1.572} {\bibfield  {journal} {\bibinfo
  {journal} {Phys. Rev. B}\ }\textbf {\bibinfo {volume} {1}},\ \bibinfo {pages}
  {572} (\bibinfo {year} {1970})}\BibitemShut {NoStop}%
\bibitem [{\citenamefont {Xia}\ \emph {et~al.}(2020)\citenamefont {Xia},
  \citenamefont {Hegde}, \citenamefont {Pal}, \citenamefont {Hua},
  \citenamefont {Gaines}, \citenamefont {Patel}, \citenamefont {He},
  \citenamefont {Aykol},\ and\ \citenamefont {Wolverton}}]{PhysRevX.10.041029}%
  \BibitemOpen
  \bibfield  {author} {\bibinfo {author} {\bibfnamefont {Y.}~\bibnamefont
  {Xia}}, \bibinfo {author} {\bibfnamefont {V.~I.}\ \bibnamefont {Hegde}},
  \bibinfo {author} {\bibfnamefont {K.}~\bibnamefont {Pal}}, \bibinfo {author}
  {\bibfnamefont {X.}~\bibnamefont {Hua}}, \bibinfo {author} {\bibfnamefont
  {D.}~\bibnamefont {Gaines}}, \bibinfo {author} {\bibfnamefont
  {S.}~\bibnamefont {Patel}}, \bibinfo {author} {\bibfnamefont
  {J.}~\bibnamefont {He}}, \bibinfo {author} {\bibfnamefont {M.}~\bibnamefont
  {Aykol}},\ and\ \bibinfo {author} {\bibfnamefont {C.}~\bibnamefont
  {Wolverton}},\ }\bibfield  {title} {\bibinfo {title} {High-throughput study
  of lattice thermal conductivity in binary rocksalt and zinc blende compounds
  including higher-order anharmonicity},\ }\href
  {https://doi.org/10.1103/PhysRevX.10.041029} {\bibfield  {journal} {\bibinfo
  {journal} {Phys. Rev. X}\ }\textbf {\bibinfo {volume} {10}},\ \bibinfo
  {pages} {041029} (\bibinfo {year} {2020})}\BibitemShut {NoStop}%
\bibitem [{\citenamefont {Guo}\ \emph {et~al.}(2024)\citenamefont {Guo},
  \citenamefont {Han}, \citenamefont {Feng}, \citenamefont {Lin},\ and\
  \citenamefont {Ruan}}]{guo2024sampling}%
  \BibitemOpen
  \bibfield  {author} {\bibinfo {author} {\bibfnamefont {Z.}~\bibnamefont
  {Guo}}, \bibinfo {author} {\bibfnamefont {Z.}~\bibnamefont {Han}}, \bibinfo
  {author} {\bibfnamefont {D.}~\bibnamefont {Feng}}, \bibinfo {author}
  {\bibfnamefont {G.}~\bibnamefont {Lin}},\ and\ \bibinfo {author}
  {\bibfnamefont {X.}~\bibnamefont {Ruan}},\ }\bibfield  {title} {\bibinfo
  {title} {Sampling-accelerated prediction of phonon scattering rates for
  converged thermal conductivity and radiative properties},\ }\href
  {https://doi.org/https://www.nature.com/articles/s41524-024-01215-8}
  {\bibfield  {journal} {\bibinfo  {journal} {npj Comput. Mater.}\ }\textbf
  {\bibinfo {volume} {10}},\ \bibinfo {pages} {31} (\bibinfo {year}
  {2024})}\BibitemShut {NoStop}%
\bibitem [{\citenamefont {Paillard}\ \emph {et~al.}(2016)\citenamefont
  {Paillard}, \citenamefont {Xu}, \citenamefont {Dkhil}, \citenamefont
  {Geneste},\ and\ \citenamefont {Bellaiche}}]{PhysRevLett.116.247401}%
  \BibitemOpen
  \bibfield  {author} {\bibinfo {author} {\bibfnamefont {C.}~\bibnamefont
  {Paillard}}, \bibinfo {author} {\bibfnamefont {B.}~\bibnamefont {Xu}},
  \bibinfo {author} {\bibfnamefont {B.}~\bibnamefont {Dkhil}}, \bibinfo
  {author} {\bibfnamefont {G.}~\bibnamefont {Geneste}},\ and\ \bibinfo {author}
  {\bibfnamefont {L.}~\bibnamefont {Bellaiche}},\ }\bibfield  {title} {\bibinfo
  {title} {Photostriction in ferroelectrics from density functional theory},\
  }\href {https://doi.org/10.1103/PhysRevLett.116.247401} {\bibfield  {journal}
  {\bibinfo  {journal} {Phys. Rev. Lett.}\ }\textbf {\bibinfo {volume} {116}},\
  \bibinfo {pages} {247401} (\bibinfo {year} {2016})}\BibitemShut {NoStop}%
\bibitem [{\citenamefont {Hukushima}\ and\ \citenamefont {Nemoto}(1996)}]{MC1}%
  \BibitemOpen
  \bibfield  {author} {\bibinfo {author} {\bibfnamefont {K.}~\bibnamefont
  {Hukushima}}\ and\ \bibinfo {author} {\bibfnamefont {K.}~\bibnamefont
  {Nemoto}},\ }\bibfield  {title} {\bibinfo {title} {Exchange $\mathrm{Monte
  Carlo}$ method and application to spin glass simulations},\ }\href
  {https://doi.org/10.1143/JPSJ.65.1604} {\bibfield  {journal} {\bibinfo
  {journal} {J. Phys. Soc. Jpn.}\ }\textbf {\bibinfo {volume} {65}},\ \bibinfo
  {pages} {1604} (\bibinfo {year} {1996})}\BibitemShut {NoStop}%
\bibitem [{\citenamefont {Wang}\ \emph {et~al.}(2015)\citenamefont {Wang},
  \citenamefont {Ren}, \citenamefont {Bellaiche},\ and\ \citenamefont
  {Xiang}}]{MC2}%
  \BibitemOpen
  \bibfield  {author} {\bibinfo {author} {\bibfnamefont {P.~S.}\ \bibnamefont
  {Wang}}, \bibinfo {author} {\bibfnamefont {W.}~\bibnamefont {Ren}}, \bibinfo
  {author} {\bibfnamefont {L.}~\bibnamefont {Bellaiche}},\ and\ \bibinfo
  {author} {\bibfnamefont {H.~J.}\ \bibnamefont {Xiang}},\ }\bibfield  {title}
  {\bibinfo {title} {Predicting a ferrimagnetic phase of $\mathrm{Zn_2FeOsO_6}$
  with strong magnetoelectric coupling},\ }\href
  {https://doi.org/10.1103/PhysRevLett.114.147204} {\bibfield  {journal}
  {\bibinfo  {journal} {Phys. Rev. Lett.}\ }\textbf {\bibinfo {volume} {114}},\
  \bibinfo {pages} {147204} (\bibinfo {year} {2015})}\BibitemShut {NoStop}%
\bibitem [{\citenamefont {Lou}\ \emph {et~al.}(2021)\citenamefont {Lou},
  \citenamefont {Li}, \citenamefont {Ji}, \citenamefont {Yu}, \citenamefont
  {Feng}, \citenamefont {Gong},\ and\ \citenamefont {Xiang}}]{PASP}%
  \BibitemOpen
  \bibfield  {author} {\bibinfo {author} {\bibfnamefont {F.}~\bibnamefont
  {Lou}}, \bibinfo {author} {\bibfnamefont {X.~Y.}\ \bibnamefont {Li}},
  \bibinfo {author} {\bibfnamefont {J.~Y.}\ \bibnamefont {Ji}}, \bibinfo
  {author} {\bibfnamefont {H.~Y.}\ \bibnamefont {Yu}}, \bibinfo {author}
  {\bibfnamefont {J.~S.}\ \bibnamefont {Feng}}, \bibinfo {author}
  {\bibfnamefont {X.~G.}\ \bibnamefont {Gong}},\ and\ \bibinfo {author}
  {\bibfnamefont {H.~J.}\ \bibnamefont {Xiang}},\ }\bibfield  {title} {\bibinfo
  {title} {{$\mathrm{PASP: Property}$ analysis and simulation package for
  materials}},\ }\href {https://doi.org/10.1063/5.0043703} {\bibfield
  {journal} {\bibinfo  {journal} {J. Chem. Phys.}\ }\textbf {\bibinfo {volume}
  {154}},\ \bibinfo {pages} {114103} (\bibinfo {year} {2021})}\BibitemShut
  {NoStop}%
\bibitem [{\citenamefont {Liu}\ \emph {et~al.}(2024)\citenamefont {Liu},
  \citenamefont {Wang}, \citenamefont {Zhang}, \citenamefont {Pang},
  \citenamefont {Cheng}, \citenamefont {Zhang},\ and\ \citenamefont
  {Ji}}]{PhysRevB.109.214422}%
  \BibitemOpen
  \bibfield  {author} {\bibinfo {author} {\bibfnamefont {N.}~\bibnamefont
  {Liu}}, \bibinfo {author} {\bibfnamefont {C.}~\bibnamefont {Wang}}, \bibinfo
  {author} {\bibfnamefont {Y.}~\bibnamefont {Zhang}}, \bibinfo {author}
  {\bibfnamefont {F.}~\bibnamefont {Pang}}, \bibinfo {author} {\bibfnamefont
  {Z.}~\bibnamefont {Cheng}}, \bibinfo {author} {\bibfnamefont
  {Y.}~\bibnamefont {Zhang}},\ and\ \bibinfo {author} {\bibfnamefont
  {W.}~\bibnamefont {Ji}},\ }\bibfield  {title} {\bibinfo {title} {Intralayer
  strain tuned interlayer magnetism in bilayer $\mathrm{CrSBr}$},\ }\href
  {https://doi.org/10.1103/PhysRevB.109.214422} {\bibfield  {journal} {\bibinfo
   {journal} {Phys. Rev. B}\ }\textbf {\bibinfo {volume} {109}},\ \bibinfo
  {pages} {214422} (\bibinfo {year} {2024})}\BibitemShut {NoStop}%
\bibitem [{\citenamefont {Tao}\ \emph {et~al.}(2014)\citenamefont {Tao},
  \citenamefont {Guo}, \citenamefont {Yang},\ and\ \citenamefont
  {Zhang}}]{a54}%
  \BibitemOpen
  \bibfield  {author} {\bibinfo {author} {\bibfnamefont {P.}~\bibnamefont
  {Tao}}, \bibinfo {author} {\bibfnamefont {H.-H.}\ \bibnamefont {Guo}},
  \bibinfo {author} {\bibfnamefont {T.}~\bibnamefont {Yang}},\ and\ \bibinfo
  {author} {\bibfnamefont {Z.-D.}\ \bibnamefont {Zhang}},\ }\bibfield  {title}
  {\bibinfo {title} {Stacking stability of $\mathrm{MoS_2}$ bilayer:
  $\mathrm{An}$ ab initio study},\ }\href
  {https://doi.org/10.1088/1674-1056/23/10/106801} {\bibfield  {journal}
  {\bibinfo  {journal} {Chinese Phys. B}\ }\textbf {\bibinfo {volume} {23}},\
  \bibinfo {pages} {106801} (\bibinfo {year} {2014})}\BibitemShut {NoStop}%
\bibitem [{\citenamefont {Sui}\ \emph {et~al.}(2024)\citenamefont {Sui},
  \citenamefont {Li}, \citenamefont {Qi}, \citenamefont {Jin}, \citenamefont
  {Lv}, \citenamefont {Wu}, \citenamefont {Liu}, \citenamefont {Zheng},
  \citenamefont {Liu}, \citenamefont {Ge} \emph {et~al.}}]{sui2024atomic}%
  \BibitemOpen
  \bibfield  {author} {\bibinfo {author} {\bibfnamefont {F.}~\bibnamefont
  {Sui}}, \bibinfo {author} {\bibfnamefont {H.}~\bibnamefont {Li}}, \bibinfo
  {author} {\bibfnamefont {R.}~\bibnamefont {Qi}}, \bibinfo {author}
  {\bibfnamefont {M.}~\bibnamefont {Jin}}, \bibinfo {author} {\bibfnamefont
  {Z.}~\bibnamefont {Lv}}, \bibinfo {author} {\bibfnamefont {M.}~\bibnamefont
  {Wu}}, \bibinfo {author} {\bibfnamefont {X.}~\bibnamefont {Liu}}, \bibinfo
  {author} {\bibfnamefont {Y.}~\bibnamefont {Zheng}}, \bibinfo {author}
  {\bibfnamefont {B.}~\bibnamefont {Liu}}, \bibinfo {author} {\bibfnamefont
  {R.}~\bibnamefont {Ge}}, \emph {et~al.},\ }\bibfield  {title} {\bibinfo
  {title} {Atomic-level polarization reversal in sliding ferroelectric
  semiconductors},\ }\href {https://doi.org/doi.org/10.1038/s41467-024-48218-z}
  {\bibfield  {journal} {\bibinfo  {journal} {Nat. Commun.}\ }\textbf {\bibinfo
  {volume} {15}},\ \bibinfo {pages} {3799} (\bibinfo {year}
  {2024})}\BibitemShut {NoStop}%
\bibitem [{\citenamefont {Fang}\ \emph {et~al.}(2021)\citenamefont {Fang},
  \citenamefont {Wang}, \citenamefont {Zhang}, \citenamefont {Guo},
  \citenamefont {Ma},\ and\ \citenamefont {Wang}}]{FANG2021110688}%
  \BibitemOpen
  \bibfield  {author} {\bibinfo {author} {\bibfnamefont {H.}~\bibnamefont
  {Fang}}, \bibinfo {author} {\bibfnamefont {B.}~\bibnamefont {Wang}}, \bibinfo
  {author} {\bibfnamefont {X.}~\bibnamefont {Zhang}}, \bibinfo {author}
  {\bibfnamefont {Y.}~\bibnamefont {Guo}}, \bibinfo {author} {\bibfnamefont
  {L.}~\bibnamefont {Ma}},\ and\ \bibinfo {author} {\bibfnamefont
  {J.}~\bibnamefont {Wang}},\ }\bibfield  {title} {\bibinfo {title}
  {Theoretical study on two dimensional group $\mathrm{IV-VI}$ ternary
  compounds with large in-plane spontaneous polarization},\ }\href
  {https://doi.org/https://doi.org/10.1016/j.commatsci.2021.110688} {\bibfield
  {journal} {\bibinfo  {journal} {Comput. Mater. Sci.}\ }\textbf {\bibinfo
  {volume} {198}},\ \bibinfo {pages} {110688} (\bibinfo {year}
  {2021})}\BibitemShut {NoStop}%
\bibitem [{\citenamefont {Peng}(2023)}]{doi:10.1021/acs.nanolett.2c04497}%
  \BibitemOpen
  \bibfield  {author} {\bibinfo {author} {\bibfnamefont {B.}~\bibnamefont
  {Peng}},\ }\bibfield  {title} {\bibinfo {title} {Stability and strength of
  monolayer polymeric $\mathrm{C60}$},\ }\href
  {https://doi.org/10.1021/acs.nanolett.2c04497} {\bibfield  {journal}
  {\bibinfo  {journal} {Nano Lett.}\ }\textbf {\bibinfo {volume} {23}},\
  \bibinfo {pages} {652} (\bibinfo {year} {2023})}\BibitemShut {NoStop}%
\bibitem [{\citenamefont {Sahoo}\ \emph {et~al.}(2013)\citenamefont {Sahoo},
  \citenamefont {Gaur}, \citenamefont {Ahmadi}, \citenamefont {Guinel},\ and\
  \citenamefont {Katiyar}}]{doi:10.1021/jp402509w}%
  \BibitemOpen
  \bibfield  {author} {\bibinfo {author} {\bibfnamefont {S.}~\bibnamefont
  {Sahoo}}, \bibinfo {author} {\bibfnamefont {A.~P.~S.}\ \bibnamefont {Gaur}},
  \bibinfo {author} {\bibfnamefont {M.}~\bibnamefont {Ahmadi}}, \bibinfo
  {author} {\bibfnamefont {M.~J.-F.}\ \bibnamefont {Guinel}},\ and\ \bibinfo
  {author} {\bibfnamefont {R.~S.}\ \bibnamefont {Katiyar}},\ }\bibfield
  {title} {\bibinfo {title} {Temperature-dependent raman studies and thermal
  conductivity of few-layer $\mathrm{MoS2}$},\ }\href
  {https://doi.org/10.1021/jp402509w} {\bibfield  {journal} {\bibinfo
  {journal} {J. Phys. Chem. C}\ }\textbf {\bibinfo {volume} {117}},\ \bibinfo
  {pages} {9042} (\bibinfo {year} {2013})}\BibitemShut {NoStop}%
\bibitem [{\citenamefont {Peimyoo}\ \emph {et~al.}(2015)\citenamefont
  {Peimyoo}, \citenamefont {Shang}, \citenamefont {Yang}, \citenamefont {Wang},
  \citenamefont {Cong},\ and\ \citenamefont {Yu}}]{peimyoo2015thermal}%
  \BibitemOpen
  \bibfield  {author} {\bibinfo {author} {\bibfnamefont {N.}~\bibnamefont
  {Peimyoo}}, \bibinfo {author} {\bibfnamefont {J.}~\bibnamefont {Shang}},
  \bibinfo {author} {\bibfnamefont {W.}~\bibnamefont {Yang}}, \bibinfo {author}
  {\bibfnamefont {Y.}~\bibnamefont {Wang}}, \bibinfo {author} {\bibfnamefont
  {C.}~\bibnamefont {Cong}},\ and\ \bibinfo {author} {\bibfnamefont
  {T.}~\bibnamefont {Yu}},\ }\bibfield  {title} {\bibinfo {title} {Thermal
  conductivity determination of suspended mono-and bilayer $\mathrm{WS_2}$ by
  raman spectroscopy},\ }\href {https://doi.org/10.1007/s12274-014-0602-0}
  {\bibfield  {journal} {\bibinfo  {journal} {Nano Res.}\ }\textbf {\bibinfo
  {volume} {8}},\ \bibinfo {pages} {1210} (\bibinfo {year} {2015})}\BibitemShut
  {NoStop}%
\bibitem [{\citenamefont {Zhou}\ \emph
  {et~al.}(2019{\natexlab{c}})\citenamefont {Zhou}, \citenamefont {Liu},
  \citenamefont {Fan}, \citenamefont {Cao},\ and\ \citenamefont
  {Sheng}}]{PhysRevB.99.085410}%
  \BibitemOpen
  \bibfield  {author} {\bibinfo {author} {\bibfnamefont {Z.~Z.}\ \bibnamefont
  {Zhou}}, \bibinfo {author} {\bibfnamefont {H.~J.}\ \bibnamefont {Liu}},
  \bibinfo {author} {\bibfnamefont {D.~D.}\ \bibnamefont {Fan}}, \bibinfo
  {author} {\bibfnamefont {G.~H.}\ \bibnamefont {Cao}},\ and\ \bibinfo {author}
  {\bibfnamefont {C.~Y.}\ \bibnamefont {Sheng}},\ }\bibfield  {title} {\bibinfo
  {title} {High thermoelectric performance in the hexagonal bilayer structure
  consisting of light boron and phosphorus elements},\ }\href
  {https://doi.org/10.1103/PhysRevB.99.085410} {\bibfield  {journal} {\bibinfo
  {journal} {Phys. Rev. B}\ }\textbf {\bibinfo {volume} {99}},\ \bibinfo
  {pages} {085410} (\bibinfo {year} {2019}{\natexlab{c}})}\BibitemShut
  {NoStop}%
\bibitem [{\citenamefont {Christensen}\ \emph {et~al.}(2008)\citenamefont
  {Christensen}, \citenamefont {Abrahamsen}, \citenamefont {Christensen},
  \citenamefont {Juranyi}, \citenamefont {Andersen}, \citenamefont {Lefmann},
  \citenamefont {Andreasson}, \citenamefont {Bahl},\ and\ \citenamefont
  {Iversen}}]{christensen}%
  \BibitemOpen
  \bibfield  {author} {\bibinfo {author} {\bibfnamefont {M.}~\bibnamefont
  {Christensen}}, \bibinfo {author} {\bibfnamefont {A.~B.}\ \bibnamefont
  {Abrahamsen}}, \bibinfo {author} {\bibfnamefont {N.~B.}\ \bibnamefont
  {Christensen}}, \bibinfo {author} {\bibfnamefont {F.}~\bibnamefont
  {Juranyi}}, \bibinfo {author} {\bibfnamefont {N.~H.}\ \bibnamefont
  {Andersen}}, \bibinfo {author} {\bibfnamefont {K.}~\bibnamefont {Lefmann}},
  \bibinfo {author} {\bibfnamefont {J.}~\bibnamefont {Andreasson}}, \bibinfo
  {author} {\bibfnamefont {C.~R.~H.}\ \bibnamefont {Bahl}},\ and\ \bibinfo
  {author} {\bibfnamefont {B.~B.}\ \bibnamefont {Iversen}},\ }\bibfield
  {title} {\bibinfo {title} {Avoided crossing of rattler modes in
  thermoelectric materials},\ }\href {http://dx.doi.org/10.1038/nmat2273}
  {\bibfield  {journal} {\bibinfo  {journal} {Nat. Mater.}\ }\textbf {\bibinfo
  {volume} {7}},\ \bibinfo {pages} {811} (\bibinfo {year} {2008})}\BibitemShut
  {NoStop}%
\bibitem [{\citenamefont {Li}\ and\ \citenamefont
  {Mingo}(2015)}]{PhysRevB.91.144304}%
  \BibitemOpen
  \bibfield  {author} {\bibinfo {author} {\bibfnamefont {W.}~\bibnamefont
  {Li}}\ and\ \bibinfo {author} {\bibfnamefont {N.}~\bibnamefont {Mingo}},\
  }\bibfield  {title} {\bibinfo {title} {Ultralow lattice thermal conductivity
  of the fully filled skutterudite $\mathrm{YbFe_4Sb_{12}}$ due to the flat
  avoided-crossing filler modes},\ }\href
  {https://doi.org/10.1103/PhysRevB.91.144304} {\bibfield  {journal} {\bibinfo
  {journal} {Phys. Rev. B}\ }\textbf {\bibinfo {volume} {91}},\ \bibinfo
  {pages} {144304} (\bibinfo {year} {2015})}\BibitemShut {NoStop}%
\bibitem [{\citenamefont {Xie}\ \emph {et~al.}(2020)\citenamefont {Xie},
  \citenamefont {Feng}, \citenamefont {Li},\ and\ \citenamefont
  {He}}]{PhysRevLett.125.245901}%
  \BibitemOpen
  \bibfield  {author} {\bibinfo {author} {\bibfnamefont {L.}~\bibnamefont
  {Xie}}, \bibinfo {author} {\bibfnamefont {J.~H.}\ \bibnamefont {Feng}},
  \bibinfo {author} {\bibfnamefont {R.}~\bibnamefont {Li}},\ and\ \bibinfo
  {author} {\bibfnamefont {J.~Q.}\ \bibnamefont {He}},\ }\bibfield  {title}
  {\bibinfo {title} {First-principles study of anharmonic lattice dynamics in
  low thermal conductivity $\mathrm{AgCrSe_{2}: Evidence}$ for a large resonant
  four-phonon scattering},\ }\href
  {https://doi.org/10.1103/PhysRevLett.125.245901} {\bibfield  {journal}
  {\bibinfo  {journal} {Phys. Rev. Lett.}\ }\textbf {\bibinfo {volume} {125}},\
  \bibinfo {pages} {245901} (\bibinfo {year} {2020})}\BibitemShut {NoStop}%
\bibitem [{\citenamefont {Xu}\ \emph {et~al.}(2022{\natexlab{b}})\citenamefont
  {Xu}, \citenamefont {Wang}, \citenamefont {Chang}, \citenamefont {Chen},
  \citenamefont {Guan},\ and\ \citenamefont {Tao}}]{a45}%
  \BibitemOpen
  \bibfield  {author} {\bibinfo {author} {\bibfnamefont {X.}~\bibnamefont
  {Xu}}, \bibinfo {author} {\bibfnamefont {X.}~\bibnamefont {Wang}}, \bibinfo
  {author} {\bibfnamefont {P.}~\bibnamefont {Chang}}, \bibinfo {author}
  {\bibfnamefont {X.}~\bibnamefont {Chen}}, \bibinfo {author} {\bibfnamefont
  {L.}~\bibnamefont {Guan}},\ and\ \bibinfo {author} {\bibfnamefont
  {J.}~\bibnamefont {Tao}},\ }\bibfield  {title} {\bibinfo {title} {Strong
  spin-phonon coupling in two-dimensional magnetic semiconductor
  $\mathrm{CrSBr}$},\ }\href {https://doi.org/10.1021/acs.jpcc.2c02742}
  {\bibfield  {journal} {\bibinfo  {journal} {J. Phys. Chem. C}\ }\textbf
  {\bibinfo {volume} {126}},\ \bibinfo {pages} {10574} (\bibinfo {year}
  {2022}{\natexlab{b}})}\BibitemShut {NoStop}%
\bibitem [{\citenamefont {Scheie}\ \emph {et~al.}(2022)\citenamefont {Scheie},
  \citenamefont {Ziebel}, \citenamefont {Chica}, \citenamefont {Bae},
  \citenamefont {Wang}, \citenamefont {Kolesnikov}, \citenamefont {Zhu},\ and\
  \citenamefont {Roy}}]{a47}%
  \BibitemOpen
  \bibfield  {author} {\bibinfo {author} {\bibfnamefont {A.}~\bibnamefont
  {Scheie}}, \bibinfo {author} {\bibfnamefont {M.}~\bibnamefont {Ziebel}},
  \bibinfo {author} {\bibfnamefont {D.~G.}\ \bibnamefont {Chica}}, \bibinfo
  {author} {\bibfnamefont {Y.~J.}\ \bibnamefont {Bae}}, \bibinfo {author}
  {\bibfnamefont {X.}~\bibnamefont {Wang}}, \bibinfo {author} {\bibfnamefont
  {A.~I.}\ \bibnamefont {Kolesnikov}}, \bibinfo {author} {\bibfnamefont
  {X.}~\bibnamefont {Zhu}},\ and\ \bibinfo {author} {\bibfnamefont
  {X.}~\bibnamefont {Roy}},\ }\bibfield  {title} {\bibinfo {title} {Spin waves
  and magnetic exchange hamiltonian in $\mathrm{CrSBr}$},\ }\href
  {https://doi.org/10.1002/advs.202202467} {\bibfield  {journal} {\bibinfo
  {journal} {Adv. Sci.}\ }\textbf {\bibinfo {volume} {9}},\ \bibinfo {pages}
  {2202467} (\bibinfo {year} {2022})}\BibitemShut {NoStop}%
\bibitem [{\citenamefont {Goodenough}(1955)}]{g1}%
  \BibitemOpen
  \bibfield  {author} {\bibinfo {author} {\bibfnamefont {J.~B.}\ \bibnamefont
  {Goodenough}},\ }\bibfield  {title} {\bibinfo {title} {Theory of the role of
  covalence in the perovskite-type manganites [$\mathrm{La,
  M}(\mathrm{II})$]$\mathrm{MnO_{3}}$},\ }\href
  {https://doi.org/10.1103/PhysRev.100.564} {\bibfield  {journal} {\bibinfo
  {journal} {Phys. Rev.}\ }\textbf {\bibinfo {volume} {100}},\ \bibinfo {pages}
  {564} (\bibinfo {year} {1955})}\BibitemShut {NoStop}%
\bibitem [{\citenamefont {Kanamori}(1960)}]{g2}%
  \BibitemOpen
  \bibfield  {author} {\bibinfo {author} {\bibfnamefont {J.}~\bibnamefont
  {Kanamori}},\ }\bibfield  {title} {\bibinfo {title} {Crystal distortion in
  magnetic compounds},\ }\href {https://doi.org/10.1063/1.1984590} {\bibfield
  {journal} {\bibinfo  {journal} {J. Appl. Phys.}\ }\textbf {\bibinfo {volume}
  {31}},\ \bibinfo {pages} {S14} (\bibinfo {year} {1960})}\BibitemShut
  {NoStop}%
\bibitem [{\citenamefont {Anderson}(1959)}]{g3}%
  \BibitemOpen
  \bibfield  {author} {\bibinfo {author} {\bibfnamefont {P.~W.}\ \bibnamefont
  {Anderson}},\ }\bibfield  {title} {\bibinfo {title} {New approach to the
  theory of superexchange interactions},\ }\href
  {https://doi.org/10.1103/PhysRev.115.2} {\bibfield  {journal} {\bibinfo
  {journal} {Phys. Rev.}\ }\textbf {\bibinfo {volume} {115}},\ \bibinfo {pages}
  {2} (\bibinfo {year} {1959})}\BibitemShut {NoStop}%
\bibitem [{\citenamefont {Wang}\ \emph
  {et~al.}(2023{\natexlab{b}})\citenamefont {Wang}, \citenamefont {Luo},
  \citenamefont {Zeng}, \citenamefont {Tang},\ and\ \citenamefont
  {Chen}}]{PRB-SSE-Path}%
  \BibitemOpen
  \bibfield  {author} {\bibinfo {author} {\bibfnamefont {Y.}~\bibnamefont
  {Wang}}, \bibinfo {author} {\bibfnamefont {N.}~\bibnamefont {Luo}}, \bibinfo
  {author} {\bibfnamefont {J.}~\bibnamefont {Zeng}}, \bibinfo {author}
  {\bibfnamefont {L.-M.}\ \bibnamefont {Tang}},\ and\ \bibinfo {author}
  {\bibfnamefont {K.-Q.}\ \bibnamefont {Chen}},\ }\bibfield  {title} {\bibinfo
  {title} {Magnetic anisotropy and electric field induced magnetic phase
  transition in the van der $\mathrm{W}$aals antiferromagnet
  $\mathrm{CrSBr}$},\ }\href {https://doi.org/10.1103/PhysRevB.108.054401}
  {\bibfield  {journal} {\bibinfo  {journal} {Phys. Rev. B}\ }\textbf {\bibinfo
  {volume} {108}},\ \bibinfo {pages} {054401} (\bibinfo {year}
  {2023}{\natexlab{b}})}\BibitemShut {NoStop}%
\bibitem [{\citenamefont {Wang}\ \emph
  {et~al.}(2020{\natexlab{c}})\citenamefont {Wang}, \citenamefont {Zhou},
  \citenamefont {Zhou}, \citenamefont {Pan}, \citenamefont {Lu}, \citenamefont
  {Wan}, \citenamefont {Wang},\ and\ \citenamefont {Ji}}]{PhysRevB.102.020402}%
  \BibitemOpen
  \bibfield  {author} {\bibinfo {author} {\bibfnamefont {C.}~\bibnamefont
  {Wang}}, \bibinfo {author} {\bibfnamefont {X.}~\bibnamefont {Zhou}}, \bibinfo
  {author} {\bibfnamefont {L.}~\bibnamefont {Zhou}}, \bibinfo {author}
  {\bibfnamefont {Y.}~\bibnamefont {Pan}}, \bibinfo {author} {\bibfnamefont
  {Z.-Y.}\ \bibnamefont {Lu}}, \bibinfo {author} {\bibfnamefont
  {X.}~\bibnamefont {Wan}}, \bibinfo {author} {\bibfnamefont {X.}~\bibnamefont
  {Wang}},\ and\ \bibinfo {author} {\bibfnamefont {W.}~\bibnamefont {Ji}},\
  }\bibfield  {title} {\bibinfo {title} {Bethe-slater-curve-like behavior and
  interlayer spin-exchange coupling mechanisms in two-dimensional magnetic
  bilayers},\ }\href {https://doi.org/10.1103/PhysRevB.102.020402} {\bibfield
  {journal} {\bibinfo  {journal} {Phys. Rev. B}\ }\textbf {\bibinfo {volume}
  {102}},\ \bibinfo {pages} {020402} (\bibinfo {year}
  {2020}{\natexlab{c}})}\BibitemShut {NoStop}%
\bibitem [{\citenamefont {Mogulkoc}\ \emph {et~al.}(2021)\citenamefont
  {Mogulkoc}, \citenamefont {Modarresi},\ and\ \citenamefont
  {Rudenko}}]{PhysRevApplied.15.064053}%
  \BibitemOpen
  \bibfield  {author} {\bibinfo {author} {\bibfnamefont {A.}~\bibnamefont
  {Mogulkoc}}, \bibinfo {author} {\bibfnamefont {M.}~\bibnamefont
  {Modarresi}},\ and\ \bibinfo {author} {\bibfnamefont {A.}~\bibnamefont
  {Rudenko}},\ }\bibfield  {title} {\bibinfo {title} {Two-dimensional chromium
  bismuthate: A room-temperature ising ferromagnet with tunable magneto-optical
  response},\ }\href {https://doi.org/10.1103/PhysRevApplied.15.064053}
  {\bibfield  {journal} {\bibinfo  {journal} {Phys. Rev. Appl.}\ }\textbf
  {\bibinfo {volume} {15}},\ \bibinfo {pages} {064053} (\bibinfo {year}
  {2021})}\BibitemShut {NoStop}%
\bibitem [{\citenamefont {Mogulkoc}\ \emph {et~al.}(2020)\citenamefont
  {Mogulkoc}, \citenamefont {Modarresi},\ and\ \citenamefont
  {Rudenko}}]{PhysRevB.102.024441}%
  \BibitemOpen
  \bibfield  {author} {\bibinfo {author} {\bibfnamefont {A.}~\bibnamefont
  {Mogulkoc}}, \bibinfo {author} {\bibfnamefont {M.}~\bibnamefont
  {Modarresi}},\ and\ \bibinfo {author} {\bibfnamefont {A.~N.}\ \bibnamefont
  {Rudenko}},\ }\bibfield  {title} {\bibinfo {title} {Two-dimensional chromium
  pnictides $\mathrm{CrX} (\text{X=P},\mathrm{As},\mathrm{Sb})$: Half-metallic
  ferromagnets with high $\mathrm{Curie}$ temperature},\ }\href
  {https://doi.org/10.1103/PhysRevB.102.024441} {\bibfield  {journal} {\bibinfo
   {journal} {Phys. Rev. B}\ }\textbf {\bibinfo {volume} {102}},\ \bibinfo
  {pages} {024441} (\bibinfo {year} {2020})}\BibitemShut {NoStop}%
\bibitem [{\citenamefont {Rudenko}\ \emph
  {et~al.}(2023{\natexlab{b}})\citenamefont {Rudenko}, \citenamefont
  {R{\"o}sner},\ and\ \citenamefont {Katsnelson}}]{rudenko2023dielectric}%
  \BibitemOpen
  \bibfield  {author} {\bibinfo {author} {\bibfnamefont {A.~N.}\ \bibnamefont
  {Rudenko}}, \bibinfo {author} {\bibfnamefont {M.}~\bibnamefont
  {R{\"o}sner}},\ and\ \bibinfo {author} {\bibfnamefont {M.~I.}\ \bibnamefont
  {Katsnelson}},\ }\bibfield  {title} {\bibinfo {title} {Dielectric tunability
  of magnetic properties in orthorhombic ferromagnetic monolayer
  $\mathrm{CrSBr}$},\ }\href
  {https://doi.org/doi.org/10.1038/s41524-023-01050-3} {\bibfield  {journal}
  {\bibinfo  {journal} {npj Comput. Mater.}\ }\textbf {\bibinfo {volume} {9}},\
  \bibinfo {pages} {83} (\bibinfo {year} {2023}{\natexlab{b}})}\BibitemShut
  {NoStop}%
\bibitem [{\citenamefont {Fu}\ \emph {et~al.}(2024)\citenamefont {Fu},
  \citenamefont {Huang}, \citenamefont {Samarawickrama}, \citenamefont
  {Watanabe}, \citenamefont {Taniguchi}, \citenamefont {Wang}, \citenamefont
  {Ackerman}, \citenamefont {Zang}, \citenamefont {Yu},\ and\ \citenamefont
  {Tian}}]{https://doi.org/10.1002/apxr.202400052}%
  \BibitemOpen
  \bibfield  {author} {\bibinfo {author} {\bibfnamefont {Z.}~\bibnamefont
  {Fu}}, \bibinfo {author} {\bibfnamefont {H.-F.}\ \bibnamefont {Huang}},
  \bibinfo {author} {\bibfnamefont {P.}~\bibnamefont {Samarawickrama}},
  \bibinfo {author} {\bibfnamefont {K.}~\bibnamefont {Watanabe}}, \bibinfo
  {author} {\bibfnamefont {T.}~\bibnamefont {Taniguchi}}, \bibinfo {author}
  {\bibfnamefont {W.}~\bibnamefont {Wang}}, \bibinfo {author} {\bibfnamefont
  {J.}~\bibnamefont {Ackerman}}, \bibinfo {author} {\bibfnamefont
  {J.}~\bibnamefont {Zang}}, \bibinfo {author} {\bibfnamefont {J.-X.}\
  \bibnamefont {Yu}},\ and\ \bibinfo {author} {\bibfnamefont {J.}~\bibnamefont
  {Tian}},\ }\bibfield  {title} {\bibinfo {title} {Anomalous tunneling
  magnetoresistance oscillation and electrically tunable tunneling anisotropic
  magnetoresistance in few-layer $\mathrm{CrPS_4}$},\ }\href
  {https://doi.org/https://doi.org/10.1002/apxr.202400052} {\bibfield
  {journal} {\bibinfo  {journal} {Adv. Phys. Res.}\ }\textbf {\bibinfo {volume}
  {3}},\ \bibinfo {pages} {2400052} (\bibinfo {year} {2024})}\BibitemShut
  {NoStop}%
\bibitem [{\citenamefont {Wu}\ \emph {et~al.}(2022)\citenamefont {Wu},
  \citenamefont {Gutiérrez-Lezama}, \citenamefont {López-Paz}, \citenamefont
  {Gibertini}, \citenamefont {Watanabe}, \citenamefont {Taniguchi},
  \citenamefont {von Rohr}, \citenamefont {Ubrig},\ and\ \citenamefont
  {Morpurgo}}]{https://doi.org/10.1002/adma.202109759}%
  \BibitemOpen
  \bibfield  {author} {\bibinfo {author} {\bibfnamefont {F.}~\bibnamefont
  {Wu}}, \bibinfo {author} {\bibfnamefont {I.}~\bibnamefont
  {Gutiérrez-Lezama}}, \bibinfo {author} {\bibfnamefont {S.~A.}\ \bibnamefont
  {López-Paz}}, \bibinfo {author} {\bibfnamefont {M.}~\bibnamefont
  {Gibertini}}, \bibinfo {author} {\bibfnamefont {K.}~\bibnamefont {Watanabe}},
  \bibinfo {author} {\bibfnamefont {T.}~\bibnamefont {Taniguchi}}, \bibinfo
  {author} {\bibfnamefont {F.~O.}\ \bibnamefont {von Rohr}}, \bibinfo {author}
  {\bibfnamefont {N.}~\bibnamefont {Ubrig}},\ and\ \bibinfo {author}
  {\bibfnamefont {A.~F.}\ \bibnamefont {Morpurgo}},\ }\bibfield  {title}
  {\bibinfo {title} {$\mathrm{Quasi-1D}$ electronic transport in a
  2$\mathrm{D}$ magnetic semiconductor},\ }\href
  {https://doi.org/https://doi.org/10.1002/adma.202109759} {\bibfield
  {journal} {\bibinfo  {journal} {Adv. Mater.}\ }\textbf {\bibinfo {volume}
  {34}},\ \bibinfo {pages} {2109759} (\bibinfo {year} {2022})}\BibitemShut
  {NoStop}%
\bibitem [{\citenamefont {Kundys}(2015)}]{10.1063/1.4905505}%
  \BibitemOpen
  \bibfield  {author} {\bibinfo {author} {\bibfnamefont {B.}~\bibnamefont
  {Kundys}},\ }\bibfield  {title} {\bibinfo {title} {Photostrictive
  materials},\ }\href {https://doi.org/10.1063/1.4905505} {\bibfield  {journal}
  {\bibinfo  {journal} {Appl. Phys. Rev.}\ }\textbf {\bibinfo {volume} {2}},\
  \bibinfo {pages} {011301} (\bibinfo {year} {2015})}\BibitemShut {NoStop}%
\bibitem [{\citenamefont {Haleoot}\ \emph {et~al.}(2017)\citenamefont
  {Haleoot}, \citenamefont {Paillard}, \citenamefont {Kaloni}, \citenamefont
  {Mehboudi}, \citenamefont {Xu}, \citenamefont {Bellaiche},\ and\
  \citenamefont {Barraza-Lopez}}]{PhysRevLett.118.227401}%
  \BibitemOpen
  \bibfield  {author} {\bibinfo {author} {\bibfnamefont {R.}~\bibnamefont
  {Haleoot}}, \bibinfo {author} {\bibfnamefont {C.}~\bibnamefont {Paillard}},
  \bibinfo {author} {\bibfnamefont {T.~P.}\ \bibnamefont {Kaloni}}, \bibinfo
  {author} {\bibfnamefont {M.}~\bibnamefont {Mehboudi}}, \bibinfo {author}
  {\bibfnamefont {B.}~\bibnamefont {Xu}}, \bibinfo {author} {\bibfnamefont
  {L.}~\bibnamefont {Bellaiche}},\ and\ \bibinfo {author} {\bibfnamefont
  {S.}~\bibnamefont {Barraza-Lopez}},\ }\bibfield  {title} {\bibinfo {title}
  {Photostrictive two-dimensional materials in the monochalcogenide family},\
  }\href {https://doi.org/10.1103/PhysRevLett.118.227401} {\bibfield  {journal}
  {\bibinfo  {journal} {Phys. Rev. Lett.}\ }\textbf {\bibinfo {volume} {118}},\
  \bibinfo {pages} {227401} (\bibinfo {year} {2017})}\BibitemShut {NoStop}%
\bibitem [{\citenamefont {Li}\ \emph {et~al.}(2024)\citenamefont {Li},
  \citenamefont {Varrassi}, \citenamefont {Yang}, \citenamefont {Franchini},
  \citenamefont {Bellaiche},\ and\ \citenamefont
  {He}}]{doi:10.1021/jacs.4c03296}%
  \BibitemOpen
  \bibfield  {author} {\bibinfo {author} {\bibfnamefont {Z.}~\bibnamefont
  {Li}}, \bibinfo {author} {\bibfnamefont {L.}~\bibnamefont {Varrassi}},
  \bibinfo {author} {\bibfnamefont {Y.}~\bibnamefont {Yang}}, \bibinfo {author}
  {\bibfnamefont {C.}~\bibnamefont {Franchini}}, \bibinfo {author}
  {\bibfnamefont {L.}~\bibnamefont {Bellaiche}},\ and\ \bibinfo {author}
  {\bibfnamefont {J.}~\bibnamefont {He}},\ }\bibfield  {title} {\bibinfo
  {title} {Ultrastrong coupling between polar distortion and optical properties
  in ferroelectric $\mathrm{MoBr_2O_2}$},\ }\href
  {https://doi.org/10.1021/jacs.4c03296} {\bibfield  {journal} {\bibinfo
  {journal} {J. Am. Chem. Soc.}\ }\textbf {\bibinfo {volume} {146}},\ \bibinfo
  {pages} {15411} (\bibinfo {year} {2024})}\BibitemShut {NoStop}%
\bibitem [{\citenamefont {Gao}\ \emph {et~al.}(2023)\citenamefont {Gao},
  \citenamefont {Jiang}, \citenamefont {Qiu},\ and\ \citenamefont
  {Zhao}}]{a50}%
  \BibitemOpen
  \bibfield  {author} {\bibinfo {author} {\bibfnamefont {Y.}~\bibnamefont
  {Gao}}, \bibinfo {author} {\bibfnamefont {X.}~\bibnamefont {Jiang}}, \bibinfo
  {author} {\bibfnamefont {Z.}~\bibnamefont {Qiu}},\ and\ \bibinfo {author}
  {\bibfnamefont {J.}~\bibnamefont {Zhao}},\ }\bibfield  {title} {\bibinfo
  {title} {Photoexcitation induced magnetic phase transition and spin dynamics
  in antiferromagnetic $\mathrm{MnPS_3}$ monolayer},\ }\href
  {https://doi.org/10.1038/s41524-023-01071-y} {\bibfield  {journal} {\bibinfo
  {journal} {npj Comput. Mater.}\ }\textbf {\bibinfo {volume} {9}},\ \bibinfo
  {pages} {107} (\bibinfo {year} {2023})}\BibitemShut {NoStop}%
\bibitem [{\citenamefont {Chen}\ \emph {et~al.}(2023)\citenamefont {Chen},
  \citenamefont {Li}, \citenamefont {Yu}, \citenamefont {Yang}, \citenamefont
  {Jin}, \citenamefont {Huang},\ and\ \citenamefont {Xiang}}]{chen2023light}%
  \BibitemOpen
  \bibfield  {author} {\bibinfo {author} {\bibfnamefont {J.}~\bibnamefont
  {Chen}}, \bibinfo {author} {\bibfnamefont {Y.}~\bibnamefont {Li}}, \bibinfo
  {author} {\bibfnamefont {H.}~\bibnamefont {Yu}}, \bibinfo {author}
  {\bibfnamefont {Y.}~\bibnamefont {Yang}}, \bibinfo {author} {\bibfnamefont
  {H.}~\bibnamefont {Jin}}, \bibinfo {author} {\bibfnamefont {B.}~\bibnamefont
  {Huang}},\ and\ \bibinfo {author} {\bibfnamefont {H.}~\bibnamefont {Xiang}},\
  }\bibfield  {title} {\bibinfo {title} {Light-induced magnetic phase
  transition in van der $\mathrm{W}$aals antiferromagnets},\ }\href
  {https://doi.org/10.1007/s11433-022-2085-x} {\bibfield  {journal} {\bibinfo
  {journal} {Sci. China Phys. Mech. Astron.}\ }\textbf {\bibinfo {volume}
  {66}},\ \bibinfo {pages} {277511} (\bibinfo {year} {2023})}\BibitemShut
  {NoStop}%
\end{thebibliography}%

\end{document}